\documentstyle[psfig,onecolumn]{mn}
\newif\ifAMStwofonts

\newcommand{\etal}{et al. }
\newcommand{\adhoc}{{\it ad hoc} }

\newcommand{\mytorus}{{\sc mytorus} }
\newcommand{\mytorusp}{{\sc mytorus}}
\newcommand{\comptt}{{\sc comptt} }
\newcommand{\compttp}{{\sc comptt}}
\newcommand{\rxte}{{\it RXTE} }

\newcommand{\chandra}{{\it Chandra} }
\newcommand{\chandrap}{{\it Chandra}}
\newcommand{\suzaku}{{\it Suzaku} }
\newcommand{\suzakup}{{\it Suzaku}}
\newcommand{\bsax}{{\it BeppoSAX} }
\newcommand{\bsaxp}{{\it BeppoSAX}}
\newcommand{\swift}{{\it Swift} }

\newcommand{\feka}{{Fe~K$\alpha$} }

\newcommand{\fekalfa}{{Fe~K$\alpha$} }
\newcommand{\fekalfap}{{Fe~K$\alpha$}}

\newcommand{\fekbeta}{{Fe~K$\beta$} }
\newcommand{\fekbetap}{{Fe~K$\beta$}}
\newcommand{\nika}{{Ni~K$\alpha$} }
\newcommand{\nikap}{{Ni~K$\alpha$}}

\newcommand{\fexxv}{Fe~{\sc xxv} }
\newcommand{\fexxvp}{Fe~{\sc xxv}}
\newcommand{\fexxvres}{Fe~{\sc xxv}(r) }

\newcommand{\fexxvi}{Fe~{\sc xxvi} }

\newcommand{\thetaobsp}{{$\theta_{\rm obs}$}}

\newcommand{\tablectratesp}{{Table~1}}

\newcommand{\tablemytdefparsp}{{Table~2}}
\newcommand{\tablemytdecoup}{{Table~3} }
\newcommand{\tablemytdecoupp}{{Table~3}}
\newcommand{\tablebatresults}{{Table~4} }
\newcommand{\tablebatresultsp}{{Table~4}}
\newcommand{\tablemytcpplresults}{{Table~5} }
\newcommand{\tablemytcpplresultsp}{{Table~5}}
\newcommand{\tablemytdcresults}{{Table~6} }
\newcommand{\tablemytdcresultsp}{{Table~6}}
\newcommand{\tablelumratios}{{Table~7} }
\newcommand{\tablelumratiosp}{{Table~7}}
\newcommand{\tableloverledd}{{Table~8} }
\newcommand{\tableloverleddp}{{Table~8}}

\ifoldfss
  \ifCUPmtlplainloaded \else
    \NewTextAlphabet{textbfit} {cmbxti10} {}
    \NewTextAlphabet{textbfss} {cmssbx10} {}
    \NewMathAlphabet{mathbfit} {cmbxti10} {} 
    \NewMathAlphabet{mathbfss} {cmssbx10} {} 
  \fi
  \ifAMStwofonts
    \ifCUPmtlplainloaded \else
      \NewSymbolFont{upmath} {eurm10}
      \NewSymbolFont{AMSa} {msam10}
      \NewMathSymbol{\upi}     {0}{upmath}{19}
      \NewMathSymbol{\umu}     {0}{upmath}{16}
      \NewMathSymbol{\upartial}{0}{upmath}{40}
      \NewMathSymbol{\leqslant}{3}{AMSa}{36}
      \NewMathSymbol{\geqslant}{3}{AMSa}{3E}

    \fi
  \fi
\fi 

\ifCUPmtlplainloaded \else
  \ifAMStwofonts \else 
    \def\upi{\pi}
    \def\umu{\mu}
    \def\upartial{\partial}
  \fi
\fi

\title{The Nature of the Compton-thick X-ray Reprocessor in NGC~4945}
\author[T. Yaqoob]{Tahir Yaqoob$^{1}$ \\
	$^1$Department of Physics and Astronomy, Johns Hopkins University, 3400 N. Charles St., Baltimore, MD 21218. \\
}
\date{
     Accepted for publication in MNRAS 18 April, 2012  }

\pagerange{\pageref{firstpage}--\pageref{lastpage}}
\pubyear{2012}

\begin{document}

\maketitle

\label{firstpage}

\begin{abstract} 

We present an exhaustive methodology for fitting Compton-thick X-ray
reprocessor models to obscured active galactic nuclei (AGNs) and for interpreting 
the results. 
We focus on the \mytorus model but also include some analysis
from other models. 
The models are applied specifically to \suzakup, \bsaxp,
and \swift BAT spectra of the Seyfert 2 galaxy  
NGC~4945 but the basic methodology is applicable to other AGNs,
including Compton-thin sources. The models overcome a major
restriction of disk-reflection models, namely the assumption of an
infinite column density. Finite column-density models produce
a rich variety of spectral shapes and characteristics that
cannot be produced by disk-reflection models, even for
Compton-thin AGN with column densities in the range $\sim 10^{23}$--$10^{24} \ \rm cm^{-2}$.
In the Compton-thick regime
we show that even though NGC~4945 is one of the brightest AGNs above 10~keV,
there are significant spectral degeneracies that correspond to very
different physical scenarios. The models that fit the data span
nearly a factor of 3 in column density ($\sim 2$ to $6 \times 10^{24} \ \rm
cm^{-2}$) and two orders of magnitude in the intrinsic 2--195~keV
luminosity. Models in which the continuum above 10~keV is dominated
by the direct (unscattered) continuum give the highest intrinsic
luminosities and column densities. Models in which the Compton-scattered
continuum dominates the spectrum above 10~keV give the lowest
intrinsic luminosities and column densities. Utilizing variability
information from other studies of NGC~4945, namely the fact that
the \fekalfa emission line does not vary whilst the continuum
above 10~keV varies significantly, we can select the solutions
in which the direct continuum dominates above 10~keV.  
The data require that
the Compton-scattered continuum and \fekalfa line emission come
predominantly from the illuminated surfaces of the X-ray reprocessor,
implying a clumpy medium with a global covering factor that is
small enough that the Compton-scattered continuum does not
dominate the spectrum above 10~keV. The line of sight may be
obscured by matter in the same distribution but a separate
ring-like structure observed edge-on is not ruled out.
The \fekalfa line-emitting region must be the same one 
recently reported to be spatially-resolved
by \chandrap, so it must be extended on a scale of $\sim 30$~pc or so.
As found in previous studies of NGC~4945, the implied intrinsic
bolometric luminosity is close to, or greater than, the Eddington
luminosity. However, a scenario that is also consistent with the data
and the models is that NGC~4945 is a strongly beamed AGN embedded in a
shell of Compton-thick (but clumpy) matter, with a covering factor
that needs less fine-tuning than the case of an isotropic
intrinsic X-ray continuum. The intensity of 
the intrinsic X-ray
continuum would be strongly aligned along or close to the line of sight,
so that the true intrinsic luminosity could 
easily be an order of magnitude less than that deduced for
an isotropic X-ray source. Beaming also appears to be consistent
with recent radio and Fermi results for NGC~4945. Such beamed Compton-thick AGNs would
be preferentially selected in hard X-ray surveys over unbeamed
Compton-thick AGNs. 

\end{abstract}


\section{Introduction}
\label{intro}

Obscured active galactic nuclei (AGNs) are of broad astrophysical
interest because it is thought that a significant fraction of the
power from accretion onto black holes is shrouded by a veil of
circumnuclear matter (e.g., Fabian 1999), and such a population of
obscured AGNs may play a significant role in contributing to
the Cosmic X-ray Background (e.g., Gilli, Comastri, \& Hasinger 2007, and
references therein). Moreover, the connection between
obscuration in AGNs and starburst activity is an area that still requires
elucidation. Modeling the properties of the obscuring structure
is also critical for understanding the unification of type~1 and type~2 AGN.
Compton-thick obscured AGN, in which the circumnuclear matter has
a Compton-scattering optical depth of $\sim 1$ or greater (at energies
of $\sim 10$~keV) have been particularly difficult to study in the 
X-ray band. This is not only because of the relative weakness of the
sources compared to unobscured AGNs, but also 
because the reprocessing of the incident
X-rays in the obscuring medium affects such a large range in energy
that the intrinsic, direct continuum may not be observed anywhere
in the observational bandpass of a given instrument or set of 
instruments. The reprocessed X-ray spectrum
is characterized by significant continuum curvature that
peaks between $\sim 10$--50~keV, and often a strong fluorescent \fekalfa
emission line. However, the detailed shape of the
X-ray spectrum depends on many factors, including the geometry and
orientation of the reprocessor, and the shape of the incident
continuum itself. 

NGC~4945 is a nearby ($z=0.001878$) Seyfert~2 galaxy that has
been known for some time to be obscured by Compton-thick matter
(see Done, Madejski, and Smith 1996) and has been observed
by every X-ray astronomy satellite since {\it Ginga}.
The source is one of the brightest AGNs above $\sim 10$~keV, yet is
an order of magnitude weaker below 10~keV, a property that is
characteristic of obscuration by material with a column density 
in the line of sight of the
order of $10^{24} \rm \ cm^{-2}$ but less than $10^{25} \rm \ cm^{-2}$.
\bsax was first able to obtain a broadband X-ray spectrum
extending up to $\sim 100$~keV with good sensitivity
(Guainazzi \etal 2000; Dadina 2007).
A recent \rxte study that included NGC~4945 observations
spanning a period of about a decade shows a consistent 2--10~keV flux level
over that period of time, which is also consistent with flux
levels in this energy band before the \rxte observations
(Rivers, Markowitz, and Rothschild 2011).
However, above 10~keV, NGC~4945 is much brighter and
highly variable, and it is one of the brightest
AGNs in the 14--195~keV \swift BAT all-sky survey (e.g., see 
Winter \etal 2008, 2009; Tueller \etal 2009, 2010). NGC~4945 is also one
of only two Seyfert 2 galaxies that is detected in the GeV band with Fermi,
the other being NGC~1068 (Lenain \etal 2011).

In 2006, \suzaku provided the best
broadband spectrum in the $\sim 1-150$~keV band, in terms of the
combination of high sensitivity above 10~keV and good spectral
resolution in the critical $\sim 6-8$~keV region that contains the 
\fekalfap, \fekbetap, \nika emission lines, and the Fe~K absorption edge,
features which can potentially provide powerful constraints on models.
Itoh \etal (2008) and Fukazawa \etal (2011) presented results of modeling the \suzaku data
using \adhoc models consisting of line-of-sight extinction that
does not correctly model the Compton-scattering cross section,
an X-ray reflection model based on a point-source illuminating a
semi-infinite slab, and discrete Gaussian components for
the fluorescent emission lines that were allowed to have arbitrary
fluxes. This type of model has been universally applied
to the X-ray spectra of both type~1 and type~2 AGN for
over 15 years. However, such a model of the Compton-thick obscuring matter is
not physical and the X-ray reflection continuum model does not
have a column density parameter because it is assumed to be infinite.
Therefore, the matter out of the line of sight responsible
for the reflection continuum (produced by Compton-scattering
and absorption) cannot be related
to the column density along the line of sight, and the physics
relating the fluorescent line fluxes to the Compton-scattered continuum
is forsaken. Moreover, the intrinsic continuum luminosity inferred
from these \adhoc models is not straightforward to interpret, and
as we shall show in the present paper, could be wrong by an order of magnitude
or more. A further sacrifice that has to be made when using a disk-reflection
model as a substitute for the true Compton-scattered continuum from
a toroidal or spherical reprocessor is that one is forced to choose
an arbitrary inclination angle for the disk. Yet, the shape of the
reflection spectrum is sensitive to geometry and to the orientation of
the reprocessor with respect to the observer. Different authors have
adopted different values for the inclination angle, so different studies
in the literature may not even be directly comparable.

More recently, Marinucci \etal (2012) have presented results from 5
new \suzaku monitoring observations of NGC~4945. 
They found that the total \fekalfa emission-line flux varies
by less than 10\% whilst the continuum above 10~keV varied by 
a factor of $\sim 2$. Marinucci \etal (2012) also showed using
\chandra data that a component of the \fekalfa line emission
is spatially resolved on a scale of at least 30~pc.
These findings are important for distinguishing between various
degenerate models.
However, it is beyond the scope of the present paper to reanalyze
the new \suzaku observations. Rather, our purpose is to lay out the
methodology in detail using the first long \suzaku observation of 2006,
the \bsax observation of 1999, and the \swift BAT all-sky survey data.
By considering a variety of possible scenarios and the associated
issues involved, the methodology will help to model and
interpret data from other obscured AGNs,
which in general will have a
lower signal-to-noise ratio than the NGC~4945 data.

In summary, the currently popular scheme for spectral-fitting analysis of Compton-thick
AGN does not extract all of the physical information contained in the data, and
what is extracted may not have a straightforward physical meaning (if any).
Murphy and Yaqoob (2009, hereafter MY09) described and made available for general use,
a toroidal model (called {\sc mytorus}) of the Compton-thick X-ray reprocessor in AGN that 
addresses some of these limitations. Applying such a model entails
many complexities because
there may be several different types of degeneracy in the data.
In the present paper we give an exhaustive account of the application of
the \mytorus model to noncontemporaneous \suzakup, \bsaxp, and \swift BAT spectra of NGC~4945.
We use the \bsax data in addition to the \suzaku data because of
the broadband coverage and good sensitivity above 10~keV of \bsaxp,
even though the spectral resolution is not as good as \suzakup. The \swift
BAT data provide information on long-term variability of the very high-energy
continuum, as well as a long-term (58-month) average of the 14--195~keV spectrum that 
serves as a useful baseline. 
We also apply the toroidal and spherical models that were made available
after \mytorus by Brightman and Nandra (2011; hereafter BN11), although
these models are more restrictive than the \mytorus model because
they do not allow for time delays between the different model components. 
Such a generalized detailed investigation is necessary to fully interpret the NGC~4945 data,
and to establish a methodology for the application and interpretation of Compton-thick
X-ray reprocessor models to other AGN.
NGC~4945 is an excellent candidate for a prototype Compton-thick AGN because
it is so bright, the column density is not too high so that the Compton-hump
is well-sampled in sensitivity by \suzakup, and the spectrum in the
$\sim 1$--10~keV band is not contaminated by numerous
emission lines from very hot gas
has as it is in NGC~1068. This means that the data in the bandpass that
includes the \fekalfa line and Fe~K edge are relatively ``clean''
(although there is an emission line due to \fexxvp).
When the modeling is applied to other AGN, such a detailed analysis will not be necessary
in most cases, and
our detailed description of the methodology for NGC~4945 is designed to
save time by enabling the researcher to establish which procedures are
not necessary for a given source and data set.

The paper is organized as follows. In \S\ref{obsdata} we describe the
basic data, and reduction procedures where relevant, from \suzakup, \bsaxp, 
and the \swift BAT. In \S\ref{formofspec} we summarize what is already
known about the general form of the X-ray spectrum of NGC~4945. In 
\S\ref{strategy} we describe the overall strategy of the analysis
that we will present, including detailed procedures for setting
up the various forms of the Compton-thick reprocessing models, and
for spectral fitting. In the following three sections 
(\S\ref{zerothorderfits} to \S\ref{decoupledmytfits}),
we give the results from fitting three classes of 
Compton-thick reprocessor models that correspond to very
distinct physical scenarios. In \S\ref{lumratios} we bring together all
the results from applying the different models and discuss the implied
intrinsic luminosities and Eddington ratios of the different spectral fits, as well as the
physical implications of each type of solution.
In \S\ref{summary} we summarize our findings. In the
appendix we present results of spectral fitting to some
\suzaku 3C~273 data in order to establish some important calibration information
pertinent to fitting the \suzaku NGC~4945 data.

\section{Observations and Data Reduction}
\label{obsdata}

\subsection{Suzaku Data}
\label{suzakudata}

The joint Japan/US X-ray astronomy satellite, \suzaku (Mitsuda \etal~2007),
was launched on 10 July, 2005.
The present study focuses on an observation campaign on NGC~4945 that
was carried out early in the life of \suzakup.
The campaign consisted of three observations of
NGC~4945 performed in 2005, August, and one in 2006, January 15.
The first two observations (in 2005, August) had relatively short exposure times ($\sim 14$~ks and 177~s),
and given the historical amplitude and spectral variability of NGC~4945
(e.g. Fukazawa \etal 2011; Itoh \etal 2008, and references therein),
in the present paper we only report results from the third observation,
which had an exposure time of nearly $100$~ks. 

\suzaku carries four X-ray Imaging Spectrometers (XIS -- Koyama \etal~2007) and
a collimated Hard X-ray Detector (HXD -- Takahashi \etal~2007).
Each XIS consists of four CCD detectors
at the focal plane of its own thin-foil X-ray telescope
(XRT -- Serlemitsos \etal~2007), and has a field-of-view (FOV) of $17.8' \times 17.8'$.
One of the XIS detectors (XIS1) is
back-side illuminated (BI) and the other three (XIS0, XIS2, and XIS3) are
front-side illuminated (FI). The bandpass of the FI detectors is
$\sim 0.4-12$~keV and $\sim 0.2-12$~keV for the BI detector. The useful
bandpass depends on the signal-to-noise ratio of the source since
the effective area is significantly diminished at the extreme ends of the
operational bandpasses. Although the BI CCD has higher effective area
at low energies, the background level across the entire bandpass is higher
compared to the FI CCDs.
Although we used the standard response matrix generator for modeling the
XIS data, 
we measured the widths of the  Mn~$K\alpha$ lines from the on-board
$^{55} \rm Fe$ calibration sources (two per XIS) using the actual
NGC~4945 observations in order to independently
check the spectral resolution degradation corrections in
the response generator. Details are given below.

The HXD consists of
two non-imaging instruments (the PIN and GSO -- see Takahashi \etal~2007)
with a combined bandpass of $\sim 10-600$~keV.
Both of the HXD instruments are background-limited,
more so the GSO, which has a
smaller effective area than the PIN. For AGNs, the source
count rate is typically much less than the background.
In order to obtain reliable background-subtracted spectra,
the background spectrum must be modeled as a function of energy and time.
The background models for the HXD/PIN and HXD/GSO have an advertised 
systematic uncertainty of 1.3\% 
\footnote{http://heasarc.gsfc.nasa.gov/docs/suzaku/analysis/pinbgd.html} and
2\% \footnote{http://heasarc.gsfc.nasa.gov/docs/suzaku/analysis/gsobgd.html}
respectively. However, the signal is background-dominated, and the
source count rate may be a small fraction of the background count rate,
so the net systematic error in the background-subtracted spectra could
be significant. The problem is worse for the GSO than it is for the PIN.
The observation of NGC~4945 was optimized for the HXD
in terms of positioning the source at the aimpoint for the HXD (the so-called
``HXD-nominal pointing'') which gives a somewhat lower count-rate in the XIS
than the ``XIS-nominal'' pointing, but gives $\sim 10\%$ higher HXD effective area.

The calibration of the relative cross-normalizations of the
three instruments involves many factors, and these are discussed
in detail in the appendix, where we derive the instrument
cross-normalization factors from a \suzaku observation of
3C~273 as a guide for the analysis of the NGC~4945 data. In the appendix we
also give details of the data reduction and
screening procedures that we used
for both the 3C~273 and NGC~4945 data, as well as details
of the background subtraction and spectral responses for all instruments.
The principal data selection and screening criteria for the XIS
were the selection of only {\it ASCA} grades
0, 2, 3, 4, and 6, the removal of
flickering pixels with the
FTOOL {\tt cleansis}, and exclusion
of data taken during satellite passages through the South Atlantic Anomaly
(SAA), as well as for time intervals less than $256$~s after passages through the SAA,
using the T\_SAA\_HXD house-keeping parameter.
Data were also rejected for Earth elevation angles (ELV) less than $5^{\circ}$,
Earth day-time elevation angles (DYE\_ELV) less than $20^{\circ}$, and values
of the magnetic cut-off rigidity (COR) less than 6 ${\rm GeV}/c^{2}$.
Residual uncertainties in the XIS energy scale
are on the order of 0.2\% or less
(or $\sim 13$~eV at 6.4~keV -- see Koyama \etal~2007). We confirmed this
from an analysis of the onboard calibration line data taken during the
NGC~4945 observations (see below for details).

The cleaning and data selection resulted in net exposure times
that are reported in \tablectratesp.
NGC~4945 is known to have many nonnuclear X-ray sources
(e.g. Schurch, Warwick, \& Roberts 2002), but most cannot be resolved
by the XIS. However, most of the nonnuclear sources are very
soft and relatively weak. We excluded the data below 0.7~keV
not only to avoid contamination from the nonnuclear sources,
but also because we found that the count rate from the background
below 0.7 keV {\it is} comparable to the source, giving
a background-subtracted spectrum that has unacceptably large systematic errors.
We did exclude two nonnuclear sources that are resolved in the XIS,
by using circular masking regions, and one of these sources was
a new transient discovered in the \suzaku data that has been
discussed in detail by Isobe \etal (2008).

We extracted XIS spectra of NGC~4945 by using
a circular extraction region with a radius of 3.5', excluding any masked
area containing the contaminating sources. 
The size of the extraction region is a trade off. If it is
too small, the XRT response function is less accurate and
there are less source counts than a larger region would provide.
A region that is too large on the other hand, has a higher background.
The selected region size is a good compromise for the NGC~4945 data.
Background XIS spectra
were made from off-source areas of the detector, after removing
a circular region with a radius of 4.5' centered on the source,
the calibration sources (using rectangular masks), and the
two prominent nonnuclear sources. There may still be residual
contamination from unresolved nonnuclear sources above 0.7~keV
and we shall bear this in mind when interpreting the spectral-fitting
results. The XIS spectra from all four detectors (1--4) were combined
into a single spectrum for spectral fitting.
The background subtraction method for the HXD/PIN and
HXD/GSO followed standard procedures, summarized in the appendix.
The spectral response matrices used for each instrument are
also described in the appendix.
The energy bandpass used for the spectrum from each instrument
was determined by the reliability of the background subtraction.
Excluding lower and upper energy ranges that gave negative counts
in the background-subtracted spectra result in the final 
energy ranges shown in \tablectratesp. In addition, the 1.83--1.93~keV
region in the XIS spectrum was excluded due to a known line-like
residual calibration feature (e.g. see Yaqoob \etal 2007).
The energy ranges, count rates, and the relative importance
of the background in the relevant energy ranges are shown in
\tablectratesp.

\begin{table}
\caption[Exposure Times and Count Rates for the NGC~4945 Suzaku Spectra]
{Exposure Times and Count Rates for the Suzaku Spectra}
\begin{center}
\begin{tabular}{lllcc}
\hline

Detector & Exposure & Energy Range & Rate$^{a}$ & Percentage of $^{b}$ \\
	 &  (ks)   & (keV)    &  (ct/s) &  On-Source Rate \\
\hline
& & & & \\
XIS & 95.1 & 0.7--1.83, 1.93--9.82 & $0.1609 \pm 0.0007$ & 85.8\% \\
PIN & 80.1 & 11.6--47.7 & $0.268\pm0.030$ & 34.7\% \\
GSO & 80.1 & 80--165 & $0.086\pm0.010$ & 2.0\% \\
& & & & \\
\hline
\end{tabular}
\end{center}
$^{a}$ Background-subtracted count rate in the energy bands specified.
$^{b}$ The background-subtracted source count rate as a percentage of the total on-source count rate,
in the utilized energy intervals.
\end{table}

Although the X-ray spectrum of NGC~4945 varies in amplitude and
shape during the \suzaku observation, we will utilize only the
time-averaged spectrum over the entire observation in order to
obtain the highest signal-to-noise ratio to model the nominal
average spectrum. Spectral variability will be kept in mind
when interpreting the results. Itoh \etal (2008) discussed in detail
the nature of the variability in NGC~4945 during the \suzaku observation,
finding that the source is variable on timescales as short as $\sim 20$~ks.
 
Binning of the spectra based upon a minimum number of counts per
energy bin (or a threshold signal-to-noise ratio per bin) was avoided
because such a procedure distorts the spectrum, especially in regions
that contain emission or absorption features. Results of spectral fitting
to such spectra can be incorrect. Instead, the XIS and PIN spectra
were examined, and uniform energy bin sizes were selected for energy
ranges that resulted in all individual bins contained more than 20
counts. Thus, the $\chi^{2}$ statistic could be used for spectral fitting.
For the GSO, the special energy bin boundaries that matched the
boundaries used for the available background files were adopted,
and these were binned by a factor of 2, as described in the appendix.

We measured the centroid energies
of the Mn~$K\alpha$ lines from the calibration sources,
using spectra from the individual XIS detectors, as well as
a spectrum combined from all four XIS detectors and both calibration 
sources.
The expected energies of the Mn~$K\alpha_{1}$ and
Mn~$K\alpha_{2}$ lines are 5.89875~keV and 5.88765~keV respectively
(Bearden 1967). Since the $K\alpha_{1}:K\alpha_{2}$ branching
ratio is 2:1, the expected centroid energy is then 5.89505~keV.
If the spectral response functions in the XIS response matrices
were perfect we should find that the calibration lines are
unresolved.
Using a single Gaussian with the centroid energy, intrinsic
width, and overall normalization allowed to be free parameters,
we found that the offset of the centroid energy (relative to
the theoretical value) and the spectral resolution was
different for the different XIS detectors. Since the final
NGC~4945 analysis was performed using data combined from all XIS
detectors, we used the calibration spectra averaged over
all XIS detectors and all calibration sources to obtain
a measurement of the line centroid offset and instrumental
broadening at 5.9~keV. We obtained an offset for the
centroid of $+12.2^{+1.0}_{-0.9}$~eV (i.e. a best-fitting
energy of $5.9073$~keV), and a Gaussian
width of $\sigma=14.1^{+3.4}_{-4.3}$~eV. The statistical errors
quoted are 90\% confidence for 3 interesting parameters.
We will use these measurements to help model the \fekalfa line emission
in NGC~4945. Although the energy scale and spectral resolution
depend somewhat on the spatial position of the source on the XIS detectors,
the data do not warrant a more sophisticated treatment.

\subsection{BeppoSAX Data}
\label{saxdata}
NGC~4945 was observed by \bsax in 1999, July 1. The \bsax
mission carried on board three imaging medium energy concentrator spectrometer
units (MECS, Boella \etal 1997), and a phoswich detector system (PDS, Frontera \etal 1997).
The MECS operated in the energy band $\sim 1.7-10$~keV, and the PDS
operated in the energy band $\sim 15-220$~keV. 
There was also a low energy concentrator spectrometer (LECS, Parmar \etal 1997) and 
a high energy gas scintillation proportional counter (HPGSPC, Manzo \etal 1997) on board,
but we did not utilize these in the present study. The LECS had a spectral
resolution that is too low for the present study, and the HPGSPC had 
insufficient signal-to-noise ratio for AGN in general.

Spectra and response matrices for the MECS and PDS data were downloaded
from HEASARC. The data, including background spectra, are already cleaned
and prepared for spectral fitting by the pipeline processing. 
For the PDS, the background spectrum that was made using the
so-called ``Variable Rise Time'' method
was used\footnote{http://heasarc.gsfc.nasa.gov/docs/sax/abc/saxabc/saxabc.html},
appropriate for weak sources. 
At the time of the NGC~4945 observation, one of the MECS detectors was already nonoperational
and the combined spectrum from MECS2 and MECS3 was utilized in the spectral
fitting. The exposure times for the MECS and PDS spectra were 46.8~ks and
43.8~ks respectively. We found that the background subtraction
for the MECS below 2 keV and above 9.5~keV was poor so we only utilized the energy range
2--9.5~keV. For the PDS it was found that background-subtraction systematics
restricted the useful energy range of the spectrum to 16.5--100~keV.
The net background-subtracted count rates in the utilized energy intervals
were $0.0478\pm0.0074$ ct/s and $2.249\pm0.023$ ct/s respectively.
The MECS and PDS spectra were binned uniformly, the bin widths being 
185~eV and 2.25~keV respectively. For both spectra, each energy bin
had at least 20 counts, validating the use of the $\chi^{2}$ statistic
for spectral fitting. 

The cross-normalization of the PDS and MECS spectra was set at 0.85:1 for
PDS:MECS, consistent with measurements  for 3C~273 in the
``BeppoSAX Cookbook''\footnote{http://heasarc.gsfc.nasa.gov/docs/sax/abc/saxabc/saxabc.html},
after correction for the ``Variable Rise Time'' method of background subtraction.
 
\subsection{Swift BAT Data}
\label{swiftdata}

\begin{figure}
\centerline{
        \psfig{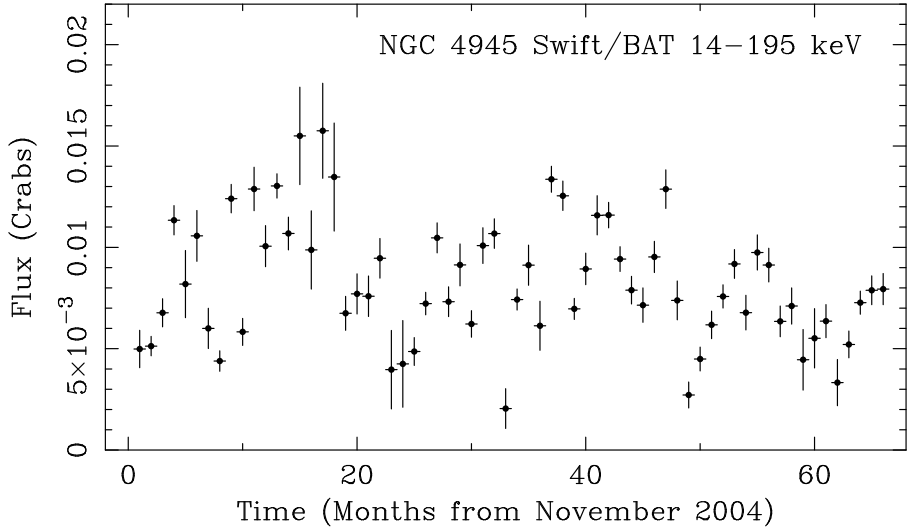}
        }
        \caption{\footnotesize The NGC~4945 \swift BAT 66-month lightcurve in the
14--195~keV band. Each time bin corresponds to a one month time interval, the
first bin corresponding to 2004, November. The flux is given in ``Crab'' units.
}
\label{fig:batlc}
\end{figure}

The {\it Burst Alert telescope} (BAT, Barthelmy \etal 2005)
aboard the \swift satellite (Geherls \etal 2004)
has been conducting an all-sky hard X-ray survey in the 14--195~keV
band since 2004, November
(e.g. see Tueller \etal 2010; 
Burlon \etal 2011). 
NGC~4945 is one of the brightest AGN that is detected in the 
\swift BAT all-sky hard X-ray survey. The 58-month catalog lists
the 14--195~keV flux of NGC~4945 as $300.97 \times 10^{-12} \ \rm ergs \ 
\ cm^{-2} \ s^{-1}$
\footnote{http://swift.gsfc.nasa.gov/docs/swift/results/bs58mon/}. 
The 58-month BAT spectrum for NGC~4945 and the associated response
matrix were downloaded from the archive
\footnote{http://swift.gsfc.nasa.gov/docs/swift/results/bs58mon/SWIFT\_J1305.4-4928}.
The standard 8-channel spectrum from the BAT is time-averaged
over a period of 58 months and was already prepared for
spectral analysis. The net count rate of the spectrum is $3.131 \pm 0.053$ ct/s.
Also available as a standard
product is a 66-month, 14--195~keV lightcurve that has time bins that
have a duration of 1 month (since November 2004), and flux units 
relative to the Crab flux. This lightcurve is shown in Fig.~\ref{fig:batlc}
and it can be seen that the 14--195~keV flux
is highly variable on timescales of months to years, showing a
dynamic range of about a factor of 8 between the highest and lowest
flux states. The straight average of the flux (in the Crabweighted
units shown in the lightcurve) is $8.24 \times 10^{-3}$. The \suzaku
spectrum corresponds to about $0.013$ in these units (obtained
by renormalizing the \suzaku spectrum to the \swift BAT data).
This puts the
\suzaku observation near the top of the range in Fig.~\ref{fig:batlc}.

\section{General Form of the X-ray Spectrum of NGC 4945}
\label{formofspec}

The general form of the broadband X-ray spectrum of NGC~4945 is
now well established (e.g. Itoh \etal 2008 and references therein).
The spectrum below $\sim 10$~keV is heavily suppressed relative
to that in the $\sim 10$--30~keV band, due to heavy obscuration.
Below $\sim 10$~keV the spectrum is dominated by scattered
emission of the intrinsic X-ray continuum in an extended optically-thin
medium (e.g., Turner \etal~1997; Marinucci \etal 2012, and
references therein). Below $\sim 1$~keV there is evidence of an additional
soft excess that is due to thermal emission from
extended circumnuclear matter (at least some of this has
been directly spatially resolved by \chandrap, as described
by Marinucci \etal 2012). Such a spectrum is typical
of obscured AGN, although the amount of suppression below 10 keV (compared
to the flux above 10 keV) varies from source to source.  
The intrinsic X-ray continuum in NGC~4945 has to be
inferred indirectly (and therefore is model-dependent) because
of the complexity and breadth of the various X-ray spectral features.
In our analysis we will adopt two forms for the intrinsic continuum,
the first being a simple power law (with a photon index of $\Gamma$), 
and a termination energy, $E_{T}$. The second continuum form
is that of a Comptonized Wien spectrum, characterized by the
temperature and optical depth of the Comptonizing plasma ($kT$ and $\tau$ respectively).
The exact form of the initial spectrum (in this case a Wien spectrum) is 
unimportant for values of the Compton-$y$ parameter that result in
a Comptonized spectrum that is roughly a power law with a high-energy rollover
(i.e. observationally relevant to NGC~4945). This model
of the intrinsic continuum is described in Titarchuck (1994), in
which it is explained
that the Wien spectrum was adopted because it is lends itself
to certain computational advantages. In XSPEC the model is known as
{\sc comptt}. The temperature of the Wien spectrum was fixed at $0.01$~keV,
below the lower end of the bandpass for any of the data sets that we 
used, and this choice does not affect the values $kT$ and $\tau$ used to
fit the data. Again, this is because for a Compton-$y$ parameter greater than
unity, the Comptonized spectrum is not sensitive to the exact form
of the initial spectrum.

A fraction of the direct continuum 
is scattered into the line of sight by optically-thin matter, extended
on the same or larger size scale as the putative torus, and we refer to this
fraction as $f_{s}$.
In the optically-thin limit, the Thomson
depth, $\tau_{\rm es}$,
of the scattering region is simply equal to
$f_{s}$ for a fully covering spherical distribution of material. 
More realistically, $f_{s} = [\Omega/4\pi] \tau_{\rm es}$,
where $[\Omega/4\pi]$ is the solid angle as a fraction of $4\pi$
subtended by the scattering zone at the X-ray source.
Only that part of the scattering zone visible to the observer should be
included in this solid angle. We assume that the optically-thin scattered continuum
has an identical spectral shape to the intrinsic continuum, although the
former is allowed to have its own uniform, Compton-thin absorption (column density $N_{\rm H1}$).
Part, or all of this column density may be due to absorption in the host galaxy.
In reality the
above two assumptions are oversimplistic, but given the complexity
of the overall baseline model, a more detailed treatment of the
optically-thin scattered continuum is not warranted.
We add to these continuum components an optically-thin thermal continuum
emission component using the {\sc apec} model with abundances
fixed at the solar values, but with the normalization
and temperature of the thermal component ($kT_{\rm apec}$ and $A_{\rm apec}$ 
respectively) allowed to float in fits.
This optically-thin emission component is of course {\it not} absorbed
by the large, Compton-thick column density (or at least, we observe
only the portion that is unobscured by it). However, 
we do include a lower, uniform column density as a 
free parameter to allow for the possibility of absorption
by material in addition to the primary 
X-ray reprocessing structure (we will refer to it as $N_{\rm H2}$). 
The column densities $N_{\rm H1}$ and $N_{\rm H2}$ are {\it not} tied
together (even though they cover a similar physical region) as this
allows for the most general situation. Preliminary spectral fitting showed
that both $N_{\rm H1}$ and $N_{\rm H2}$ are less than $\sim 10^{22} \ \rm cm^{-2}$
(i.e., Compton-thin) and in fact similar in value to each other.
Since these column densities are so small (Thomson depth $\ll 0.01$)
it is not necessary to model Compton scattering or fluorescent line
emission from these absorption components.

In \S\ref{strategy} we describe in detail how the
Compton-thick absorption, scattering and 
fluorescent \fekalfa and \fekalfa line emission in NGC 4945 are modeled.
The \nika line emission is weak in the NGC 4945 data and is 
not yet included in the \mytorus model but it is included in 
some of the other Compton-thick reprocessor models that we used.
In the former case, the \nika is modeled
as a separate unresolved Gaussian component (characterized by
a centroid energy and line flux denoted by $E_{\rm Ni K\alpha}$ and
$I_{\rm Ni K\alpha}$ respectively). The \suzaku data also show
line emission centered around $6.7$~keV, which we ascribe
to \fexxvp, and this is also modeled by a separate Gaussian component.
It very likely originates in the ionized extended zone 
and does not originate in the Compton-thick X-ray reprocessor which 
produces the \fekalfa line centered at 6.4~keV in neutral matter.
The centroid energy, intrinsic Gaussian width, and flux of the line
centered at $\sim 6.7$~keV are donated by $E_{\rm Fe XXV}$, $\sigma_{\rm Fe XXV}$,
and $I_{\rm Fe XXV}$ respectively. We note that individual lines of the
\fexxv triplet cannot be resolved by \suzaku and the emission that is
observed may even have some contribution from lower ionization states
so the line width should not be interpreted as necessarily due entirely
to velocity broadening.

All spectral fits to the \bsax data will have fewer free model parameters than
the models fitted to the \suzaku data because
the \bsax MECS spectrum only extends down to  2~keV. Our approach is
to freeze the normalization
and temperature of the soft X-ray, optically-thin, thermal continuum
at values obtained from the \suzaku fits. The same is done for the column density
associated with the optically-thin thermal emission ($N_{\rm H2}$).
In addition, since the spectral resolution of the
MECS is greater than 600~eV at the energies of the Gaussian emission-line components
(i.e., in the 6--8~keV range),
the centroid energies of the lines were fixed at 6.700~keV and 7.472~keV, corresponding
to the theoretical energies of the \fexxvres and \nika lines respectively.
Moreover, the intrinsic width of the \fexxvres line was fixed at a best-fitting
value obtained from the \suzaku data,
and the intrinsic widths of the \feka and \nika lines were fixed at $100 \ \rm km \ s^{-1}$ FWHM
because these lines in the \suzaku data were unresolved.
Whereas the \suzaku fits have two relative normalization parameters
(which we refer to as  $C_{\rm PIN:XIS}$ and $C_{\rm GSO:XIS}$),
the \bsax data only have one (which we refer to as $C_{\rm PDS:MECS}$).
The value of $C_{\rm PIN:XIS}$ was fixed at 1.12 (see appendix), $C_{\rm GSO:XIS}$ was allowed
to float unless stated otherwise, and $C_{\rm PDS:MECS}$ was fixed at 0.85 (see \S\ref{saxdata}).

Models for the \swift BAT data of course always have fewer model components than those
for the \suzaku and \bsax data because the soft X-ray emission component, the optically-thin
scattered continuum, the associated column densities, and the fluorescent emission lines
can omitted.

\section{Analysis Strategy}
\label{strategy}
There are a number of critical steps in the analysis scheme that we will
pursue, and in this section we outline the key motivators and drivers at each stage of
the complex analysis.
These steps in the analysis will be described in more detail in later sections 
when necessary.

\subsection{Constraints from Variability}
\label{variabilityconstraints}

The large-amplitude, short and long-timescale continuum variability above 10 keV, as
demonstrated by the \swift BAT (Fig.~\ref{fig:batlc}) and other data (Itoh \etal 2008; Marinucci \etal 2012, and
references therein), 
imposes a constraint on the size of the Compton-thick region responsible 
for producing the narrow \fekalfa emission and the associated Compton-scattered
continuum. Since the flux of the \fekalfa line in NGC~4945 does not respond 
to the variable continuum above 10~keV, it is a robust inference that the
continuum above 10~keV cannot be dominated by the Compton-scattered 
component of the continuum since that latter component has to originate in the same region as
the \fekalfa line. Therefore, the continuum above 10~keV must be dominated
by the direct (unscattered) continuum transmitted through the Compton-thick medium
in the line of sight
(we refer to this component in general as the zeroth-order continuum).
However, since we will be describing a methodology that should be applicable to
AGNs in general, we will also consider cases in which the high-energy continuum
above 10~keV is dominated instead by the Compton-scattered continuum.
For example, as we shall see, an AGN shrouded in a fully-covering Compton-thick
spherical distribution of matter will have a high-energy continuum that is
dominated by the Compton-scattered continuum (this is reversed if
the source is Compton-thin). For most AGNs there may not be
sufficient data (in quantity and/or quality) to determine the variability properties
of the high-energy continuum and \fekalfa line, so both cases (zeroth-order
continuum dominating and not dominating above 10~keV) would need to be considered
if they may provide degenerate spectral solutions.
However, we note that if an AGN is Compton-thin in the line of sight, with
a column density in the range $\sim 10^{23}$ to $10^{24} \ \rm cm^{-2}$, there will
be much less ambiguity because the zeroth-order continuum will
impose strong and characteristic 
features on the spectrum in terms of the detailed shape of the Fe~K edge and the 
continuum below it. The EW of the \fekalfa line with respect to the total
continuum then narrows down the parameter space even further. Even spectra of weak AGNs
in this regime of intermediate Compton depth may yield less ambiguity in 
the model solutions than much brighter AGNs that are Compton thick. 
(Sources with column densities less than $\sim 10^{23} \ \rm cm^{-2}$ are of course
far less complex, since they are dominated
by the zeroth-order continuum at all energies.) 

\subsubsection{Separability of the Zeroth-Order Continuum and the Compton-scattered Continuum}
\label{continuumseparation}

In our analysis we will explore and derive results for several scenarios, including
those that are ruled out by the variability properties of NGC~4945.
The purpose here is to illustrate how the different scenarios can give
degenerate spectral solutions even for the high signal-to-noise ratio
of the NGC~4945 data. For weaker AGNs, and those that lack vital
variability information, being aware of the spectral degeneracies and
correctly interpreting them will be important.
We point out that in order to adequately model NGC 4945,
and any other source for which time variability of the Compton-thick scattered
continuum needs to be considered, ideally we
need a Compton-thick reprocessor model in which the zeroth-order continuum
and the Compton-scattered continuum are separable (i.e. they must be allowed
to have independent normalizations for the purpose of spectral fitting).
In the Compton-thick spectral-fitting models of BN11
the two continuum components are not separable. The only model that is
currently publicly available that has the required capabilities to handle
the possibility of different timescales of variability of the Compton-scattered and
zeroth-order continuum components
is the \mytorus model (see MY09).
Nevertheless, we will still make use of the BN11 models. 

\subsection{Constraints from the Intrinsic Fe~K$\alpha$ Line Width}
\label{ironlineconstraints}

In principle,
we can estimate the light-crossing time across any
spatially unresolved X-ray reprocessor that might be present, using measurements
of the \fekalfa intrinsic line width. If the FHWM velocity of the line is
$V_{\rm FWHM}$, using a virial estimate of the velocity dispersion of
$\sqrt{3}V_{\rm FWHM}/2$ (following Netzer 1990), and a simple Keplerian
assumption, gives a light-travel time from the X-ray source to the reprocessor
of $t=(r/c) \sim 67M_{6}[3/V_{1000}]^{2}$~ks, where $V_{1000}$ is the measured
FWHM in units of $1000 \ \rm km \ s^{-1}$, and $M_{6}$ is the central black-hole
mass in units of $10^{6}M_{\odot}$. 
The \chandra high-energy grating (HEG) measured an intrinsic width of the narrow \fekalfa
line in NGC~4945 of $\sim 2780 \ \rm km \ s^{-1}$~FWHM, with a two-parameter, 99\% confidence range
of $\sim 1700$ to $\sim 5200 \ \rm km \ s^{-1}$~FWHM (Shu, Yaqoob, \& Wang 2011).
The HEG has the best spectral resolution in the Fe~K band currently available.
Using $M_{6}=1.4$ (Greenhill \etal 1997), we
see that for a nominal value of $V_{1000}$ of $3$, the light-crossing timescale is $\sim 100$~ks,
and for a value of $5$ (around the upper limit for $V_{1000}$), the timescale is
$\sim 33$~ks. The fastest timescale for high-energy continuum variability in NGC 4945 reported so far
is $\sim 20$~ks for a flux doubling (Itoh \etal 2008). 
However, it is likely that the line broadening is due to the component
of the \fekalfa line that is spatially extended, because grating spectrometers
cannot distinguish between spatial and true spectral broadening. Nevertheless,
we refer to the calculation of velocity broadening for the sake of application
to AGNs other than NGC 4945, for which the line emission is spatially unresolved.
We note in passing that
the HEG centroid energy of the \fekalfa line in NGC~4945 is extremely well constrained
($6.389^{+0.007}_{-0.008}$~keV), confirming its origin in neutral Fe or nearly neutral Fe.

We should also bear in mind that the case of NGC~4945 does not rule
out the possibility of two physically distinct regions of Compton-thick matter.
Although the \fekalfa line emission that is spatially resolved and
extended, on a scale of 30~pc or greater, could account for the
bulk of the total \fekalfa line emission (Marinucci \etal 2012), it is
possible that there could be a smaller Compton-thick region that obscures the 
line of sight but has a sufficiently small covering factor that it does
not make a significant contribution to the \fekalfa line.

\subsection{Intrinsic Luminosities}
\label{intrinsicluminosities}

It is important to understand that, in addition to finding solutions
that fit the various X-ray spectra of NGC~4945 (and other AGNs), we must keep a check
on the implied intrinsic continuum luminosities because they can
differ by an order of magnitude or more for different degenerate scenarios.
In fact, we shall see that it is possible for different
spectrally degenerate models applied to two observations of a
source to predict opposite senses of variability under some
circumstances. In other words, one set of models may imply
a decreasing intrinsic luminosity going from one observation to
the next, whilst a different set of models may imply an
increasing intrinsic luminosity going from one observation to
the next.
Here we simply point out a general characteristic of the various models,
namely that the higher the contribution of the zeroth-order
continuum relative to the Compton-scattered continuum,  
the greater the intrinsic luminosity is.
This is because Compton scattering shifts more of the intrinsic
continuum into the observer's line of sight compared to the case when
the observer receives only the zeroth-order continuum. In other
words, if there is a Compton-scattered continuum component 
observed in the spectrum, it can only {\it decrease} the burden
on the intrinsic continuum to produce the observed luminosity for
a given column density. 
To put it another way, the intrinsic continuum must lie above
the zeroth-order continuum by a very specific amount that 
depends only on the line-of-sight column density, {\it regardless
of geometry and regardless of the level of the Compton-scattered
continuum}. There are two
particular corollaries of this. One is
that if the observed spectrum can be well-fitted by {\it only}
the zeroth-order continuum (e.g., if the
reprocessor has a negligible global covering factor), such a fit yields
the maximum possible intrinsic continuum luminosity for a given
line-of-sight column density. The second corollary is that 
for a given line-of-sight column density, a distribution
of matter with full covering (such as a spherically-symmetric
distribution) will give the {\it minimum} possible intrinsic luminosity.
In the Compton-thin limit, these minimum and maximum luminosities will
of course be equal to each other because in that limit there is only
the zeroth-order continuum regardless of covering factor.
As the Compton depth of the matter distribution approaches unity
($N_{\rm H} \sim 1.2 \times 10^{24} \ \rm cm^{-2}$),
the Compton-scattered continuum from a fully-covered
source dominates over the zeroth-order
continuum in the observed spectrum. As we will see, the two extremes give
implied intrinsic luminosities that can differ by an order of magnitude or more.
In all of the tables in which we will
give the results of spectral fitting, we will give observed fluxes and
luminosities in various energy bands. However,
the intrinsic luminosities and their ratios with respect to the
Eddington luminosity will not be discussed until all of the spectral
fitting results have been presented. Discussion of the intrinsic luminosities
and their implications will be presented in \S\ref{lumratios}.

\subsection{Procedure for NGC~4945}
\label{analysisprocedure}

Our analysis procedure for NGC~4945 is then as follows.

We begin with the simplest and extreme scenario, namely that in which the
spectrum of NGC~4945 above 10 keV consists {\it only} of the zeroth-order
continuum. 
\\

\par\noindent
(i)We examine
all of the three data sets above 10 keV only (i.e. utilizing only
HXD data for \suzakup, only PDS data for \bsaxp, and the \swift BAT data).
The model above 10~keV is very simple because no emission lines
or soft X-ray emission components need to be included. \\
(ii) These spectral fits are described in \S\ref{zerothorderfits}, 
and we will find that, in addition to a simple power law,
we need to consider intrinsic spectra that rollover in the instrument
bandpasses. For this we use a thermally Comptonized intrinsic continuum
(details in \S\ref{formofspec}). 
\\

Obviously, the Compton-scattered continuum
cannot be zero since there has to be a specific flux of this continuum 
that is associated 
with the fluorescent \fekalfa and \fekbeta emission lines.
However, these extremal fits will be useful for four reasons:
\\

\par\noindent
(i) One is that
the fits can guide
spectral fitting with more complex models across the full instrumental
bandpasses. \\
(ii) Another reason is that our results can be used to assess
the limitations of deriving key physical parameters (such as column 
density and intrinsic luminosity) when a
particular source {\it only} has \swift BAT data available.
Even though the \suzaku and \bsax observations are not contemporaneous
with the \swift BAT data, the latter contain information about long-term
variability spanning a period of over 5 years, and about the average spectrum over
a similarly long time period. \\
(iii) A third reason is that
the zeroth-order continuum fits will provide an indicator
of the maximum possible intrinsic continuum luminosities. \\
(iv) A fourth
reason is that this is the {\it only} model in which an arbitrary
intrinsic continuum can be used, with any of the intrinsic 
continuum parameters allowed to be free (without having to generate
grids of tables at finite parameter-value intervals).

We then proceed to obtain full solutions for the \suzaku data and \bsax data, including
the \fekalfa emission line and other model components in the fits.
\\

\par\noindent
(i) Spectral fitting is then divided into two classes of models, one in which
the Compton-thick reprocessor has the zeroth-order and Compton-scattered
continuum components ``coupled'' to each other, and one in which they are ``decoupled.''
 \\
(ii) Only the \mytorus model allows decoupling, and when used in this mode, 
one interpretation is that the model
mimics a patchy (or clumpy) X-ray reprocessor in which the ``holes'' allow 
unobscured
observation of some of the reflection and fluorescence from the far-side, inner
surface of the structure. In this mode, the geometry is 
not necessarily strictly toroidal, and the global covering factor
is unspecified  because the spectrum is
dominated by matter observed through the ``holes'' and by matter in
the line of sight.
Another scenario is that a subset of the
decoupled models describes two distinct X-ray reprocessing regions,
one that obscures the central X-ray source with less than full global
covering, and another more
extended reflection region, such as that which has
been spatially resolved in NGC~4945 (Marinucci \etal 2012). \\
(iii) The decoupled mode of the \mytorus model is in fact
closest to the procedure that has been universally used in the literature for
AGNs in general for 
over 15 years: a disk-reflection continuum modeled with 
{\sc pexrav} or {\sc pexmon} (or equivalent), combined with decoupled
line-of-sight extinction modeled with {\sc cabs} or just {\sc zphabs} (absorption only). 
However, disk-reflection models assume an infinite column density for
the material responsible for
producing the Compton-scattered continuum so they cannot produce spectral features in the
data that are characteristic of scattering in a finite column-density medium.
Combined with the fact that the geometry of disk-reflection models may be inappropriate,
the \mytorus model, even in the decoupled mode, 
yields column densities and intrinsic continuum luminosities
that have more straightforward physical interpretations. Also, the line-of-sight
extinction in the \mytorus model
employs a fully relativistic Compton-scattering cross section. \\
(iv) The detailed setup and parameters of the coupled and 
decoupled modes of the \mytorus model are given in \S\ref{mytorusmodel}.
The results of fitting the coupled \mytorus model are given in \S\ref{coupledmytfits}, and
those obtained from fitting the decoupled \mytorus model are given in \S\ref{decoupledmytfits}.
A particular feature of the decoupled model is that it allows
a Compton-thick structure observed edge-on to produce a
large fluorescent \fekalfa line EW whilst still allowing the zeroth-order
continuum to dominate above 10~keV. \\
(v) We will also utilize the toroidal and fully-covering spherical models of BN11.
Since the zeroth-order and Compton-scattered continua cannot be decoupled
in these models, their application is described along with the
other coupled models in \S\ref{coupledmytfits}.

\subsection{The MYTORUS Model}
\label{mytorusmodel}

The toroidal Compton-thick X-ray reprocessor model, {\sc mytorus}, has
been described in detail in MY09, and Yaqoob \& Murphy (2011a). The geometry
consists of a torus with a circular cross section, whose diameter
is characterized by the equatorial column density, $N_{\rm H}$.
The model is currently restricted to a configuration in which the
global covering factor of the reprocessor is $0.5$, corresponding to
a solid angle subtended by the structure at the central X-ray source of
$2\pi$. However, spectral fits with the zeroth-order continuum only
(see \S\ref{zerothorderfits}) correspond to the limit of a negligible
covering factor, and fits with the BN11 fully-covering spherical model
(see \S\ref{coupledmytfits}) correspond to the other extremal limit
in the covering factor. We will also utilize the toroidal model
of BN11, which does allow the covering factor to vary between
0.1 and 0.9, but, like the BN11 spherical model,
this model does not allow separation of the
zeroth-order and Compton-scattered continua.

The practical implementation of the \mytorus model allows
free relative normalizations between different components of
the model in order to accommodate differences in the actual geometry
(compared to the specific model assumptions used in the original calculations), and time delays
between direct, scattered, and fluorescent line photons
\footnote{See http://mytorus.com/manual/ for details}.
The zeroth-order component of the model is essentially an
energy-dependent multiplicative factor
that is independent of the geometry and independent of the intrinsic continuum.
The multiplicative factor is then implemented with a single XSPEC table
for all applications of the model (it is {\tt mytorus\_Ezero\_v00.fits}).
The Compton-scattered continuum is implemented as an XSPEC additive
table model, utilizing different tables for a power-law input continuum
and a Comptonized thermal ({\sc comptt}) input continuum (see Titarchuk 1994). For the
former, tables with a termination energy of 200~keV were used, and
for the latter, each table has a unique, fixed value of the Comptonizing
plasma temperature. Tables with different temperatures were utilized, and
details will be given in the appropriate places in the descriptions of the 
data analysis procedures. The \fekalfa and \fekbeta line emission
is implemented with another XSPEC additive
table model that is selected from line tables made with a range of
energy offsets for best-fitting the peak energies of the emission lines.
Preliminary fitting showed that an offset of $+20$~eV is optimal for
the NGC~4945 \suzaku data (this offset covers both instrumental energy offsets,
as well as any intrinsic offset due to very mild ionization).
Different sets of emission-line tables are used for the 
power-law and Comptonized thermal intrinsic continua (i.e.
each Compton-scattered continuum table has a corresponding emission-line table
for a given offset energy).
Again, details of the actual tables used will be given at the appropriate
places in the data analysis descriptions. 
The \fekalfa and \fekbeta lines are broadened by two Gaussian convolution model
components ({\sc gsmooth} in XSPEC). One of these accounts for the
residual instrumental broadening that is not in the response matrix, and is
normalized to the broadening of the calibration source data (see \S\ref{suzakudata}),
with a $\sqrt{E}$ dependence of the Gaussian width, and the other is the actual velocity broadening
(Gaussian width proportional to line energy). 

\subsubsection{Default (Coupled) Model}
\label{mytcoupled}

In this mode of use (regardless of the form of the intrinsic continuum),
the angle made by the axis of the torus with the observer's line of sight
(\thetaobsp) is coupled to the column density that is intercepted
by the zeroth-order continuum. In other words, the effective geometry of
the X-ray reprocessor is {\it precisely} that assumed in the original
Monte Carlo calculations (MY09). 
We denote the relative normalization between the scattered 
continuum and the 
direct, or zeroth-order continuum, by $A_{S}$, which has
a value of 1.0 for the assumed geometry with the assumption that either the
intrinsic X-ray continuum flux is constant,
or, for 
a variable intrinsic X-ray continuum, that the X-ray reprocessor is
compact enough for the Compton-scattered flux to respond
to the intrinsic continuum
on timescales much less than the integration time for the spectrum.
It is important to note that $A_{S}$ is {\it not} simply
related to the covering factor of the X-ray reprocessor because
the detailed shape of the Compton-scattered continuum varies with 
covering factor.
Analogously to $A_{S}$, the parameter
$A_{L}$ is the relative normalization of the \fekalfa line
emission, with a value of 1.0 having a similar meaning to that for $A_{S}=1.0$.
In our analysis we will set $A_{L}=A_{S}$ unless otherwise stated.

For the sake of reproducibility, we give below the exact model expression
that we used in XSPEC (for the case of an intrinsic power-law continuum). 
The numbered model components and model parameters
are described in \tablemytdefparsp.
This example is for a power-law intrinsic continuum with a termination energy,
$E_{T}$, of 200~keV, and an offset of $+20$~eV for the emission-line tables. 
Substitution of the scattered continuum and emission-line
tables with corresponding tables for a different intrinsic continuum is straightforward
(for example, see \S\ref{mytdecoupled}).
In the case of the {\sc comptt} model substituting for the power-law continuum,
the photon index, $\Gamma$, is replaced by the Comptonizing plasma optical depth ($\tau$)
for a given table selected for the Comptonizing plasma temperature, $kT$
(and the normalization, $A_{\rm PL}$, replaced by $A_{\rm comptt}$).
The parameters $C_{\rm PIN:XIS}$, $C_{\rm GSO:XIS}$, and $C_{\rm PDS:MECS}$
are instrumental cross-normalizations (clearly, one or more of these will be absent
in the model if data from the corresponding instruments are not utilized).
The value of $C_{\rm PIN:XIS}$ was fixed at 1.12 (see appendix), $C_{\rm GSO:XIS}$ was allowed
to float unless stated otherwise, and $C_{\rm PDS:MECS}$ was fixed at 0.85 (see \S\ref{saxdata}).
Hereafter, Compton-scattered continuum components in the {\sc mytorus} model
will be referred to generically by the label MYTS and any
energy-dependent multiplicative factor that is used to obtain the
zeroth-order continuum will be referred to generically by the label MYTZ.

\begin{eqnarray}
\rm
model \  = \  constant<1>*phabs<2>( & & \nonumber \\ 
\rm zpowerlw<3>*etable\{mytorus\_Ezero\_v00.fits\}<4> & + & \nonumber \\
\rm constant<5>*atable\{mytorus\_scatteredH200\_v00.fits\}<6> & + & \nonumber \\
\rm constant<7>*gsmooth<8>(gsmooth<9>(atable\{mytl\_V000010pEp020H200\_v00.fits\}<10>)) & + & \nonumber \\
\rm zgauss<11> + zgauss<12>) & + & \nonumber \\
\rm constant<13>*zphabs<14>*zpowerlw<15> + zphabs<16>*apec<17>) \nonumber 
\end{eqnarray}

\begin{table}
\caption[Parameters for the Baseline model.]
{Parameters for the Baseline Coupled {\sc mytorus} Model with a Power-law 
Intrinsic Continuum}
\begin{center}
\begin{tabular}{llllll}
\hline
& & & & & \\
Par.$^{a}$ & Comp. $^{b}$ & Symbol & Units & Status & Description \\
& & & & & \\
\hline
& & & & & \\
1 & 1 & $C_{\rm PIN:XIS}$, $C_{\rm PDS:MECS}$ & \ldots & fixed & PIN:XIS or PDS:MECS instrumental cross-normalization ratio. \\  
2 & 1 & $C_{\rm GSO:XIS}$ & \ldots & free & GSO:XIS instrumental cross-normalization ratio. \\
3 & 2 & $N_{\rm H,Gal}$ & $10^{21} \rm \ cm^{-2}$  & fixed (1.57) & Galactic column density. \\
4 & 3,6,10,15 & $A_{\rm PL}$ & ph. $\rm cm^{2} \ s^{-1} \ keV^{-1}$ & free & Intrinsic power-law normalization at 1~keV. \\
5 & 3,6,10,15 & $\Gamma$ & & free & Photon index of the power-law continuum. \\
6 & 4,6,10 & $N_{\rm H}$ & $10^{24} \rm \ cm^{-2}$  & free & Equatorial column density of the torus. \\
7 & 4,6,10 & $\theta_{\rm obs}$ & degrees & free & Inclination angle of the torus. \\
8 & 5 & $A_{S}$ & \ldots & free & Scaling factor for the scattered continuum from the torus. \\
9 & 7 & $A_{L}$ & \ldots & $=A_{S}$ & Scaling factor for the fluorescent line emission from the torus. \\
10 & 8 & $\sigma_{\rm det}$ & eV & fixed$^{c}$ (14) & Line width accounting for XIS resolution degradation. \\ 
11 & 9 & FWHM[\fekalfap,\fekbetap] & $\rm km \ s^{-1}$ & fixed$^{d}$ (100) & Intrinsic width of the \fekalfa and \fekbeta emission lines. \\
12 & 11 & $E_{\rm Fe~XXV}$ & keV & free$^{d}$ & Gaussian centroid energy of the line emission at $\sim 6.7$~keV. \\
13 & 11 & FWHM[\fexxvp] & $\rm km \ s^{-1}$ & free$^{d}$ & Intrinsic width of the line emission at $\sim 6.7$~keV. \\
14 & 11 & $I_{\rm Fe~XXV}$ & ph. $\rm cm^{2} \ s^{-1}$ & free & Flux of line emission at $\sim 6.7$~keV. \\
15 & 12 & $E_{\rm Ni~K\alpha}$ & keV & free$^{d}$ & Gaussian centroid energy of the \nika line. \\
16 & 10 & FWHM[\nikap] & $\rm km \ s^{-1}$ & fixed$^{d}$ (100) & Intrinsic width of the \nika line. \\
17 & 12 & $I_{\rm Ni~K\alpha}$ & ph. $\rm cm^{2} \ s^{-1}$ & free & Flux of the \nika line emission. \\
18 & 13 & $f_{s}$ & \ldots & free & Scattering fraction due to optically-thin matter. \\
19 & 14 & $N_{\rm H1}$ & $10^{21} \rm \ cm^{-2}$ & free & Absorber covering optically-thin scattering zone. \\
20 & 16 & $N_{\rm H2}$ & $10^{21} \rm \ cm^{-2}$ & free & Absorber covering extended thermal emission zone. \\
21 & 17 & $A_{\rm apec}$ & ph. $\rm cm^{2} \ s^{-1} \ keV^{-1}$ & free & Normalization of the {\sc apec} 
model. \\
22 & 17 & $kT_{\rm apec}$ & keV & free & Temperature of the {\sc apec} model. \\
& & & & & \\
\hline
\end{tabular}
\end{center}
$^{a}$ Parameter number. 
$^{b}$ Model component(s) that the parameter
appears in (corresponding to the model component numbers
in the model expression shown in \S\ref{mytcoupled}).
$^{c}$ Determined from fitting calibration source data       
(see \S\ref{suzakudata}).
$^{d}$ The Gaussian emission-line centroid energies, and the
intrinsic width of the \fexxv line  were initially
allowed to float but they were then frozen at their best-fitting
values in order to obtain robust statistical errors on the
other parameters of interest. A single intrinsic width is
associated with the \fekalfa
and \fekbeta lines, and this width, along with that of the 
\nika line was fixed at 
$100 \ \rm km \ s^{-1}$~FWHM since the lines were found to be unresolved.
Upper limits were derived by allowing the two line widths 
to float in turn (see \S\ref{coupledmytfits} and \S\ref{decoupledmytfits}).
\end{table}

\subsubsection{Decoupled Model}
\label{mytdecoupled}

A problem with all Monte-Carlo simulations of the Compton-scattered continuum
from a toroidal structure observed edge-on is that the spectrum is
extremely sensitive to the geometry of the ``edges'' of the structure.
This is because the column density presented to incident continuum photons
is small at the edges, and the Compton-scattered flux from these regions
can easily dominate the entire spectrum if the equatorial part of the reprocessor
is Compton-thick. This is not just a problem with the \mytorus model but it is
a general problem with similar models observed edge-on (e.g., the toroidal models
of Ghisellini, Haardt, \& Matt 1994, Ikeda, Awaki, \& Terashima 2009, and BN11).
In reality, the geometry of the toroidal structure, particularly at the edges,
may not be well represented by the exact model geometry that is used.
For example, a structure that is  cylindrical 
should produce an edge-on Compton-scattered spectrum that is much
weaker than that produced by the particular toroidal geometry of the \mytorus model.
Another problem is that the finite size of the inclination-angle bins
means that even when the inclination angle is $90^{\circ}$ in the model,
there is always some ``leakage'' contribution from photons with angles
less than $90^{\circ}$.
We can mitigate these problems in the \mytorus model by decoupling the
zeroth-order continuum from the inclination angle (\thetaobsp), fixing this
angle at $90^{\circ}$, and allowing the relative normalization of the Compton-scattered
continuum to be free, and much less than 1.0 if necessary. In this context,
we refer to the relative normalization as $A_{S90}$ (it was just $A_{S}$ in the coupled model).
Since the zeroth-order continuum is independent of geometry (being purely a line-of-sight
quantity), the inclination angle associated with this component becomes a dummy
parameter and it is fixed at $90^{\circ}$ so that the zeroth-order column density
is literally equal to the value of $N_{H}$ for this model component. (In the
coupled mode, the equatorial $N_{H}$ is not equal to the line-of-sight column density
for general values of \thetaobsp.) The column density associated with the
zeroth-order continuum and the Compton-scattered continuum remain coupled to a single
value in our implementation for NGC 4945, although they could be decoupled if
we wanted to mimic a more complex structure for the reprocessor.

Yet another issue with a Compton-thick
structure observed edge-on is that the
zeroth-order and Compton-scattered continua could be so weak that even
a small amount of patchiness or clumpiness could give rise to an observed spectrum
that is actually dominated by a small amount of reflection of the intrinsic continuum by
the far side of the structure. In other words, even if a few percent of the
reflection from the inner far side of the reprocessor is unobscured by material
on the near side of the structure, the observed spectrum may be dominated
by the far-side reflection, at least below $\sim 10$~keV, at energies
relevant for the \fekalfa and \fekbeta line emission. 
We can mimic this far-side reflection spectrum with the face-on
reflection spectrum in \mytorus (and in fact the shape of the reflection
spectrum from the inner faces of the torus is similar for all lines of sight
that do not intercept the torus). In practice, to implement this, we simply add another 
Compton-scattered continuum table to the XSPEC model, but this time we fix the
value of the inclination angle at $0^{\circ}$ for that component only. This component has its own
normalization, which we call $A_{S00}$. Thus, we have two Compton-scattered
continua and each can be varied independently relative to the
zeroth-order continuum with the parameters $A_{S00}$ 
and $A_{S90}$. Each component has its own \fekalfa and \fekbeta emission-line
table, each of which has all of its parameters tied to the corresponding continuum 
model parameters. 

\begin{figure}
\centerline{
        \psfig{figure=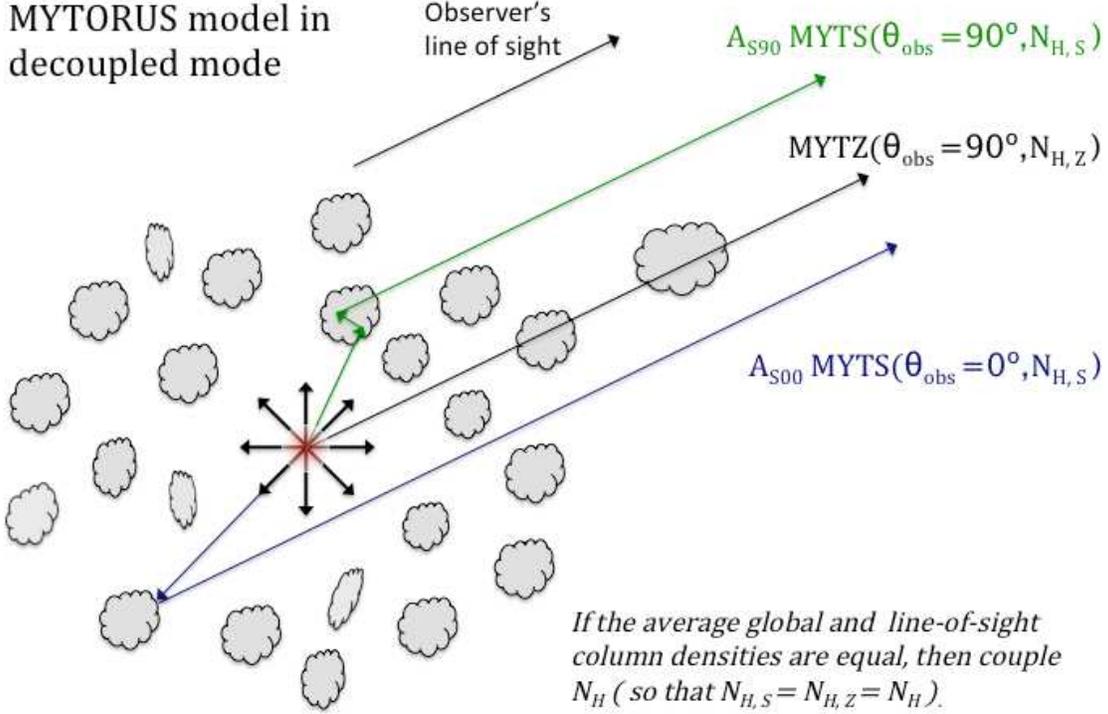,width=15cm,angle=0}
        }
        \caption{\footnotesize Illustration of the physical scenario that is
approximately emulated when the \mytorus model is used in ``decoupled mode'' 
(as described in \S\ref{mytdecoupled}). The X-ray reprocessing material,
represented as clouds, is in the form
of a clumpy distribution with an arbitrary covering factor, and
has an average, global radial column density
denoted by $N_{\rm H,S}$. The X-ray continuum source, shown
with radial lines ending in arrows, is smaller than any of the individual
matter clumps. The spectrum from back-side reflection that reaches
the observer without further interception by any absorbing matter is
approximated using a $0^{\circ}$ MYTS component, and scattered emission
that is not from back-side reflection is approximated using
a $90^{\circ}$ MYTS component. The direct, zeroth-order continuum
that reaches the observer after intercepting line-of-sight material is
modeled with an MYTZ component ($\theta_{\rm obs}$ is set to $90^{\circ}$ in
order for the $N_{H}$ parameter to correspond directly to the
line-of-sight column density). This latter component can have a different
column density, $N_{\rm H,Z}$, if for example, there
is additional material in the line of sight, so that $N_{\rm H,Z}$ is
different to the global average. This is useful for modeling variable
sources in which matter can come into or out of the the line of sight.
Multiple clumps in the line of sight, including those
in the main distribution, can be modeled with a single MYTZ component.
}
\label{fig:mytdecoupled}
\end{figure}

The generalized geometrical setup that is approximated
by the decoupled model is illustrated in Fig.~\ref{fig:mytdecoupled}.
In addition to decoupling of the inclination angle,
there is a provision for the line-of-sight column density ($N_{\rm H,Z}$) to be
decoupled from the global average column density ($N_{H,S}$), and this
is particularly useful for modeling sources in which there is
spectral variability due to matter moving in and out of the line sight.
Only a single line-of-sight component needs to be utilized for a given
observation, corresponding to the total line-of-sight column density.
Individual clumps that move in and out of the line of sight are
assumed to be too small to make a measurable difference to the 
Compton-scattered continuum and fluorescent line emission (if this
is not the case, the differences would manifest themselves in
the existing MYTS components).

Note that neither $A_{S00}$ nor $A_{S90}$ should be interpreted as covering factors.
In a scenario in which the high-energy spectrum is dominated
by the zeroth-order continuum, the global covering factor cannot be constrained
even in principle. This is because any Compton-scattered and fluorescent line flux that 
{\it is} observed in the patchy reprocessor scenario
would be dominated by back-side reflection from the inner far side of
the X-ray reprocessor, through the unobscured patches. The amount of ``leakage'' due
to these patches is not related to the bulk global covering factor. In fact, the
parameter $A_{S00}$ in this case would be more closely related to the
fraction of the total solid angle subtended by the X-ray reprocessor that is
punctured by ``holes.'' Alternatively, in a dual reprocessor scenario, 
the illumination pattern of the extended reprocessor
by the central X-ray source and the exact geometry of circumnuclear
material are uncertain so $A_{S00}$ cannot be interpreted as a simple
covering factor. It should also be remembered that both of the parameters
$A_{S00}$ and $A_{S90}$ also include the effects of any time delays between
the intrinsic continuum and the response of the reflection spectra.  


The resulting decoupled \mytorus model, as implemented in XSPEC, is written out in full
below, and its parameters are summarized in \tablemytdecoupp. The intrinsic X-ray spectrum
here is the \comptt Comptonized thermal spectrum but could just as well be replaced
by the power-law intrinsic continuum (as was used to illustrate the usage of the
coupled \mytorus model in \S\ref{mytcoupled}). The specific XSPEC tables used
in the example expression correspond to a Comptonizing plasma temperature of 
$22$~keV, and an offset of $+20$~eV for the emission-line tables.

\begin{eqnarray}
\rm
model \  = \  constant<1>*phabs<2>( & & \nonumber \\ 
\rm comptt<3>*etable\{mytorus\_Ezero\_v00.fits\}<4> & + & \nonumber \\
\rm constant<5>*atable\{mytorus\_scatteredkT022\_v00.fits\}<6> & + & \nonumber \\
\rm constant<7>*atable\{mytorus\_scatteredkT022\_v00.fits\}<8> & + & \nonumber \\
\rm gsmooth<9>(gsmooth<10>(constant<11>*atable\{mytl\_V000010pEp020kT022\_v00.fits\}<12> & + & \nonumber \\
\rm constant<13>*atable\{mytl\_V000010pEp020kT022\_v00.fits\}<14>)) & + & \nonumber \\
\rm zgauss<15> + zgauss<16>) & + & \nonumber \\
\rm constant<17>*zphabs<18>*comptt<19> + zphabs<20>*apec<21>) \nonumber 
\end{eqnarray}

\begin{table}
\caption[Parameters for the Baseline model with compTT.]
{Parameters for the Baseline Model with {\sc mytorus} Used in Decoupled Mode with a
thermal Comptonized Intrinsic Continuum}
\begin{center}
\begin{tabular}{llllll} 
\hline
& & & & & \\
Par.$^{a}$ & Comp. $^{b}$ & Symbol & Units & Status & Description \\
& & & & & \\
\hline
& & & & & \\
1 & 1 & $C_{\rm PIN:XIS}$, $C_{\rm PDS:MECS}$ & \ldots & fixed & PIN:XIS or PDS:MECS cross-normalization ratio. \\                 
2 & 1 & $C_{\rm GSO:XIS}$ & \ldots & free & GSO:XIS instrumental cross-normalization ratio. \\
3 & 2 & $N_{\rm H,Gal}$ & $10^{21} \rm \ cm^{-2}$  & fixed & Galactic column density, fixed at 1.57. \\
4 & 3,6,8,12,14,19 & $A_{\rm comptt}$ & ph. $\rm cm^{2} \ s^{-1} \ keV^{-1}$ & free & Normalization of the intrinsic continuum (\compttp). \\
5 & 3,6,8,12,14,19 & $kT$ & keV & fixed & Temperature of the Comptonizing plasma. \\
6 & 3,6,8,12,14,19 & $\tau$ & & free & Optical depth of the Comptonizing plasma. \\
7 & 4,6,8,12,14 & $N_{\rm H}$ & $10^{24} \rm \ cm^{-2}$  & free & Column density of all reprocessor components. \\
8 & 5 & $A_{S00}$ & \ldots & free & Scaling factor for a face-on scattered continuum. \\
9 & 7 & $A_{S90}$ & \ldots & free & Scaling factor for an edge-on scattered continuum. \\
10 & 9 & $\sigma_{\rm det}$ & eV & fixed$^{c}$ (eV) & Line width accounting for XIS resolution degradation. \\ 
11 & 10 & FWHM[\fekalfa,\fekbetap] & $\rm km \ s^{-1}$ & fixed$^{d}$ (100) & Intrinsic width of the \fekalfa and \fekbeta lines. \\
12 & 11 & $A_{L00}$ & \ldots & $=A_{S00}$ & Scaling factor for fluorescent line emission. \\
12 & 13 & $A_{L90}$ & \ldots & $=A_{S90}$ & Scaling factor for fluorescent line emission. \\
14 & 15 & $E_{\rm Fe~XXV}$ & keV & free$^{d}$ & Gaussian centroid energy of the line emission at $\sim 6.7$~keV. \\
15 & 15 & FWHM[\fexxvp]  & $\rm km \ s^{-1}$ & free$^{d}$ & Intrinsic width of the line emission at $\sim 6.7$~keV. \\
16 & 15 & $I_{\rm Fe~XXV}$ & ph. $\rm cm^{2} \ s^{-1}$ & free & Flux of line emission at $\sim 6.7$~keV. \\
17 & 16 & $E_{\rm Ni~K\alpha}$ & keV & free$^{d}$ & Gaussian centroid energy of the \nika line. \\
18 & 16 & FWHM[\nika]  & $\rm km \ s^{-1}$ & fixed$^{d}$ & Intrinsic width of the \nika line. \\
19 & 16 & $I_{\rm Ni~K\alpha}$ & ph. $\rm cm^{2} \ s^{-1}$ & free & Flux of the \nika line emission. \\
20 & 17 & $f_{s}$ & \ldots & free & Scattering fraction due to optically-thin matter. \\ 
21 & 18 & $N_{\rm H1}$ & $10^{21} \rm \ cm^{-2}$ & free & Absorber covering optically-thin scattering zone. \\
22 & 20 & $N_{\rm H2}$ & $10^{21} \rm \ cm^{-2}$ & free & Absorber covering extended thermal emission zone. \\
23 & 21 & $A_{\rm apec}$ & ph. $\rm cm^{2} \ s^{-1} \ keV^{-1}$ & free & Normalization of the {\sc apec} 
model. \\
24 & 21 & $kT_{\rm apec}$ & keV & free & Temperature of the {\sc apec} model. \\
& & & & & \\
\hline
\end{tabular} 
\end{center}
$^{a}$ Parameter number. 
$^{b}$ Model component(s) that the
parameter appears in (corresponding to the model component numbers
in the model expression shown in \S\ref{mytdecoupled}).
$^{c}$ Determined from fitting calibration source data       
(see \S\ref{suzakudata}).
$^{d}$ The Gaussian emission-line centroid energies, and the 
intrinsic width of the \fexxv line  were initially
allowed to float but they were then frozen at their best-fitting
values in order to obtain robust statistical errors on the
other parameters of interest. A single intrinsic width is 
associated with the \fekalfa
and \fekbeta lines, and this width, along with that of the
\nika line was fixed at  
$100 \ \rm km \ s^{-1}$~FWHM since the lines were found to be unresolved.
Upper limits were derived by allowing the two line widths    
to float in turn (see \S\ref{coupledmytfits} and \S\ref{decoupledmytfits}).
\end{table}

Although the decoupled \mytorus model involves some \adhoc parameterization
in order to account for the unknown details of the exact geometry of the
X-ray reprocessor and its unknown clumpiness, the procedure is different
to the current practice of using the inappropriate geometry of disk reflection 
(in which the disk has an infinite column density), plus zeroth-order 
continuum extinction that has incorrect physics, along with \fekalfa line emission
that cannot be related to a physical reprocessor. The line-of-sight column density and
intrinsic luminosity derived from the
latter model may not have simple physical interpretations. Moreover,
the proper use of a finite column density for the 
material responsible for the Compton-scattered continuum produces
a rich variety of spectral shapes that cannot be produced by standard
disk-reflection spectra. Residuals in spectral fits using the disk-reflection
models might therefore be misinterpreted and erroneously identified with a
different origin.

\subsection{Other Models of the Compton-thick X-ray Reprocessor}
\label{bnctmodels}

We will also refer to spectral fits using the toroidal and spherical
X-ray reprocessor models of BN11, implemented using the XSPEC tables
{\tt torus1006.fits} and {\tt sphere0708.fits} respectively (see BN11 for details).
The model parameter setup for the BN11 toroidal model is essentially
similar to that of the \mytorus model in \tablemytdefparsp, with two exceptions.
One is that there are no parameters corresponding to $A_{S}$ and $A_{L}$,
because neither the Compton-scattered continuum nor the fluorescent line
spectrum can be varied with respect to the zeroth-order continuum.
The second difference is that there is an extra parameter in the BN11 torus model
that corresponds to the half-opening angle of the toroidal structure, which
therefore controls the global covering factor (which can be varied between
0.1 and 0.9). For the BN11 spherical model, the parameter setup is again
similar to that for the BN11 toroidal model, except that the covering
factor is effectively fixed at 1.0, and there is no inclination angle
parameter because of the spherical symmetry. Two extra parameters
allow element abundances to be free.
The spherical model is currently the only Compton-thick reprocessor
model available that allows element abundances to be free
parameters. We call the two parameters $X_{\rm Fe}$ and $X_{\rm M}$,
where the former is the Fe abundance relative to the
adopted solar value, and the latter is a single abundance multiplier for 
C, O, Ne, Mg, Si, S, Ar, Ca, Cr, and Ni relative to their respective solar values.
The solar abundances adopted in the BN11 spherical
model are the same as those used in the \mytorus model (Anders \& Grevesse 1989).
The column density in the spherical model, $N_{\rm H}$, is the radial column density.
A technical issue with the both of the BN11 models is that the energies of
the fluorescent emission lines cannot be varied. This is problematic because
the signal-to-noise ratio of the NGC~4945 \suzaku data is so high that
instrumental calibration errors and/or mild ionization cannot be ignored.
The spectral fitting is highly sensitive to offsets as small as 10~eV. To
accommodate the inflexibility of the BN11 models, we first allowed the redshift
parameter to float and let the \fekalfa line drive the spectral fits to 
the optimal redshift. The redshift was then frozen permanently at the best-fitting value
before proceeding with the full spectral-fitting analysis.

\subsection{Spectral Fitting}
\label{spfitting}

We used XSPEC (Arnaud 1996) v12.6\footnote{http://heasarc.gsfc.nasa.gov/docs/xanadu/xspec/}
for spectral fitting (Arnaud 1996).
Galactic absorption with a  column density of $1.57 \times 10^{21} \ \rm cm^{-2}$
(Heiles \& Cleary 1979) was included in all of the models
described hereafter and its inclusion will be implicitly assumed.
For all absorption components including the Galactic one, we used
photoelectric cross sections given by Verner \etal (1996).
Element abundances from Anders \& Grevesse (1989) were used throughout.
All astrophysical model parameter values will be given in the rest frame of NGC~4945,
unless otherwise stated.
Due to the large number of spectral fits and the large variety of
combinations of model components and parameters, for the sake of brevity,
certain quantities and details pertaining to particular spectral fits will be given
in the tables of results and not repeated again in the text, unless
it is necessary. Specifically, we are referring to the number of free parameters
in a fit, the number of interesting parameters,
the number of degrees of freedom, the null hypothesis probability,
and the $\Delta \chi^{2}$ criteria for the derivation of statistical errors.
In most cases we give 68\% confidence multiparameter errors, since
these $1\sigma$ errors can be used in future statistical analyses of samples of AGN. However, in some
cases, certain parameters have to be frozen for the sake of stability of
a spectral fit, and in that case those parameters are allowed to be free,
one parameter at a time, in order to derive statistical errors. 
For these, one-parameter, 90\% confidence errors
are given. Further details will be given on a case-by-case basis.

For each spectral fit we will give the observed fluxes, $F_{\rm obs}$, and
the observed (rest-frame) luminosities, $L_{\rm obs}$, in various energy
bands, in the tables of spectral-fitting results. These quantities are
not corrected for absorption or Compton scattering in either
the line-of-sight material, or in the circumnuclear material. 
These fluxes and luminosities will not be discussed until all of the spectral fits
have been presented. At that point, in \S\ref{lumratios}, we will present and
discuss
the calculated intrinsic continuum luminosities for all of the spectral fits,
as well as the ratio of the luminosities to the Eddington luminosity, $L_{\rm Edd}$.
For a black-hole mass of $1.4 \times 10^{6} \ M_{\odot}$, $L_{\rm Edd}
= 1.77 \times 10^{44} \rm \ erg \ s^{-1}$.
We use a standard cosmology of $H_{0} = 70 \
\rm km \ s^{-1} \ Mpc^{-1}$, $\Lambda = 0.73$, $q_{0} = 0$ throughout the paper.

\section{Spectral Fits with Only a Zeroth-Order Continuum}
\label{zerothorderfits}

\begin{figure}
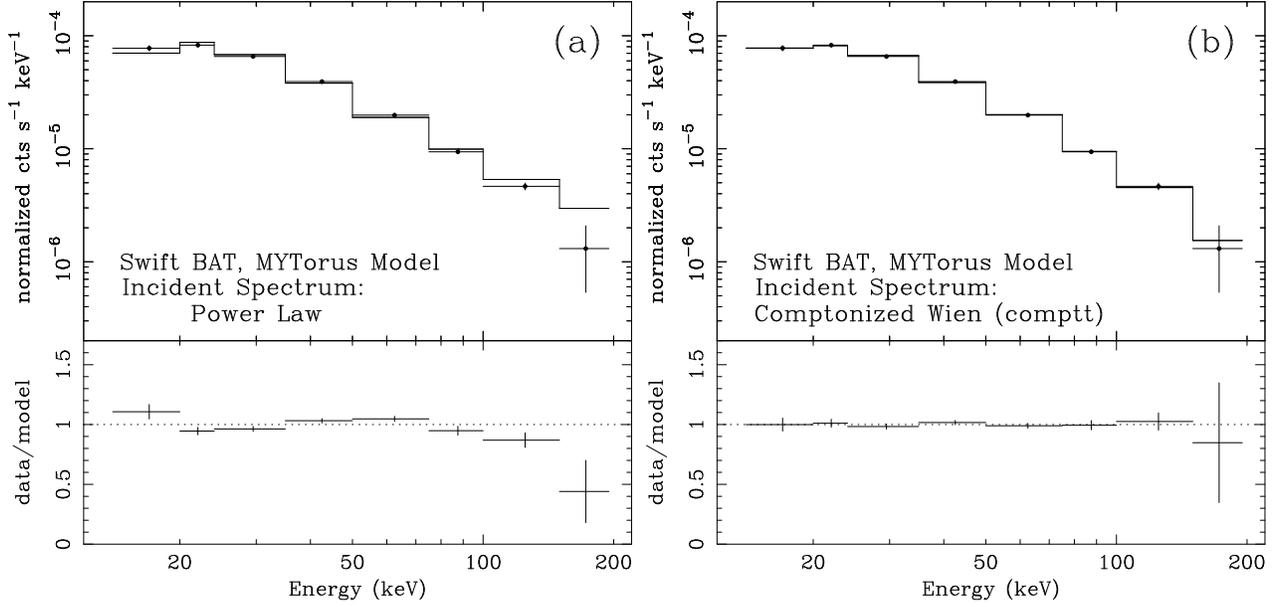

\centerline{
 \psfig{figure=f3a.ps,height=8cm,angle=270}
 \psfig{figure=f3b.ps,height=8cm,angle=270}
}
\caption{\footnotesize Spectra and data/model
ratios for the \swift BAT 58-month data for NGC~4945 where the model 
consists of only the zeroth-order component of the \mytorus model. 
Shown in (a) are the results for
an intrinsic power-law continuum (the photon index and column density were free
parameters). Shown in (b) are the results for a thermally Comptonized incident
spectrum (the plasma temperature and column density were free parameters).
The power-law incident continuum gives a poor fit. See \tablebatresults and
\S\ref{zerothorderfits} for details. 
}
\label{fig:batonlyfits}
\end{figure} 

\begin{figure}
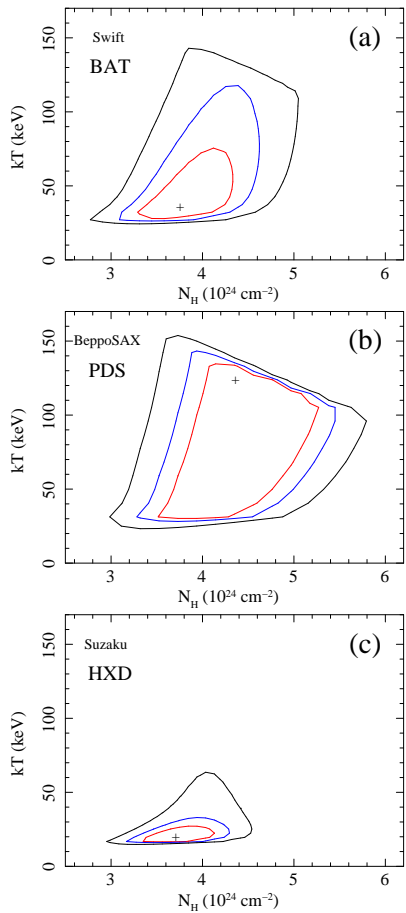

\centerline{
 \psfig{figure=f4a.ps,height=4cm,angle=270}}
\centerline{
 \psfig{figure=f4b.ps,height=4cm,angle=270}}
\centerline{
 \psfig{figure=f4c.ps,height=4cm,angle=270}}
\caption{\footnotesize 
Confidence contours of the intrinsic continuum Comptonizing plasma temperature ($kT$)
versus column density of the X-ray reprocessor. The contours were obtained from spectral
fits to NGC~4945 data where the model consisted only of the zeroth-order continuum
for  a thermally Comptonized spectrum incident on a Compton-thick X-ray reprocessor.
For these fits only data above 10~keV were utilized
because in that regime the Compton-scattered continuum in NGC~4945 may be negligible
in some physical scenarios, such as a ring-like Compton-thick patchy X-ray reprocessor observed edge-on.
The three sets of contours shown in (a), (b), and (c) correspond to spectral fits to
data from the \swift BAT, \bsax PDS, and \suzaku HXD (PIN and GSO) respectively.
The red, blue, and black lines correspond
to two-parameter,
68\%, 90\%, and 99\% confidence levels respectively. See \S\ref{zerothorderfits} for details.
}
\label{fig:ktvsnhcont}
\end{figure}

\begin{figure}
\centerline{
 \psfig{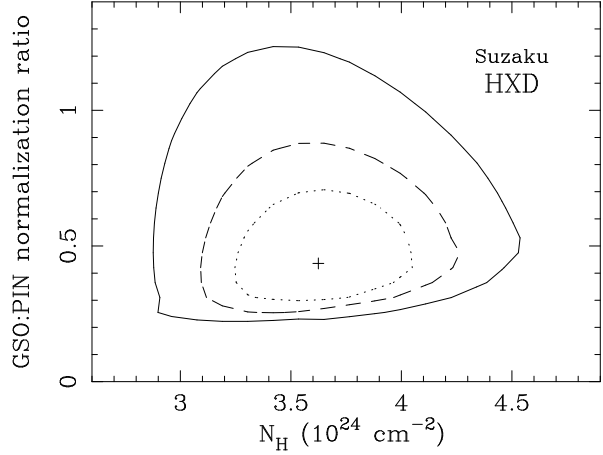}
}
\caption{\footnotesize
Confidence contours obtained from spectral
fits to the NGC~4945 PIN and GSO data using a thermally Comptonized spectrum incident on a Compton-thick reprocessor.
Only the zeroth-order continuum component of the \mytorus model was utilized,
modeling physical scenarios in which the Compton-scattered continuum
is negligible above 10~keV.
The relative normalization factor for the PIN and GSO was a free parameter in the fits,
and the contours show the value of this parameter versus the line-of-sight column density, $N_{\rm H}$, of the
reprocessor (the Comptonizing plasma temperature, $kT$ was also a free parameter).
The dotted, dashed, and solid lines correspond
to two-parameter,
68\%, 90\%, and 99\% confidence levels respectively. It can be seen that $N_{\rm H}$ is
uncorrelated with the GSO:PIN normalization ratio. See \S\ref{zerothorderfits} for details.
}
\label{fig:gsopinvsnh}
\end{figure}

\begin{figure}
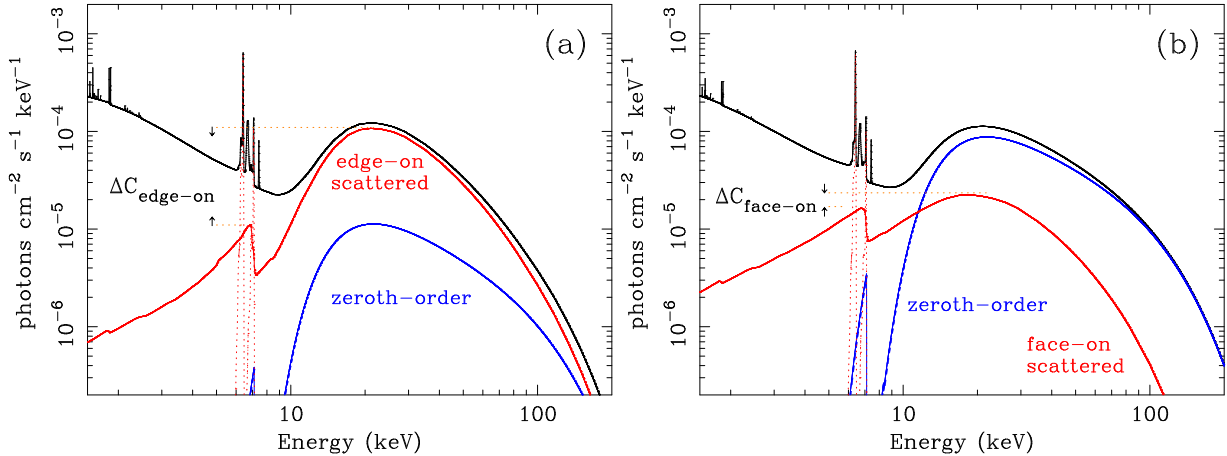

\centerline{
 \psfig{figure=f6a.ps,height=6cm,angle=270}
 \psfig{figure=f6b.ps,height=6cm,angle=270}
}
\caption{\footnotesize A demonstration of why an edge-on Compton-scattered spectrum forces
a solution in which the magnitude of the zeroth-order continuum is substantially
diminished compared to the Compton-scattered continuum. Shown in (a) is a \mytorus model fit
to the \suzaku data of NGC~4945 with a decoupled, edge-on scattered continuum. In this case,
the contrast, $\Delta C_{\rm edge-on}$, between the peak of the ``Compton hump'' and the scattered continuum in the Fe~K band,
is so large that in order to fit the \fekalfa line, the magnitude of the scattered continuum
is forced to be much larger than the zeroth-order continuum. In (b) the edge-on scattered
continuum is replaced by a decoupled face-on scattered continuum, and in this case the
contrast, $\Delta C_{\rm face-on}$, is much less than $\Delta C_{\rm edge-on}$, and
the \fekalfa line flux is much larger, so a solution in which the zeroth-order continuum
dominates is possible. Note that the black curves show the total spectra including
all of the model components shown in \tablemytdecoupp, but not all of the individual
components are shown separately for the sake of clarity. See \S\ref{zerothorderfits} for details.} 
\label{fig:dfzdspectra}
\end{figure}

In this section we investigate spectral solutions in which the 
spectra above 10~keV are fitted {\it only} with the zeroth-order continuum.
The Compton-scattered continuum cannot of course be zero because
the model \fekalfa emission-line flux cannot be zero. However, if we are looking
for solutions in which the Compton-scattered continuum is negligible compared
to the zeroth-order continuum, we can work with the approximation of {\it no}
Compton-scattered continuum and {\it no} fluorescent line emission when we
are fitting data above 10~keV. We can even omit the optically-thin scattered
continuum and we can certainly omit the soft X-ray thermal emission components
(and any absorption only associated with those components). We can then
use the results from these simplified fits as a guide for setting up the
correct, full-bandpass model that {\it will} include the Compton-scattered
continuum and fluorescent line emission. Moreover, since the zeroth-order
continuum is obtained from the intrinsic X-ray continuum simply by multiplying
by an energy-dependent factor, we can use an arbitrary incident continuum
model with all of its key parameters allowed to float (as opposed to having to
generate tables that only allow a restricted number of free parameters).

\begin{table}
\caption[NGC~4945 {\it Swift} BAT Spectral-Fitting Results.]
{NGC~4945 {\it Swift} BAT Spectral-Fitting Results.}
\begin{center}
\begin{tabular}{lccccc}
\hline
& & & & & \\
Parameter & zeroth-order &  \mytorus & \mytorus & BN11 (torus) & BN11 (sphere) \\
&              only &  (coupled)  & (coupled) & & \\ 
& & & & & \\
\hline
& & & & & \\
Intrinsic Continuum & \comptt & {\sc power law} & \comptt & {\sc power law} & {\sc power law} \\
Free Parameters & 4 & 4 & 4 & 4 & 3 \\
Interesting Parameters & 3 & 3 & 3 & 3 & 2 \\
$\chi^{2}$ & $2.05$ & $2.38$ & $0.99$ & $1.90$ & $4.60$ \\
Degrees of Freedom & 4 & 4 & 4 & 4 & 5 \\
Reduced $\chi^{2}$ & $0.51$ & 0.596 & $0.250$ & 0.474 & 0.920 \\
Null probability & $0.727$ & $0.666$ & $0.910$ & $0.755$ & $0.467$ \\
$\Delta\chi^{2}$ criterion (68\% confidence) & $3.50$ & $3.50$ & $3.50$ & $3.50$ & $2.28$ \\
$\Gamma$ & \ldots &  $1.83^{+0.15}_{-0.22}$ & \ldots &  $1.59^{+0.27}_{-0.03}$ & $1.75^{+0.02}_{-0.03}$ \\
$\tau$ & $0.94^{+0.55}_{-0.76}$ & \ldots & $0.99^{+0.18}_{-0.20}$ & \ldots & \ldots \\
$kT$ & $36^{+62}_{-10}$ & \ldots & $51^{+16}_{-17}$ & \ldots & \ldots \\
$N_{\rm H} \rm \ (10^{24} \ cm^{-2})$ & $3.78^{+0.71}_{-0.63}$ & $>3.37$ & 
 $3.13^{+0.77}_{-0.58}$ & $>2.95$ & $2.74^{+0.27}_{-0.25}$  \\
$\theta_{\rm torus}$ (degrees)$^{a}$ & \ldots & $60$(f) & $60$(f) & $59.7^{+0.3}_{-3.8}$ & \ldots \\
$\theta_{\rm obs}$ (degrees)$^{b}$ & \ldots & $78(>60)$ & $90$(f) & $87.1$ (f) & \ldots \\
& & & & & \\
$F_{\rm obs}$[10-100  keV] ($10^{-11} \rm \ erg \ cm^{-2} \ s^{-1}$) & $16.8$ & $18.2$ & $17.8$ & $18.4$ & $18.2$ \\
$F_{\rm obs}$[14-195  keV] ($10^{-11} \rm \ erg \ cm^{-2} \ s^{-1}$) & $24.6$ & $25.1$ & $25.0$ & $24.8$ & $25.6$ \\
$L_{\rm obs}$[10--100 keV] ($10^{42} \rm \ erg \ s^{-1}$) & $1.30$ & $1.41$ & $1.38$ & $1.42$ & $1.41$ \\
$L_{\rm obs}$[14--195 keV] ($10^{42} \rm \ erg \ s^{-1}$)  & $1.91$ & $1.94$ & $1.94$ & $1.92$ & $1.98$ \\
& & & & & \\
\hline
\end{tabular}
\end{center}
Results for the \swift BAT data for NGC~4945 from spectral-fitting
with Compton-thick reprocessor models. Details can be found in
\S\ref{zerothorderfits} for the fit with only the zeroth-order 
continuum, and for the remaining fits in \S\ref{coupledmytfits}.
The BN11 torus and spherical fits refer to the models of
Brightman \& Nandra (2011).
All fluxes and luminosities are in the observed frame,
uncorrected for absorption or Compton scattering.
$^{a}$ Torus opening half-angle (the maximum allowed
for the BN11 model is $87.1$ degrees). $^{b}$ Inclination angle
of the observer with respect to the azimuthal symmetry axis of the torus.
\end{table}

\subsection{Zeroth-order Continuum Fits to the Swift BAT Data}
\label{batzerothfits}
 
We begin by first reporting the results of fitting the 58-month \swift BAT spectrum using
a power-law incident continuum.
There were only three free parameters,
$A_{\rm PL}$, $\Gamma$, and $N_{\rm H}$ (see \S\ref{mytorusmodel}) and 8 spectral channels
in the spectrum (i.e., 5 degrees of freedom). It can be seen 
from Fig.~\ref{fig:batonlyfits}(a) that the fit is
very poor (the {\it reduced} $\chi^{2}$ value is 5.65), and the 
data/model residuals in Fig.~\ref{fig:batonlyfits}(a) show significant
deviations across the entire 14--195~keV band, but in particular show a
high-energy rollover compared to a power-law.
The probability of obtaining the measured $\chi^{2}$ value or higher is
$3.2 \times 10^{-5}$, meaning that the model can be formally rejected at 
greater than the $4.1\sigma$ level. We conclude that if the high-energy
X-ray spectrum in NGC~4945 is to be dominated by the zeroth-order continuum,
the intrinsic X-ray continuum cannot be a simple power law, and the
spectrum has to start rolling over below $\sim 100$~keV.
Therefore, we replaced the power-law continuum with the Comptonized
thermal continuum model, \comptt (as described in \S\ref{mytdecoupled}). 
The model now had 4 free parameters, namely the overall normalization, plasma 
optical depth ($\tau$) and temperature ($kT$), and the line-of-sight
column density ($N_{H}$). An excellent fit was obtained, with $\chi^{2}=2.05$
for 4 degrees of freedom, as can be seen in Fig.~\ref{fig:batonlyfits}(b), which
shows the data, model, and data/model ratio. We obtained best-fitting
parameters $\tau=0.94^{+0.55}_{-0.76}$, $kT=36^{+62}_{-10}$~keV, and $N_{\rm H}=3.78^{+0.71}_{-0.63}
\ \rm 10^{24} \ \rm cm^{-2}$ (see \tablebatresultsp).
The best-fitting parameters
are also given in \tablebatresultsp, where they can be compared with
parameters from other models that were fitted to the
\swift BAT (to be described below). In Fig.~\ref{fig:ktvsnhcont}(a) we show the two-parameter,
68\%, 90\%, and 99\% confidence contours of $kT$ versus $N_{\rm H}$. We
see that the 99\% confidence lower limit on $kT$ is flat (at $\sim 20$~keV) over most of
the allowed range of $N_{\rm H}$, and the 99\%, two-parameter upper limit on $kT$ is $\sim 140$~keV.

\subsection{Zeroth-order Continuum Fits to the Suzaku and BeppoSAX Data}
\label{zerothszbsax}

We repeated the exercise, this time
fitting high-energy data from \bsax and \suzakup, with the same simple
model that includes only the zeroth-order continuum (for an intrinsic 
continuum modeled by \compttp).
Confidence contours of $kT$ versus $N_{\rm H}$
are shown in Fig.~\ref{fig:ktvsnhcont}(b) and Fig.~\ref{fig:ktvsnhcont}(c) for the \bsax and \suzaku data
respectively. Since we were explicitly examining the high-energy spectra, we used
only the PDS data for \bsax
(see \S\ref{saxdata}) and only the PIN and GSO data for \suzaku
(see \S\ref{suzakudata}). All of the data sets
gave a consistent, albeit large, range in $kT$ and $N_{\rm H}$. 
The \suzaku data gave the tightest constraints on $kT$ and $N_{\rm H}$,
with best-fitting values of 22~keV and $3.7 \times 10^{24} \ \rm cm^{-2}$ respectively.
We do not give the exact parameters and statistical errors derived from the \bsax and
\suzaku fits because these parameters will change when 
they are derived from full model fits that will
include the Compton-scattered continua and fluorescent lines (and the data below 10~keV).
Since the \mytorus Compton-scattered continuum tables have to be calculated for
fixed values of $kT$, we will use the contours in Fig.~\ref{fig:ktvsnhcont} as a guide
for the full-band spectral fits (see \S\ref{decoupledmytfits}).
We note that for the \suzaku confidence contours in Fig.~\ref{fig:ktvsnhcont}, the GSO to XIS 
cross-normalization ratio was a free parameter in the spectral fit.
Fig.~\ref{fig:gsopinvsnh} shows two-parameter, 68\%, 90\%, and 99\% confidence contours
for this cross-normalization versus $N_{\rm H}$, and we see
that the ratio is uncorrelated with the derived column density. The range
in the ratio is fairly large, but nevertheless consistent with the results for 3C 273
given in the appendix. 

\subsection{A Requirement of the Compton-scattered Continuum If the High-energy Spectrum
is Dominated By the Zeroth-order Continuum}
\label{faceonvsedgeoncscont}

We now demonstrate that if we are looking for Compton-thick solutions in which the
high-energy spectrum is dominated by the zeroth-order continuum, then the 
Compton-scattered continuum and \fekalfa line emission must be dominated by
photons originating from back-illumination
and then reaching the observer along paths that do not intercept the Compton-thick
structure. In other words, the structure must be clumpy, allowing 
a reflection continuum to reach the observer either
from the far inner side of a toroidal structure, or from an extended
and dispersed distribution of matter (as in NGC~4945). The reason for
this constraint is that the \fekalfa line emission and Compton-scattered
continuum below 10~keV, relative to the Compton-hump peak flux in the $\sim 10-30$~keV
range, is much larger for unobstructed reflection than it is from reflection from
matter intercepting the line of sight (e.g., see Yaqoob \etal 2010). If the
reprocessor were not patchy, then raising the \fekalfa line flux high enough to account
for the observed spectral data also raises the Compton-scattered flux so high that
the Compton hump necessarily swamps the zeroth-order continuum. 
This is illustrated in Fig.~\ref{fig:dfzdspectra}, which shows model spectra in which 
the \mytorus Compton-scattered continuum and fluorescent lines are decoupled from the 
zeroth-order continuum (see \S\ref{mytdecoupled}), comparing the two cases in which
the reflection features are observed edge-on (Fig.~\ref{fig:dfzdspectra}(a)) and
face-on (Fig.~\ref{fig:dfzdspectra}(b)). Both of these extreme cases fit the
full-band (three-instrument) \suzaku data, but Fig.~\ref{fig:dfzdspectra} shows that only the
face-on case allows the zeroth-order continuum to dominate above 10~keV.
The intrinsic continuum in Fig.~\ref{fig:dfzdspectra} was the \comptt model with $kT=22$~keV.
We do not give the remaining model parameters here because  
detailed fits with variations on this decoupled \mytorus model will be given below, in \S\ref{decoupledmytfits}.
The purpose of Fig.~\ref{fig:dfzdspectra} is only to illustrate the need for a clumpy X-ray reprocessor
if the hard X-ray continuum variability in NGC 4945 is to be produced by the zeroth-order continuum.
In practice there is likely to be
some contribution from both the far-side unobscured reflection, and the near-side
emission from the obscuring material,
but the former component must dominate over the latter component. Even when this condition is
met, the Compton-scattered continuum spectrum cannot be too high, otherwise it would again
dominate over the zeroth-order continuum and we are not looking for those 
solutions here (which will be discussed in \S\ref{coupledmytfits}).
\section{Spectral Fits with Coupled Reprocessor Models}
\label{coupledmytfits}

\begin{figure}
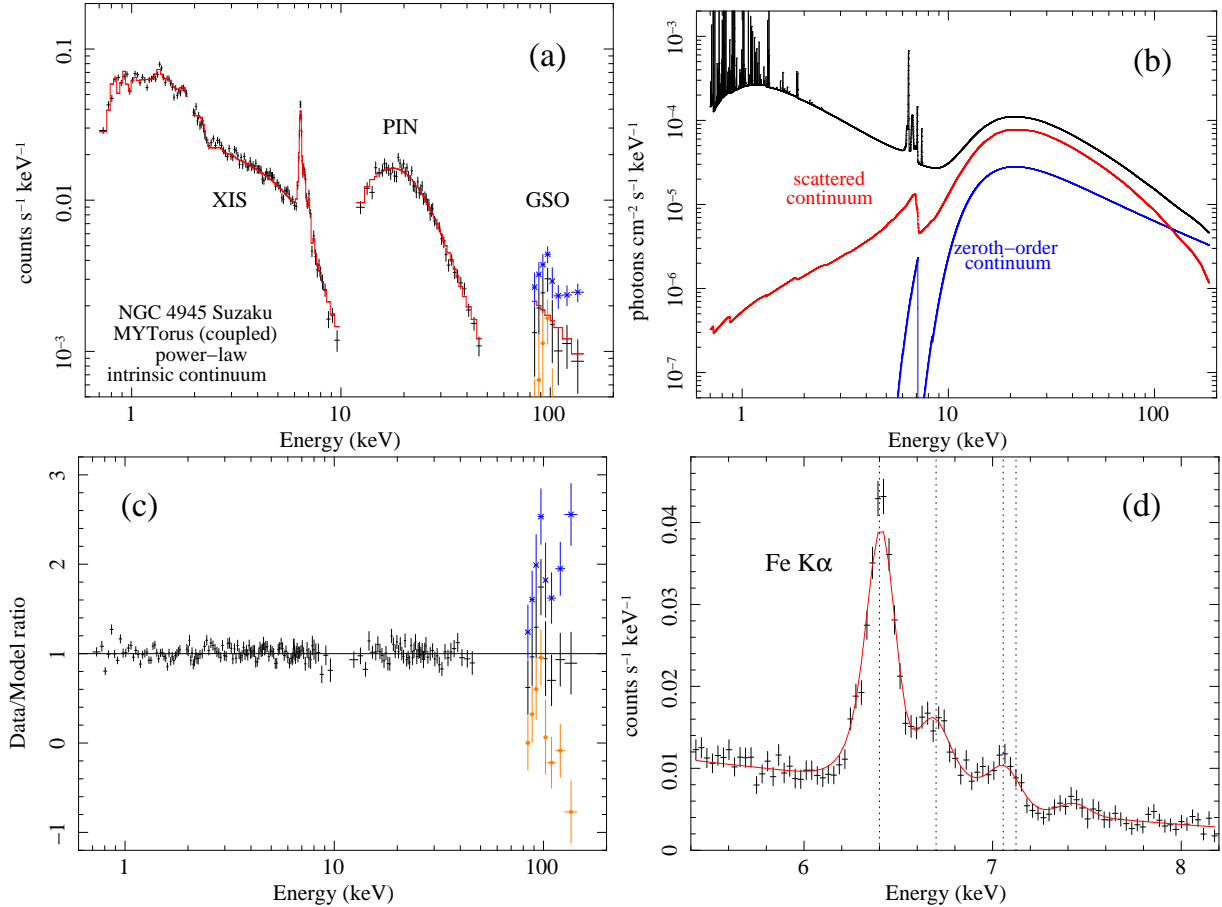

\centerline{
 	\psfig{figure=f7a.ps,height=6cm,angle=270}	
        \psfig{figure=f7b.ps,height=6cm,angle=270}
	}
\centerline{
        \psfig{figure=f7c.ps,height=6cm,angle=270}
        \psfig{figure=f7d.ps,height=6cm,angle=270}        
        }
        \caption{\footnotesize Results of spectral fitting to the
NGC~4945 \suzaku data with the default \mytorus X-ray reprocessor model
(see \S\ref{coupledmytfits} for details). 
In the default configuration, the zeroth-order continuum is linked to
the scattered continuum (and fluorescent lines) by virtue of the 
specific geometry adopted for the \mytorus model. 
In this configuration, the best fit gives a Compton-scattered continuum 
that dominates over the zeroth-order continuum below $\sim 100$~keV. 
In this scenario, the X-ray reprocessor
must be compact enough for the Compton-scattered continuum to respond to
intrinsic continuum variability. 
(a) The counts spectra overlaid with
the folded best-fitting model.
The three data groups from left to right 
correspond to data from the XIS, HXD/PIN, and the HXD/GSO. 
The data from XIS0, 1, 2, and 3 are combined into a single spectrum.
For clarity, for some energy ranges, 
the data shown in the plot are binned more coarsely 
than the binning used for the actual spectral fitting,
For the GSO, the effect of a systematic error in the background
model of $-2\%$ and $+2\%$ is shown in blue and brown respectively (see text).
(b) The total best-fitting model photon spectrum (black), with the contribution from the
Compton-scattered continuum and the zeroth-order continuum shown
in red and blue, respectively. Other constituent components are not shown
for clarity (but are included in the total spectrum).
(c) The ratios of the data to the best-fitting model.
Again, two additional ratios are shown for the GSO, corresponding to
the background-subtraction systematic errors used in (a).
(d) A close-up zoom of the Fe~K region in panel (a).
From left to right, the dotted lines correspond to the energies
of Fe~{\sc i}~K$\alpha$, \fexxv(r), Fe~{\sc i}~K$\beta$, and
the neutral Fe~K edge. The bump above the Fe~K edge is the \nika line.}
\label{fig:mytszcp}
\end{figure}

In this section we present the results of fitting the \suzakup,
\bsaxp, and \swift BAT spectra with coupled models of the Compton-thick
X-ray reprocessor (see \S\ref{strategy}).
We fitted three sets of models, using \mytorusp, the BN11 torus, and
the BN11 sphere, for the Compton-thick reprocessor. We tried using
both a simple power-law and Comptonized thermal intrinsic continuum
in the case of the \mytorus models. The BN11 models do not allow
any other intrinsic continuum aside from a power law.

\subsection{BN11 toroidal model}
We found that we could not obtain an acceptable fit with the BN11 toroidal model to
either the \suzaku data or the \bsax data, even with all parameters of the
model allowed to float, including the toroidal opening angle.
(The other free parameters are 
the same as those of the \mytorus model.)
The best values of the reduced $\chi^{2}$ that could be
obtained were $1.82$ and $4.40$ for the \suzaku and \bsax data respectively.
For the \suzaku fit, the probability for obtaining such a bad fit or worse, due
to statistical fluctuations alone, is $3.5 \times 10^{-19}$. For the \bsax data,
the null hypothesis probability is 10 orders of magnitude worse.
One reason for the poor fits is that
the geometry of the toroidal BN11 model
is rather peculiar in that all rays from the intrinsic X-ray continuum
are presented with the same zeroth-order column density, whereas in
the \mytorus model, rays incident at the equator are presented with a larger
column density than those incident away from the equator. This results
in very different emergent spectra from the BN11 model compared to the \mytorus model.
The shape of the $\sim 10-50$~keV spectral hump in the BN11 model severely mismatches the data,
and the \fekalfa line fluxes associated with the reprocessed continuum
cannot match the data. 

The \swift BAT data, having a lower statistical quality than the other
data sets, did yield an acceptable fit, but the orientation of the
torus had to be fixed at the highest inclination allowed by the model,
and only a lower limit could be obtained on $N_{\rm H}$. The results
for this fit are shown in \tablebatresultsp.

\subsection{\suzaku Fits with the Coupled \mytorus Model}

For the \mytorus model, good fits were obtained for the \suzaku data
with both intrinsic continuum models, and the results are shown in
\tablemytcpplresultsp. The difference in $\chi^{2}$ between the two fits
is not significant, bearing in mind the complexity of the models and
systematic errors that we do not account for. 
Note that the parameter $A_{S}$ is free in both fits and the best-fitting
values center around unity, which corresponds to the default, steady-state
configuration of \mytorusp. The orientation of the torus is essentially edge-on
for both intrinsic continua. We will discuss the column density after
discussing the remaining fits. 

The fit to the \suzaku data with
the power-law intrinsic continuum is shown in Fig.~\ref{fig:mytszcp}.
The folded model and counts spectra are shown in Fig.~\ref{fig:mytszcp}(a),
the best-fitting photon spectrum is shown in Fig.~\ref{fig:mytszcp}(b),
and the data/model ratios are shown in Fig.~\ref{fig:mytszcp}(c).
It can be seen that an excellent fit is obtained over the
whole bandpass. In this fit, the continuum above 10~keV is dominated
by the Compton-scattered continuum, not the zeroth-order continuum.
In order to assess the importance of the GSO 
background-subtraction systematics,
we applied systematic offsets to the background spectrum of $-2\%$ and
$+2\%$
(the extreme values for the advertised systematic errors in
the GSO background model). The results are shown in Fig.~\ref{fig:mytszcp},
panels (a) and (c), from which it can be seen that the negative
offset can result in an error in the background-subtracted spectrum
of up to $\sim 150\%$, and the positive offset can result
in negative counts in the background-subtracted spectrum.
The GSO:XIS fitted normalization parameter very likely includes
some compensation for any background-subtraction systematic errors.

The folded model and counts spectrum zoomed in on the
Fe~K region is shown in Fig.~\ref{fig:mytszcp}(d), and it
can be seen that the detailed fit is excellent. All four emission lines 
(\fekalfap, \fexxvp, \fekbetap, and \nikap), and the Fe~K
edge are very well modeled. The line emission at $\sim 6.7$~keV
must originate in a region that is distinct from the
one that produces the \fekalfap, \fekbetap, and \nika lines
because these latter three lines, and the Fe~K edge are
modeled with neutral matter. It also appears that no \fexxvi line
emission is required: including an additional 
unresolved Gaussian emission-line
component at 6.966~keV gives a two-parameter, 90\% confidence upper limit on
the EW of 34~eV.

\subsection{\bsax and \swift BAT Fits with the Coupled \mytorus Model}

Next, we describe the fits to the \bsax data. In this case, we were not able to
obtain an acceptable fit with a power-law intrinsic continuum and \mytorusp.
The best value of the reduced $\chi^{2}$ that we obtained was
$1.91$, with a null probability of $8.5 \times 10^{-6}$. However, when we
used the Comptonized thermal model for the intrinsic continuum we did obtain
an acceptable fit (see \tablemytcpplresultsp) but with one very important
concession. That is, the parameter $A_{S}$ is required to be an order-of-magnitude
smaller than unity ($A_{S}=0.101^{+0.149}_{-0.068}$, $A_{L} \equiv A_{S}$), which means that
the Compton-scattered continuum is heavily suppressed (but the fluorescent
line spectrum is still high enough to fit the \fekalfa line).

Spectral-fitting results to the \swift BAT spectrum
with the \mytorus model, using both intrinsic
continuum models, are shown in \tablebatresultsp, and acceptable
fits were obtained in both cases. The parameter $A_{S}$ was fixed at $1.0$
in both cases.

\subsection{Spectral Fits with a Spherical Model}
\label{sphericalfits}

\begin{figure}
\centerline{
     \psfig{figure=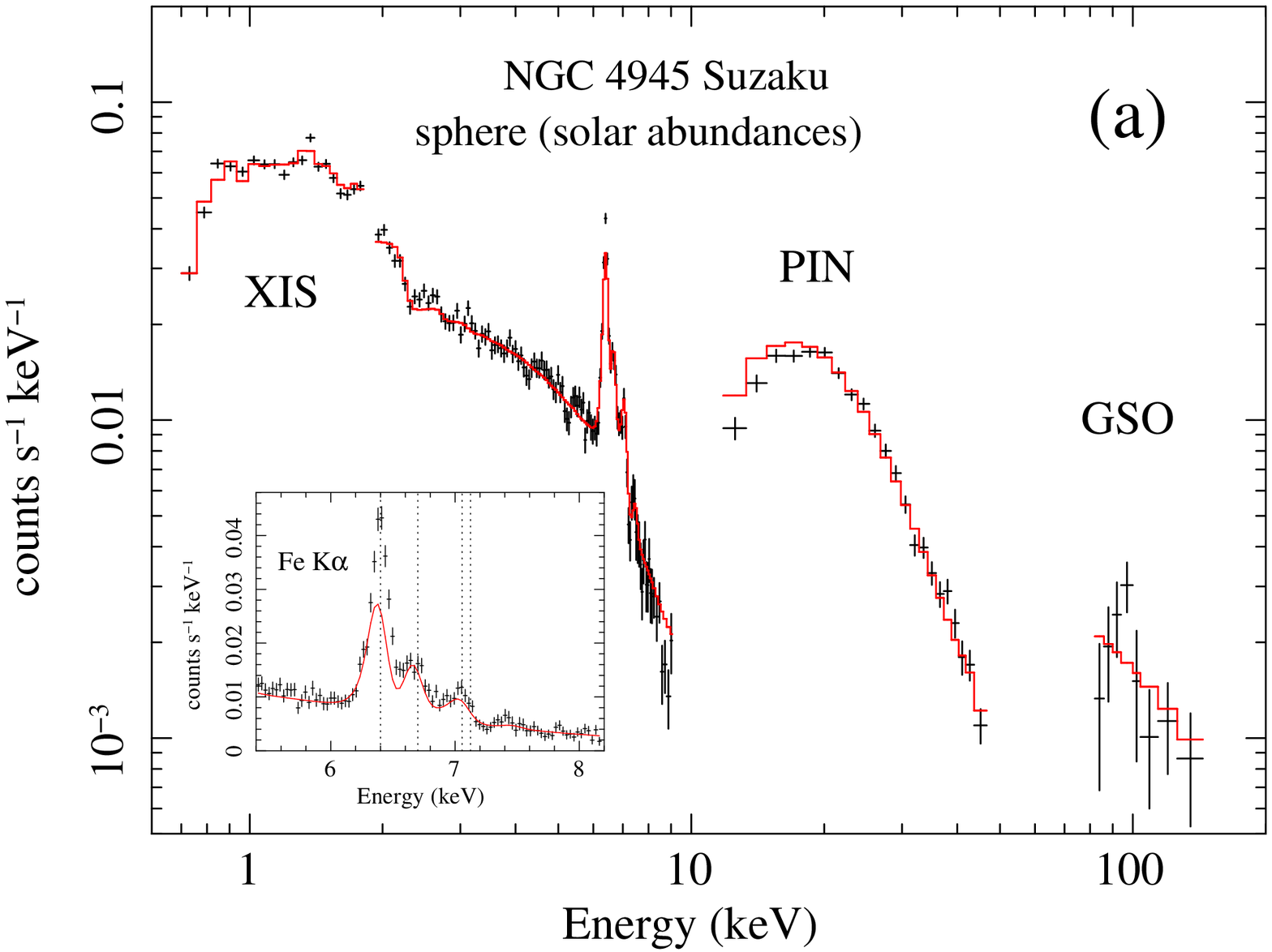,height=6cm,angle=0}
     \psfig{figure=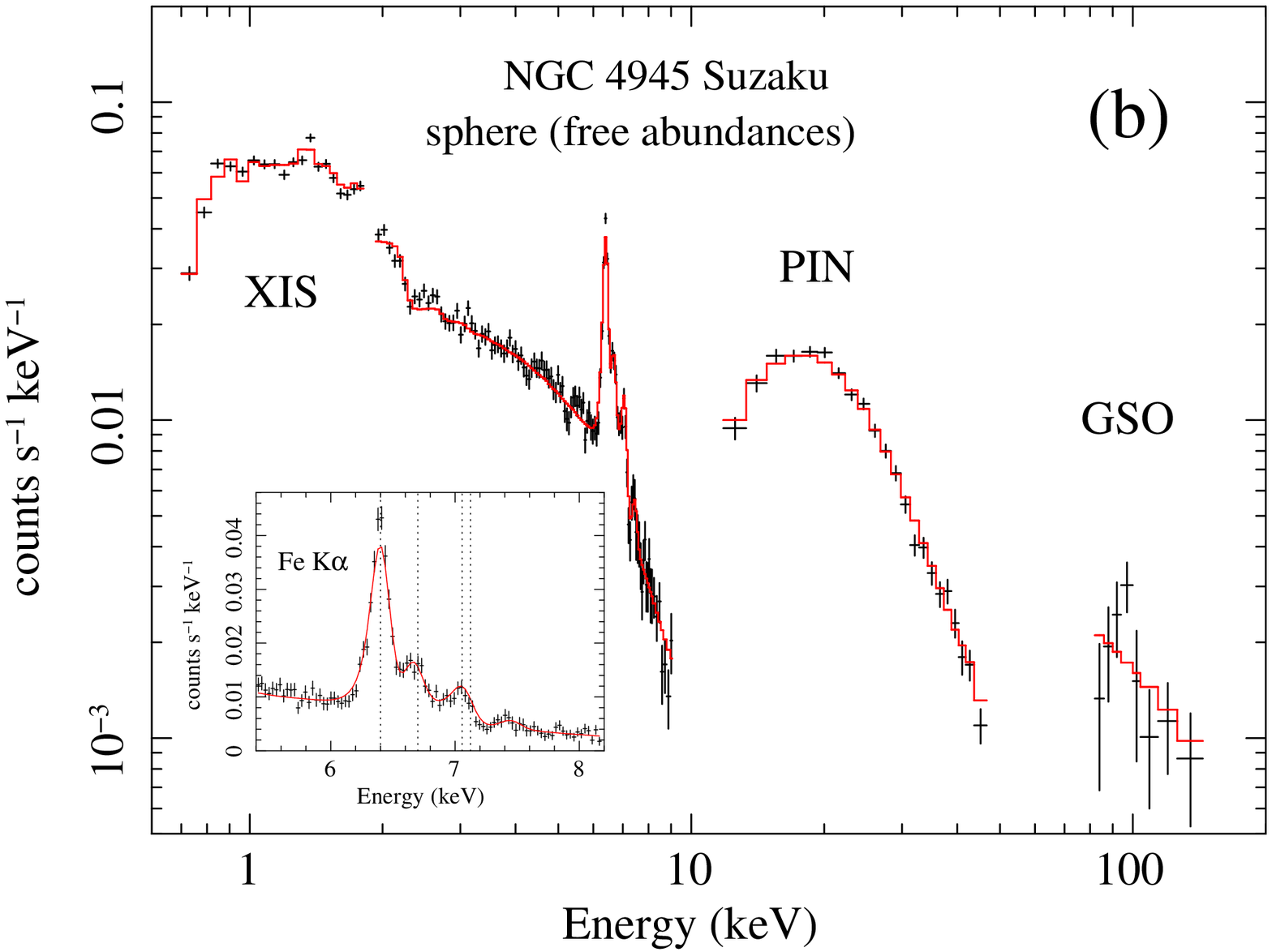,height=6cm,angle=0}
}
        \caption{\footnotesize Spectral fits to the \suzaku data for NGC~4945,
showing the BN11 spherical model overlaid on the counts spectra for
(a) solar abundances, and (b) the relative abundance parameters 
for Fe ($X_{\rm Fe}$) and the other metals ($X_{\rm M}$) free (see \S\ref{sphericalfits}). 
In each case a zoomed view is shown of the spectral region containing
the \fekalfap, \fekbetap, \nika lines, and the Fe~K absorption edge.
The fit with solar abundances illustrates how the column density required to
fit the $\sim 10-50$~keV spectral peak underpredicts the \fekalfa emission line. This fit
was performed by omitting the 6--8~keV data and finding the best fit, and
then including the 6--8~keV data for the plot (after inserting the required \fexxv
emission line at $\sim 6.7$~keV). Allowing the abundances to float
then enables a good fit to both the \fekalfa line and the Compton hump, as shown
in Fig.~\ref{fig:szsphabun}(b). See \S\ref{sphericalfits} and \tablemytcpplresults for
details.
}
\label{fig:szsphabun}
\end{figure}

\begin{figure}
\centerline{
     \psfig{figure=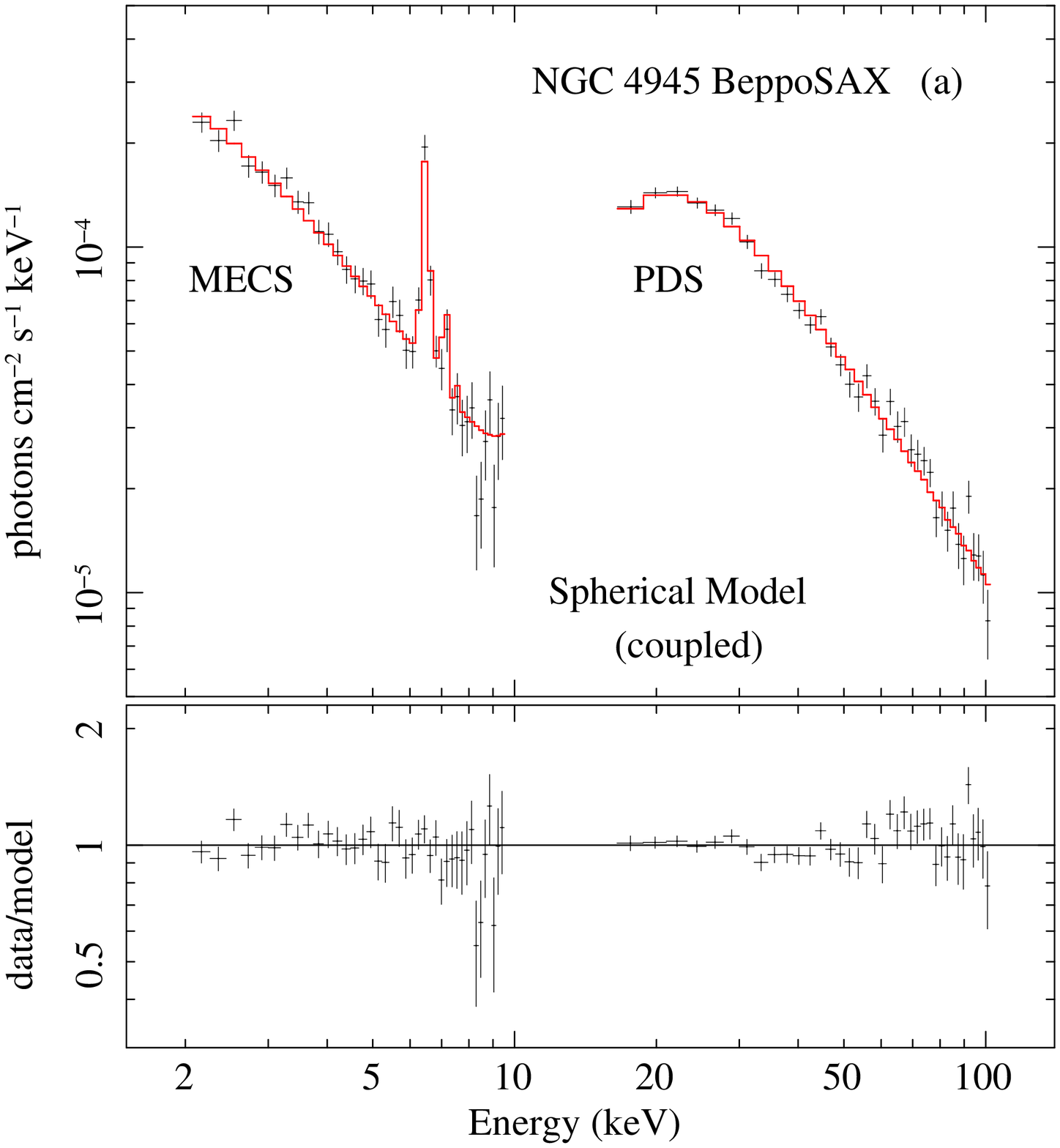,height=8cm,angle=0}
     \psfig{figure=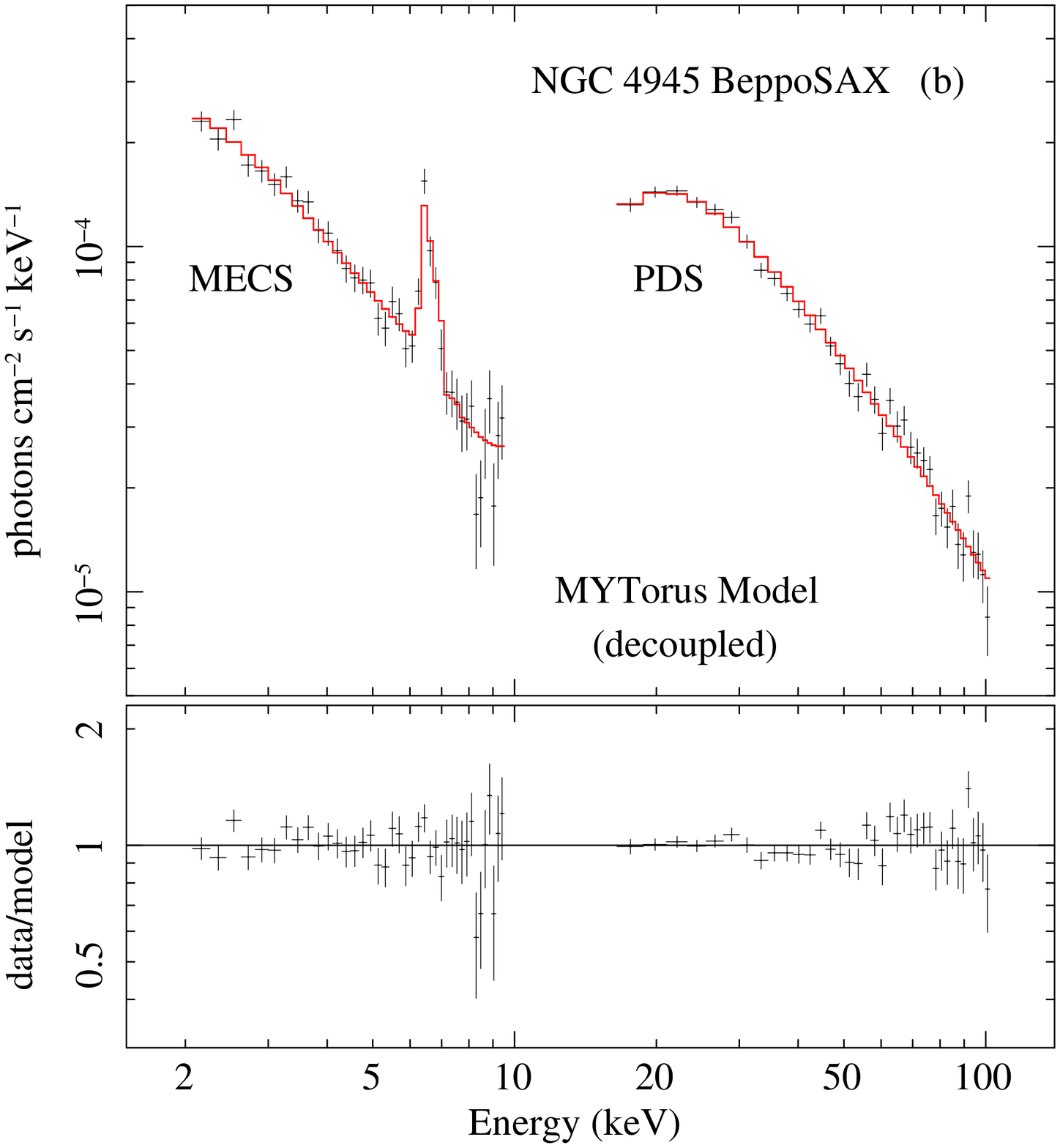,height=8cm,angle=0}
}
        \caption{\footnotesize Spectral fits to the \bsax data for NGC~4945,
showing two models overlaid on the unfolded spectra (top panels), and the
data/model ratios (bottom panels). 
(a) A fit with the spherical model of Brightman and Nandra (2011).
The model parameters are shown in \tablemytcpplresults 
(see \S\ref{coupledmytfits} for details).
(b) A fit with the decoupled \mytorus model (mimicking a patchy reprocessor).
The incident, intrinsic continuum for
the model fit for this fit has a thermally Comptonized spectrum with 
the plasma temperature, $kT$, fixed at $80$~keV.
The model parameters are shown in \tablemytdecoup (see \S\ref{decoupledmytfits} for
details). 
For clarity, the data in some energy ranges is binned more coarsely than
the binning used for the actual spectral fitting.
Note that the unfolded spectra are for illustration purposes only
and are by their nature model-dependent so should be interpreted
with caution (the fitting is performed on the counts spectra not
the unfolded spectra). 
}
\label{fig:bsaxmytsph}
\end{figure}

\begin{figure}
\centerline{
        \psfig{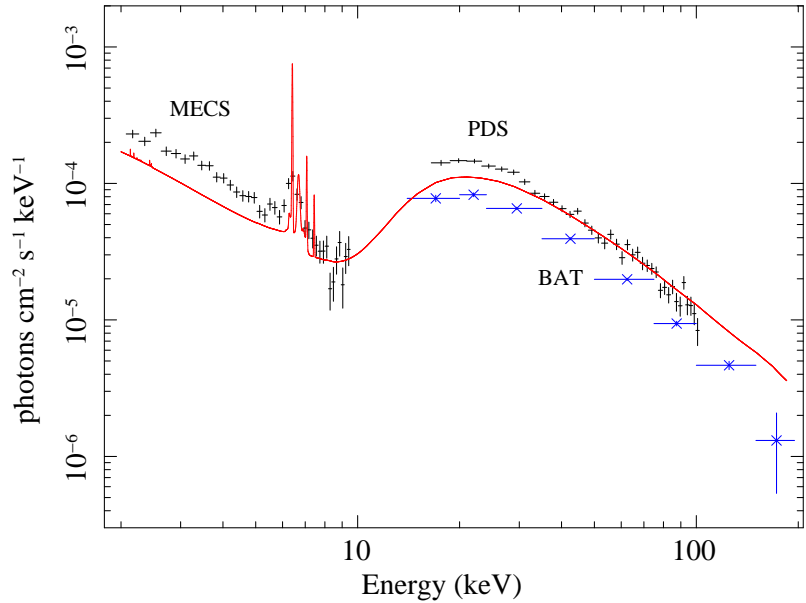}
        }
        \caption{\footnotesize The best-fitting coupled \mytorus model (red curve)
with a power-law intrinsic continuum that was fitted
to the NGC~4945 \suzaku data (taken in 2006, January),
overlaid on the noncontemporaneous \bsax data (black), and on the \swift BAT data (blue).
The \bsax data were
taken in 1999, July, and the \swift BAT data are averaged over 58 months.
The best-fitting \suzaku model is the same as the one shown in Fig.~\ref{fig:mytszcp}(b) and
the model parameters are shown in
\tablemytcpplresults (see \S\ref{coupledmytfits} for details). No corrections have been made
for the different cross-normalizations of the different instruments for this plot.
Also, the \bsax and \swift BAT photon spectra were only crudely constructed
by unfolding using only a simple power-law
continuum with absorption, so they should be interpreted with caution. The purpose of the plot is
only to show the approximate relative magnitudes of the different spectra in the
Fe~K band ($\sim 6-8$~keV), and around the Compton hump (between $\sim 10$--$50$~keV).
}
\label{fig:saxbatoverlay}
\end{figure}

Finally, we fitted the BN11 spherical model to all the data sets
(see \S\ref{bnctmodels} for details of this spherical model).
We could not obtain an acceptable fit to the \suzaku with solar abundances. 
The best-fitting reduced $\chi^{2}$ value was $1.51$, with a null hypothesis
probability of $1.7 \times 10^{-9}$. The poor fit is illustrated in
Fig.~\ref{fig:szsphabun}(a), which shows that the shape of the Compton hump
is not correctly reproduced and the \fekalfa line flux is not well-fitted. 
Essentially, the column density that is required to correctly fit the Compton
hump, severely underpredicts the \fekalfa line flux, and this is shown in the
zoomed-in panel in Fig.~\ref{fig:szsphabun}(a). However, when we allowed
the relative Fe abundance, $X_{\rm Fe}$, to float, the best-fitting
$\chi^{2}$ value dropped by 83.0 for the addition of a single free parameter, 
and a good fit was obtained.
On the other hand, a reduction in $\chi^{2}$ of less than 4 was obtained by allowing the relative
abundance of the other elements ($X_{\rm M}$) to float (see \S\ref{bnctmodels} for
details of the meaning of the abundance parameters).
The data and model, including a zoomed view of the Fe~K spectral region,
are shown in Fig.~\ref{fig:szsphabun}(b), and the best-fitting parameters
and statistical errors are shown in \tablemytcpplresultsp. It can be seen
that the required enhancement of the Fe abundance relative to solar
is very modest ($X_{\rm Fe}=1.44^{+0.25}_{-0.22}$), yet the difference in the goodness
of fit is substantial. 
It is also important to realize that varying the Fe abundance
in a Compton-thick medium has effects well beyond the Fe~K spectral
region because of the high absorption optical depth over a broad
energy range, and the indirect effect on Compton scattering.
The best-fitting relative abundance of the
other elements, $X_{\rm M}=0.94^{+0.11}_{-0.07}$, was consistent with unity.
The best-fitting $\chi^{2}$ value of $447.2$ for $354$ degrees of freedom
indicates a marginally worse fit than the coupled \mytorus model, and
this is due to a small underprediction of the \fekalfa line flux,
as can be seen in Fig.~\ref{fig:szsphabun}(b). However, less than 
full covering of the X-ray reprocessor (or equivalently, some patchiness)
would alleviate the problem.

We have shown the spherical model solutions mainly
in the interest of application to AGN other than NGC~4945. In the case
of NGC~4945, in principle, the deficit in the \fekalfa line flux in the
solar-abundance model could be made up by the \fekalfa line emission
from the extended region spatially resolved by \chandra (and indeed
this component should be included). However, none of the spherical
model solutions are applicable to NGC~4945 because the continuum
above 10~keV is dominated by the Compton-scattered continuum, overwhelming
the zeroth-order continuum, and this is inconsistent with the
variability of the high-energy continuum and the lack of simultaneous
variability of the \fekalfa line flux (as reported in Marinucci \etal 2012).
Note that when applying the spherical model to other sources,
it should be remembered that it is not quite self-consistent
because it is {\it required} that the X-ray
reprocessor is patchy or has less than full
covering, otherwise there would be no
optically-thin scattered continuum, so that the X-ray continuum below
$\sim 10$~keV would be more suppressed than it actually is,
and it would be decreasing in magnitude with
decreasing energy.

We also obtained an acceptable fit with the spherical model to the \bsax data, again
with the relative Fe abundance allowed to float, but we had to
fix the relative abundance of the other heavy elements
at their solar values in order to obtain a stable spectral fit. 
The results are given in \tablemytcpplresultsp,
and the data, model, and data/model ratios are shown in Fig.~\ref{fig:bsaxmytsph}(b).
The best-fitting relative Fe abundance is $1.03^{+0.52}_{-0.37}$, and
this is statistically consistent with the corresponding  
result obtained from the \suzaku data.
The \mytorus solution to the \bsax data is 
physically very different to the spherical model solution. The former
is dominated by the zeroth-order continuum, whereas the latter is dominated
by the Compton-scattered continuum.
However, the two solutions cannot
be distinguished statistically (the $F$-test shows that the probability of obtaining
the slightly higher $\chi^{2}$ value, or higher, for the spherical model than the \mytorus model
is about 38\%). 

The BN11 spherical model gave an acceptable fit to the \swift BAT spectrum,
and the results for solar abundances are shown in \tablebatresults
(the abundance parameters could not be constrained).

\subsection{Overview of the Parameters for the Coupled Models}

In summary, we can say that for coupled models of the X-ray reprocessor,
the \suzaku data for NGC~4945 prefer a spectrum in which
the Compton-scattered continuum dominates over the zeroth-order continuum, and these
data prefer a medium that has a substantial global covering
factor (a few tenths to nearly full covering, or full covering with some patchiness). 
However, models in which the Compton-scattered continuum dominates are
inconsistent with the continuum above 10~keV in NGC~4945 varying
independently of the \fekalfa line (Marinucci \etal 2012).
On the other hand, two very different degenerate solutions can describe the \bsax data.
One solution has the spectrum dominated by the zeroth-order continuum, consistent
with a ring-like geometry in which the X-ray reprocessor has a small
covering factor. The other solution instead has the Compton-scattered continuum
dominate the spectrum over the zeroth-order continuum, achieved by the
full covering of the spherical model. 
It is true to say that both the \bsax and \suzaku data are consistent with
a nearly fully-covering (or patchy) spherical model, in which case no changes in covering factor
are required going from the \bsax observation to the \suzaku observation,
but such a scenario can be ruled out for NGC~4945 because of the
variability constraints discussed earlier. 
In the interest of modeling other AGNs, this scenario
would require a global decrease of the column density
by $\sim 50\%$ between observations (see \tablemytcpplresultsp), and an Fe abundance up to $\sim 50\%$
higher than solar (for both observations).
We shall also see in \S\ref{lumratios} that
the spherical model has an intrinsic luminosity that is more than an order of
magnitude less than the zeroth-order-dominated spectrum. We defer to \S\ref{lumratios}
for further discussion (after presenting the decoupled model fits).
In the remainder of this section, we summarize some of the other results in 
\tablebatresults (for the \swift BAT data) and \tablemytcpplresults
(for \suzaku and \bsaxp).
Statistical errors for all of the parameters can be found in \tablebatresults and
 \tablemytcpplresults but are not always quoted in the text.
To aid the comparison between the \suzakup, \bsaxp, and \swift BAT spectra, we have
overlaid the best-fitting \mytorus model for the \suzaku spectra
(with an intrinsic power-law continuum) on
the \bsax and \swift BAT photon spectra in Fig.~\ref{fig:saxbatoverlay}.
 
\begin{table}
\caption[Spectral-fitting results with the ${\sc mytorus}$ model.]
{NGC~4945 Spectral-Fitting Results with the Coupled Reprocessor Models}
\begin{center}
\begin{tabular}{lcccccc}
\hline
& & & & & & \\
Parameter & \suzaku & \suzaku & \suzaku & \bsax & \bsax \\
& & & & & & \\
\hline
& & & & & & \\
Model & \mytorus & \mytorus & B11 sphere & \mytorus & BN11 sphere  \\
Intrinsic Continuum & {\sc power law} & \comptt & {\sc power law} & \comptt & {\sc power law} \\
Free Parameters & 13 & 14 & 13 & 8 & 7 \\
Interesting Parameters & 11 & 12 & 11 & 7 & 6 \\
$\chi^{2}$ / degrees of freedom & 440.6/354 & 435.1/353  & 447.2/354 & 85.8/70 & 92.5/70 \\
Reduced $\chi^{2}$ & $1.24$ & $1.23$ & $1.26$ & $1.23$ & $1.32$ \\
Null Probability & $1.3 \times 10^{-3}$ & $1.9 \times 10^{-3}$ & $5.6 \times 10^{-4}$ & $0.095$ & $0.037$ \\
$\Delta\chi^{2}$ criterion (68\% confidence) & 12.6 & 13.70 & 12.6 & 8.15 & 7.01 \\ 
GSO:XIS normalization ratio & $0.446^{+0.084}_{-0.084}$ & $0.567^{+0.110}_{-0.107}$ & 
 $0.404^{+0.076}_{-0.076}$ & \ldots & \ldots \\

$\Gamma$ or $\tau$ & $1.596^{+0.044}_{-0.035}$ & $1.20^{+0.09}_{-0.12}$ & 
 $1.489^{+0.037}_{-0.031}$ & $0.31^{+0.09}_{-0.10}$ & $1.789^{+0.149}_{-0.026}$   \\
$kT$ (keV) & \ldots & $50^{+4}_{-4}$ & \ldots & 80(f) & \ldots \\ 

$N_{H} \rm \ (10^{24} \ cm^{-2})$ & $3.48^{+0.22}_{-0.17}$ & $3.62^{+0.26}_{-0.23}$ & 
 $2.17^{+0.18}_{-0.15}$ & $4.55^{+1.38}_{-0.80}$ & $3.05^{+0.48}_{-0.41}$ \\
$\theta_{\rm obs}$ (degrees) & $85.4^{+0.8}_{-0.6}$ & $85.7^{+0.6}_{-0.7}$ & \ldots & $89.3^{+0.7}_{-3.6}$ & \ldots \\
$A_{S}$  & $1.00^{+0.10}_{-0.26}$ & $1.00^{+0.15}_{-0.17}$ & \ldots & $0.101^{+0.149}_{-0.068}$ & \ldots \\
$A_{L}$ & $=A_{S}$ & $=A_{S}$ & \ldots & $=A_{S}$ & \ldots \\
$X_{\rm Fe}$ & 1.0(f) & 1.0(f) & $1.44^{+0.25}_{-0.22}$ & 1.0(f) & $1.03^{+0.52}_{-0.37}$ \\
$X_{\rm M}$ & 1.0(f) &  1.0(f)  & $0.94^{+0.11}_{-0.07}$  & 1.0(f) & 1.0(f) \\  
FWHM [\fekalfap, \fekbetap] ($\rm km \ s^{-1}$) & $<1215$(f) & $<1215$(f) &  $<1215$(f) & $100$(f) & $100$(f) \\
$I_{\rm Fe~K\alpha}$ ($\rm 10^{-5} \ photons \ cm^{-2} \ s^{-1}$)
& $3.17$ & $3.18$ & \ldots &  $1.83$ & \ldots \\
EW$_{\rm Fe~K\alpha}$ (eV) & $730$ & $749$ & \ldots & $367$ & \ldots \\
$E_{\rm Fe~XXV}$ (keV) & $6.686^{+0.014}_{-0.016}$ & $6.686^{+0.014}_{-0.016}$ & 
 $6.680^{+0.015}_{-0.012}$ & $6.700$(f) & $6.700$(f) \\
FWHM [\fexxv] ($\rm km \ s^{-1}$) & $4950_{+2535}^{-2735}$ & $4950_{+2535}^{-2735}$ & $<6025$(f) & $4950$(f) & $4950$(f)  \\
$I_{\rm Fe~XXV}$ ($\rm 10^{-5} \ photons \ cm^{-2} \ s^{-1}$) &
$0.98^{+0.26}_{-0.29}$ & $0.98^{+0.26}_{-0.30}$ & $0.88^{+0.20}_{-0.26}$ & $1.7^{+1.3}_{-1.3}$ & $<2.2$ \\
EW$_{\rm Fe~XXV}$ (eV) & $200^{+53}_{-51}$ & $245^{+65}_{-65}$ & $215^{+49}_{-64}$ & $240^{+120}_{-180}$ & $<249$ \\
$E_{\rm Ni~K\alpha}$ (keV) & $7.434^{+0.066}_{-0.041}$ & $7.434^{+0.066}_{-0.041}$ & 
$7.450^{+0.048}_{-0.073}$ & $7.472$(f) & $7.472$(f) \\
FWHM [\nika] ($\rm km \ s^{-1}$) & $<7410$(f) & $<7410$(f) & $<8600$(f) &  \ldots & \ldots \\
$I_{\rm Ni~K\alpha}$ ($\rm 10^{-5} \ photons \ cm^{-2} \ s^{-1}$) & $<0.45$ & $<0.45$ & $<0.41$ & $<0.88$ & $<0.46$ \\
EW$_{\rm Ni~K\alpha}$ (eV) & $<150$ & $<150$ & $<140$ & $<252$ & $<115$ \\
$10^{3}f_{s}$ (optically-thin scattered fraction) & $7.9^{+1.1}_{-0.5}$ & $7.3^{+1.7}_{-0.7}$ & $38.5^{+3.2}_{-2.2}$ &
 $0.61^{+0.99}_{-0.47}$ & $20.5^{+1.4}_{-5.8}$ \\
$N_{H1} \rm \ (10^{21} \ cm^{-2})$ & $2.8^{+0.7}_{-0.9}$ & $2.6^{+0.8}_{-0.9}$ & 
 $2.3^{+0.7}_{-1.2}$ & $13.0^{+8.5}_{-6.6}$ & $7.6^{+4.2}_{-4.2}$ \\

$A_{\rm apec}$ ($10^{-3} \ \rm phot. \ cm^{-2} \ s^{-1} \ keV^{-1}$) &
 $1.8^{+0.6}_{-0.6}$ & $1.7^{+0.6}_{-0.6}$ & $1.8^{+0.6}_{-0.6}$ & $1.7$(f) & $1.7$(f) \\
$kT_{\rm apec}$ (keV) & $0.260^{+0.046}_{-0.035}$ & $0.260^{+0.046}_{-0.035}$ & 
 $0.260^{+0.048}_{-0.038}$ & $0.260$(f) & $0.260$(f) \\
$N_{H2} \rm \ (10^{21} \ cm^{-2})$ & $3.8^{+1.1}_{-0.3}$ & $3.8^{+1.0}_{-0.3}$ & 
 $3.8^{+1.0}_{-0.2}$ & $3.8$(f) & $3.8$(f) \\
$L_{\rm APEC}$ [Intrinsic] ($10^{40} \rm \ erg \ s^{-1}$) & $2.9$ & $2.9$ & $2.9$ & \ldots & \ldots \\
&  \\
$F_{\rm obs}$[0.7--2 keV] ($10^{-11} \rm \ erg \ cm^{-2} \ s^{-1}$) & $0.08$ & $0.08$ & $0.08$ & \ldots & \ldots \\
$F_{\rm obs}$[2--10 keV] ($10^{-11} \rm \ erg \ cm^{-2} \ s^{-1}$)  & $0.39$ & $0.39$ & $0.39$ & $0.48$ & $0.49$ \\
$F_{\rm obs}$[10--100 keV] ($10^{-11} \rm \ erg \ cm^{-2} \ s^{-1}$) & $30.0$ & $27.6$ & $31.1$ & $34.7$ & $34.5$ \\
$F_{\rm obs}$[14--195 keV] ($10^{-11} \rm \ erg \ cm^{-2} \ s^{-1}$) & $45.8$ & $38.7$ & $49.3$ & $47.5$ & $46.7$ \\
$L_{\rm obs}$[0.7--2  keV] ($10^{42} \rm \ erg \ s^{-1}$) & $0.0058$ & $0.0058$ & $0.0058$ & \ldots & \ldots \\
$L_{\rm obs}$[2--10  keV] ($10^{42} \rm \ erg \ s^{-1}$)  & $0.030$ & $0.030$ & $0.030$ & $0.037$ & $0.038$  \\
$L_{\rm obs}$[10--100  keV]  ($10^{42} \rm \ erg \ s^{-1}$)  & $2.3$ & $2.1$ & $2.4$ & $2.7$ & $2.7$  \\
$L_{\rm obs}$[14--195  keV] ($10^{42} \rm \ erg \ s^{-1}$)  & $3.5$  & $3.0$ & $3.8$ & $3.7$ & $3.6$ \\
& \\
\hline
\end{tabular}
\end{center} 
Spectral-fitting results obtained from fitting the NGC~4945 \suzaku 
and \bsax data with various coupled models of the Compton-thick X-ray reprocessor,
as described in \S\ref{coupledmytfits}.
A parameter value that is followed
by ``(f)'' indicates that the parameter was fixed during the fitting but the 
statistical errors were obtained by allowing the fixed parameter to be free
only for obtaining the statistical errors. The FWHM values of the \fekalfap, \fekbetap, and \nika lines were fixed
at $100 \ \rm km \ s^{-1}$ during the fitting. The $\Delta \chi^{2}$ criterion used
for these fixed parameters, and for the GIS:XIS normalization ratio, was
$2.706$, corresponding to 90\% confidence for interesting parameter.
The statistical errors for the remaining free parameters 
used a $\Delta \chi^{2}$ criterion shown in the appropriate column of the table.
No statistical errors
are given for the \fekalfa line flux and EW because these emission-line parameters
are determined completely by the other parameters of the \mytorus model
(but see \S\ref{coupledmytfits} for an estimate of the statistical errors). The 
continuum fluxes ($F_{\rm obs}$) and
luminosities ($L_{\rm obs}$) are all observed values and uncorrected for
absorption and Compton scattering.
\end{table}

\subsubsection{Continua}

The temperature of the optically-thin thermal emission
component is $kT_{\rm apec}=0.260^{+0.046}_{-0.035}$~keV for both \suzaku fits
with the \mytorus model
(\tablemytcpplresultsp), so it is not sensitive
to the different models for the intrinsic continuum. The same temperature
is obtained with the spherical model, albeit with slightly different
statistical errors. The
intrinsic luminosity of the optically-thin thermal component
is $2.9 \times 10^{40} \rm \ erg \ s^{-1}$ and this is not sensitive
to the differences in any of the three coupled models fitted to the
\suzaku data. The same is
true for the column density associated with the region producing
the thermal emission ($N_{\rm H2}$), and is $3.8 \times 10^{21} \rm \ cm^{-2}$. 
This optically-thin thermal emission continuum component
is below the bandpass of both the \bsax and \swift BAT data so its
parameters cannot be
constrained by those data. In the case of the \bsax fits, the
parameters of this continuum component, along with $N_{\rm H2}$,
were fixed at the values obtained from the \suzaku fits. 
(For the \swift BAT fits, this continuum component was omitted altogether.)
 
In the case of the power-law model for the intrinsic continuum
applied to the \suzaku data with the \mytorus model,
the photon index is $\sim 1.6$.
The power-law continuum is much flatter with the spherical reprocessor
model ($\Gamma < 1.5$), but it is steeper when the spherical
model is applied to the \bsax data ($\Gamma \sim 1.8$).
For the Comptonized thermal continuum we
tried fits with \mytorus tables for a range of values of $kT$
that were consistent with the 99\% confidence contours shown in Fig.~\ref{fig:ktvsnhcont},
and found that $kT=50$~keV gave the best fit for the \suzaku data. The parameter $kT$ was
then allowed to be free, although strictly speaking it is only correct to do
this
for the zeroth-order continuum. However, the statistical error obtained in this
way is sufficiently small ($kT=50\pm4$~keV), that the procedure is
justified in the present application. Interestingly, the \swift BAT spectrum,
which is averaged over 58 months, gave a consistent value of $kT=51^{+16}_{-17}$~keV
for the \mytorus coupled model fit (\tablebatresultsp). However, the
statistical errors are larger so should be regarded as only approximate.
For the \bsax data we found that fits with $kT$ in the range 50--100~keV yielded
insignificant differences in the best-fitting values of $\chi^{2}$. In order
to get stable fits, $kT$ had to be frozen, so we performed
fits for $kT$ fixed at $50$, $80$, $100$~keV. For the sake of brevity, the
full set of parameters and statistical errors are given in \tablemytcpplresults only
for the fit with $kT=80$~keV, but parameter values 
and luminosities for the broader range of
$kT$ will be given where appropriate. The plasma optical depth,
$\tau$, is $\sim 1$--$1.2$ for the \suzaku and \swift BAT data, and $\sim 0.3$ for
the \bsax data.  

The percentage of the intrinsic continuum scattered into
the line of sight by optically-thin circumnuclear material
is highly model-dependent because the luminosity of the intrinsic 
continuum is highly model-dependent. However, models with similar
intrinsic luminosities will give similar values of $f_{s}$
as is the case for the two models shown in \tablemytcpplresults for
the \suzaku data ($\sim 7-8 \times 10^{-3}$). Both of these models are dominated
by the Compton-scattered continuum so they have similar intrinsic continuum
luminosities. On, the other hand, for the \bsax data we get values of
$f_{s}$ of $6.1 \times 10^{-4}$ and $2 \times 10^{-2}$ for the
\mytorus and spherical models respectively. The spherical model
applied to the \suzaku data gives the highest value of $f_{s} \sim 3.9 \times 10^{-2}$.
The smallest and largest values differ by
nearly a factor of $64$ because the \mytorus fit to the \bsax data is dominated by
the zeroth order continuum, whereas the spherical fit is dominated
by the Compton-scattered continuum (see discussion in \S\ref{coupledmytfits}).  

The column density associated with the optically-thin scattering
region ($N_{\rm H1}$) is $\sim 2$--$3 \times 10^{21} \ \rm cm^{-2}$ for
the \suzaku fits, and $\sim 7-13 \times 10^{21} \ \rm cm^{-2}$ for
the \bsax fits.

\subsubsection{Column Density of the X-ray Reprocessor}

The toroidal equatorial column density for the coupled \mytorus fit to
the \suzaku data 
with a power-law intrinsic continuum is $N_{\rm H} = 3.49^{+0.19}_{-0.17} \times 10^{24} \
\rm \ cm^{-2}$. The \mytorus \suzaku fit with a Comptonized thermal intrinsic continuum 
gave a consistent value of $N_{\rm H}$ (see \tablemytcpplresultsp).
On the other hand, the column density for the \bsax fit with the coupled \mytorus
model is somewhat higher than the values from the \suzaku fits, at
$N_{\rm H} = 4.55^{+1.38}_{-0.80} \times 10^{24} \ \rm \ cm^{-2}$. What is also
important is that the column density obtained from the spherical (BN11) model
fitted to the \bsax data is significantly less than that from the \mytorus fit,
at $3.06^{+0.33}_{-0.34} \times 10^{24} \rm \ cm^{-2}$. This value is also less than that
obtained from the \suzaku \mytorus fits. However, the spherical model
fitted to the \suzaku data gave the lowest column density out of
all the spectral fits ($N_{\rm H} = 2.17^{+0.18}_{-0.15} \times 10^{24} \ \rm \ cm^{-2}$). The reason for the very different
values of $N_{\rm H}$ for the spherical fits compared to the \mytorus fit to the
\bsax spectra is that the latter
is dominated by the zeroth-order continuum, but in the spherical model fits the
spectrum is dominated by the Compton-scattered continuum.

From \tablebatresultsp, which shows results from spectral fitting to the
58-month time-averaged \swift BAT spectrum, we see that the coupled \mytorus 
model fits gave values of $N_{\rm H}$ that are statistically consistent
with those obtained from the \suzaku and \bsax fits, since the statistical
errors on the \swift BAT and \bsax values are large. The spherical 
model fit gave a lower value of $N_{\rm  H}$ than the \mytorus fits, as
would be expected ($2.74^{+0.27}_{-0.25} \times 10^{24} \ \rm cm^{-2}$),
and this value is statistically consistent with that obtained from
the spherical model fitted to the \bsax data.

\subsubsection{Inclination Angle}

In all of the spectral fits to the \suzaku and \bsax data
with the coupled \mytorus model, the
orientation of the X-ray reprocessor is very tightly constrained
to be edge-on, or nearly edge-on (see \tablemytcpplresultsp).
The inclination angle is not less than $84^{\circ}$ in any 
of the spectral fits to the \suzaku and \bsax data, even accounting
for the statistical errors (which are small). The physical
driver behind this is that the observer's view of the inner surface
of the reprocessor will give an \fekalfa line flux that is
too large if the inclination angle is too small, whilst at the
same time the column density (which also affects the \fekalfa line
flux) cannot be too small because the Compton hump
must be prominent enough to fit the high-energy ($>10$~keV) part
of the spectrum. In fact, we can see that for the \bsax \mytorus fit,
the inclination angle is driven fully to an edge-on orientation,
and the parameter $A_{S}$ is driven down to $\sim 0.1$ in order to
keep the \fekalfa line flux in check.

\subsubsection{Fe K$\alpha$ Line Flux and Equivalent Width}
\label{coupledfekalpha}

The \fekalfa line flux is not explicitly
an adjustable parameter because the line is produced self-consistently
in all of the Compton-thick reprocessor models that we applied.
However, by subtracting the continuum-only flux in the 5--7~keV band
measured from the best-fitting model (by temporarily
setting $A_{L}=0$),
from the total flux in the same band measured with $A_{S}=A_{L}$, we
can estimate the \fekalfa line flux.
In this way we obtained  
$3.2 \times 10^{-5} \ \rm photons \ cm^{-2} \ s^{-1}$ for both
of the \mytorus fits to the \suzaku data and an EW
of $730$~eV (see \tablemytcpplresultsp). 
Note that the \fekalfa line flux and EW include both the zeroth-order and the Compton shoulder
components of the \fekalfa line.
The \fekalfa line is not separable from the continuum in the BN11 models
so we could not obtain \fekalfa line flux measurements for these models.

The statistical errors on the \fekalfa line flux cannot be obtained
in the usual way.
However, in the \mytorus fits the parameter $A_{L}$ can be temporarily
``untied'' from $A_{S}$ in order to 
crudely estimate the statistical errors on the line flux.
By perturbing $A_{L}$ either side of the best-fitting value we estimated
that the 90\% confidence, one-parameter statistical error on the
\fekalfa line flux is of the order of $5\%$ for the \suzaku fits. This is also a reasonable,
rough estimate on the statistical error on the EW. An empirical, Gaussian
fit to the \fekalfa emission line is in agreement with these
estimates (e.g., see Itoh \etal 2008).

A similar procedure applied to the \bsax \mytorus spectral fit (see \tablemytcpplresultsp) 
yielded $1.8 \times 10^{-5} \ \rm photons \ cm^{-2} \ s^{-1}$ (EW~$\sim 370$~eV), with an
estimated, 90\% confidence, one-parameter
statistical error of $\sim 35\%$. 
However, because of the poor spectral resolution of the \bsax MECS, there
is considerable blending of the 6.4~keV and 6.7~keV lines. We found that
empirically fitting the MECS data with a simple power-law and two
narrow Gaussian emission lines returned a flux for the 6.4~keV line
of $2.5^{+0.7}_{-0.7} \times 10^{-5} \ \rm photons \ cm^{-2} \ s^{-1}$, statistically
consistent with the \fekalfa line flux from the \suzaku spectral fits.
In the interest of general analysis, we point out that 
the \fekalfa line parameters obtained from the \bsax 
spectral fit in \tablemytcpplresults 
are not sensitive to $kT$. Specifically, for fits with $kT$ in the
range 50 to 100~keV, the line flux only varied between 
$\sim 1.7 \times 10^{-5}$ and $\sim 1.9 \times 10^{-5} \ \rm photons \ cm^{-2} \ s^{-1}$.

\subsubsection{The Line Emission at 6.7~keV}

The line emission in the \suzaku data centered at $\sim 6.67$--$6.70$~keV 
(see \tablemytcpplresultsp) is 
consistent with an identification
of the \fexxv He-like triplet, but the individual members of
the triplet cannot be resolved with the spectral resolution
of CCDs. The peak energy of the unresolved He-like triplet
depends on the relative fluxes in the forbidden, intercombination,
and resonant components of
the triplet, at expected energies of 6.637, 6.675, and 6.700~keV
\footnote{NIST, http://physics.nist.gov/PhysRefData/}.
As well as members of the triplet, the line emission
profile between the peak energies of the \fekalfa and \fekbeta lines
could have a contribution from intermediate ionic species
of Fe as well. These factors mean that the intrinsic width
of the line shown in \tablemytcpplresults should not be interpreted
as purely a velocity width, since some of the broadening could be due 
due to the presence of several line components.
In fact, we found that
the overall line complex is unresolved
at a confidence level of 90\% for two parameters.

The \bsax data have a factor of 4 worse spectral resolution
than the \suzaku data so we had to fix the centroid of the line
emission at 6.7~keV, and the intrinsic line width at the value
from the \suzaku fits. The problem of the blending of the 6.4~keV and 6.7~keV
line emission in the \bsax MECS data
has already been discussed above (\S\ref{coupledfekalpha}).

\subsubsection{The Ni K$\alpha$ Line}

From \tablemytcpplresultsp, it can be seen that
the \nika line energy measured from the \suzaku data, 
namely $7.434^{+0.066}_{-0.041}$~keV, is consistent with the 
theoretical value of $7.472$~keV (e.g., Bearden 1967).
The \nika line fluxes in \tablemytcpplresults have lower limits
that are zero, but the 11 and 12-parameter, 68\% confidence statistical errors
are likely to be too conservative (since the \nika line is a relatively
isolated, narrow feature). Using instead, a one-parameter, 90\% confidence
criterion for the statistical errors, we obtained a
\nika line flux of $2.0^{+1.2}_{-1.2} \times 10^{-6}
\ \rm photons \ cm^{-2} \ s^{-1}$ from the \mytorus fits.
The EW corresponding to these line flux measurements and errors is
$46\pm28$~eV, somewhat less than the calculated Monte Carlo value (see Yaqoob \& Murphy 2011b),
but the difference is less than the uncertainty in the cosmic Ni abundance.
We tried letting the intrinsic width of the \nika line component
float in the \suzaku fits but found that the line is unresolved,
obtaining only a loose upper limit on the width of 
$\sim 7410 \rm \ km \ s^{-1}$~FWHM
for the \mytorus fits 
(for 90\% confidence, one interesting parameter).
Due to the weakness of the \nika line and the poor spectral resolution of
the \bsax data, we did not obtain upper limits on the intrinsic width of
the line for the \bsax data. Note that the BN11 spherical model
already includes \nika line emission so the \nika line parameters shown in
\tablemytcpplresults for the spherical model fits correspond to
additional line emission, which was included to allow for the 
considerable uncertainty in the Ni cosmic abundance (e.g., see discussion
in MY09).

\section{Spectral Fits with Decoupled Reprocessor Models}
\label{decoupledmytfits}

\footnotesize
\begin{table}
\caption[Spectral-fitting results with the decoupled ${\sc mytorus}$ model.]
{NGC~4945 Spectral-Fitting Results with the Decoupled ${\sc mytorus}$ Model}
\begin{center}
\begin{tabular}{lcccccc}
\hline
& & & & & & \\
Parameter & \suzaku & \suzaku & \suzaku & \suzaku & \bsax & \bsax \\
& & & & & & \\
\hline
& XIS & XIS & XIS & XIS & & \\
& PIN & PIN & PIN & PIN & & \\
& GSO &     & GSO &     & & \\
Free Parameters & 12 & 11 & 12 & 11 & 8 & 8 \\
Interesting Parameters & 10 & 10 & 10 & 10 & 7 & 7 \\
$\chi^{2}$ & 444.1 & 433.4 & 447.1 & 437.7 & 85.8 & 81.4 \\
degrees of freedom & 355 & 348 & 355 & 348 & 70 & 70 \\
Reduced $\chi^{2}$ & 1.25 & 1.25 & 1.26 & 1.26 & 1.23 & 1.16 \\
Null probability & $8.9\times10^{-4}$ & $1.2\times10^{-2}$ & $6.4\times10^{-4}$ & $7.5\times10^{-4}$
 & $9.6\times10^{-2}$ & $0.167$\\
$\Delta\chi^{2}$ criterion (68\% confidence) & 11.50 & 11.50 & 11.50 & 11.50 & 8.15 & 8.15 \\ 
PIN:XIS normalization & 1.12 (f) & 1.12 (f) & 1.12 (f) & 1.12 (f) & \ldots & \ldots \\
GSO:XIS normalization & $0.73^{+0.15}_{-0.14}$ & \ldots & $0.45^{+0.08}_{-0.09}$ & \ldots & \ldots & \ldots \\
$kT$ & 22~keV & 22~keV & 50~keV & 50~keV & 50~keV & 100~keV \\
$\tau$ & $2.09^{+0.08}_{-0.09}$ & $2.08^{+0.09}_{-0.08}$ & $0.95^{+0.04}_{-0.06}$ & $0.95^{+0.04}_{-0.06}$ & $0.50^{+0.12}_{-0.38}$ & $0.15^{+0.05}_{-0.06}$  \\

$N_{H} \rm \ (10^{24} \ cm^{-2})$ & $4.00^{+0.10}_{-0.07}$ & $4.00^{+0.10}_{-0.07}$ 
& $4.21^{+0.07}_{-0.07}$ & $4.21^{+0.07}_{-0.07}$ 
& $4.76^{+1.06}_{-0.80}$ & $4.94^{+0.98}_{-0.83}$ \\

$10^{2}A_{S90} (=A_{L90})$ & $5.2^{+3.8}_{-3.3}$ (f) 
 & $5.2^{+3.9}_{-3.2}$ (f) & $<3.4$ (f) & $<3.4$ (f) & $<23$(f) & $<23$(f) \\

$10^{2}A_{S00} (=A_{L00})$ & $1.67^{+0.19}_{-0.17}$ 
 & $1.67^{+0.18}_{-0.17}$ & $1.39^{+0.13}_{-0.11}$  & $1.40^{+0.12}_{-0.12}$ & $0.21^{+0.28}_{-0.13}$ & $0.18^{+0.24}_{-0.11}$ \\ 

$10^{3}f_{s}$ (optically-thin scattered fraction) & $1.50^{+0.18}_{-0.10}$ 
 & $1.50^{+0.18}_{-0.10}$ & $0.94^{+0.05}_{-0.05}$
 & $0.94^{+0.05}_{-0.05}$ & $0.29^{+0.43}_{-0.21}$ & $0.20^{+0.34}_{-0.13}$ \\

$N_{H1} \rm \ (10^{21} \ cm^{-2})$  & $3.0^{+0.6}_{-0.8}$  
 & $3.0^{+0.6}_{-0.8}$ & $3.1^{+0.7}_{-0.8}$ & $3.2^{+0.6}_{-0.9}$ 
 & $12.9^{+7.5}_{-6.7}$ & $16.2^{+7.7}_{-6.9}$ \\

& & & & & & \\
FWHM [\fekalfap, \fekbetap] ($\rm km \ s^{-1}$) & $<1235$ (f) & $<1235$ (f) & $<1235$ (f) & $<1235$ (f) 
 & $100$(f) & $100$(f) \\

$I_{\rm Fe~K\alpha}$ ($\rm 10^{-5} \ photons \ cm^{-2} \ s^{-1}$) & 
 $3.11$ & $3.11$ & $3.09$ & $3.09$ & $1.77$ & $1.95$ \\

EW$_{\rm Fe~K\alpha}$ (eV) & $708$ & $708$ & $704$ & $704$ & $341$ & $376$ \\
& & & & & & \\

$E_{\rm Fe~XXV}$ (keV)  & $6.687^{+0.015}_{-0.014}$ & $6.687^{+0.015}_{-0.014}$
 & $6.687^{+0.015}_{-0.014}$ & $6.687^{+0.015}_{-0.014}$ & $6.700$(f) & $6.700$(f) \\

FWHM [\fexxv] ($\rm km \ s^{-1}$) & $4970_{-3070}^{+2320}$ & $4970_{-3070}^{+2320}$
 & $4970_{-3385}^{+2210}$ & $4970_{-3385}^{+2210}$ & $4970$(f) & $4970$(f) \\ 

$I_{\rm Fe~XXV}$ ($\rm 10^{-5} \ photons \ cm^{-2} \ s^{-1}$) & $0.96^{+0.24}_{-0.24}$  
& $0.96^{+0.24}_{-0.24}$ & $0.94^{+0.24}_{-0.24}$ & $0.94^{+0.24}_{-0.24}$ 
& $1.6^{+1.3}_{-1.3}$ & $1.5^{+1.4}_{-1.3}$ \\

EW$_{\rm Fe~XXV}$ (eV) & $219^{+55}_{-55}$ & $219^{+55}_{-55}$ 
 & $214^{+55}_{-55}$ & $214^{+55}_{-55}$ & $320^{+260}_{-260}$ & $305^{+285}_{-265}$ \\
& & & & & & \\
$E_{\rm Ni~K\alpha}$ (keV) & $7.434^{+0.057}_{-0.042}$ & $7.434^{+0.057}_{-0.042}$ 
 & $7.434^{+0.057}_{-0.042}$ & $7.434^{+0.057}_{-0.042}$ & $7.472$(f) & $7.472$(f) \\

FWHM [\nika] ($\rm km \ s^{-1}$) & $<7285$(f) & $<7305$(f) & $<7285$(f) & $<7330$(f) & $100$(f) & $100$(f) \\
$I_{\rm Ni~K\alpha}$ ($\rm 10^{-5} \ photons \ cm^{-2} \ s^{-1}$) & $<0.43$ 
 & $<0.43$ & $<0.43$ & $<0.43$ & $<0.86$ & $<0.98$ \\

EW$_{\rm Ni~K\alpha}$ (eV) & $<146$ & $<146$ & $<146$ & $<146$ & $<261$ & $<307$ \\
& & & & & & \\

$N_{H2} \rm \ (10^{21} \ cm^{-2})$ & $3.8^{+0.7}_{-0.3}$ & $3.8^{+0.7}_{-0.3}$ 
 & $3.8^{+0.5}_{-0.4}$ & $3.8^{+0.5}_{-0.4}$ & 3.8(f) & 3.8(f)\\

$A_{\rm apec}$ ($\rm 10^{-3} \ phot. \ cm^{-2} \ s^{-1} \ keV^{-1}$) 
 & $1.77^{+0.60}_{-0.55}$ & $1.77^{+0.60}_{-0.55}$ & $1.77^{+0.63}_{-0.59}$ & $1.77^{+0.56}_{-0.55}$
 & $1.77$(f) & $1.77$(f) \\ 
$kT_{\rm apec}$ (keV) & $0.260^{+0.046}_{-0.035}$ & $0.260^{+0.046}_{-0.035}$ & $0.260^{+0.046}_{-0.035}$
 & $0.260^{+0.046}_{-0.035}$ & $0.260$(f) & $0.260$(f) \\

$L_{\rm apec}$ [Intrinsic] ($10^{40} \rm \ erg \ s^{-1}$) & $2.9$ & $2.9$ & $2.9$ & $2.9$ & \ldots & \ldots \\

$F_{\rm obs}$[0.7--2  keV] ($10^{-11} \rm \ erg \ cm^{-2} \ s^{-1}$) & $0.075$ 
 & $0.075$ & $0.075$ & $0.075$  & \ldots & \ldots \\
$F_{\rm obs}$[2--10 keV] ($10^{-11} \rm \ erg \ cm^{-2} \ s^{-1}$)  & $0.39$ 
 & $0.39$ & $0.39$ & $0.39$ & $0.48$ & $0.48$ \\
$F_{\rm obs}$[10--100 keV] ($10^{-11} \rm \ erg \ cm^{-2} \ s^{-1}$) & $26.5$ 
 & $26.5$ & $29.2$ & $29.2$ & $34.7$ & $34.7$ \\
$F_{\rm obs}$[14--195 keV] ($10^{-11} \rm \ erg \ cm^{-2} \ s^{-1}$) & $31.8$
 & $31.8$ & $45.2$ & $45.2$ & $46.2$ & $47.8$ \\

$L_{\rm obs}$[0.7--2 keV] ($10^{42} \rm \ erg \ s^{-1}$) & $0.0057$ & $0.0058$ 
 & $0.0058$ & $0.0058$ & \ldots & \ldots \\
$L_{\rm obs}$[2--10 keV] ($10^{42} \rm \ erg \ s^{-1}$)  & $0.030$ 
 & $0.030$ & $0.030$ & $0.030$ & $0.032$ & $0.037$ \\
$L_{\rm obs}$[10--100 keV] ($10^{42} \rm \ erg \ s^{-1}$)  & $2.0$ 
 & $2.0$ & $2.3$ & $2.3$ & $2.7$ & $2.7$ \\
$L_{\rm obs}$[14--195 keV] ($10^{42} \rm \ erg \ s^{-1}$)  & $2.5$ 
 & $2.5$ & $3.5$ & $3.5$ & $3.6$ & $3.7$ \\
\hline
\end{tabular}
\end{center}
Spectral-fitting results obtained from fitting the NGC~4945 \suzaku and \bsax data with the
decoupled \mytorus model, using a thermally Comptonized incident intrinsic
continuum, as described in \S\ref{mytdecoupled} and \S\ref{decoupledmytfits}. 
The first two columns for the \suzaku fits compare the same model 
(in which $kT=22$~keV) with and without the
GSO, as indicated. The second pair of columns for the \suzaku fits compare a model
with $kT=50$~keV, with and without the GSO.
The parameters
of the model are also summarized in \tablemytdecoupp. A parameter value that is followed
by ``(f)'' indicates that the parameter was fixed during the fitting but the
statistical errors were obtained by allowing the fixed parameter to be free
only for obtaining the statistical errors. 
The FWHM values
of the \fekalfap, \fekbetap, and \nika lines were fixed
at $100 \ \rm km \ s^{-1}$ during the fitting. The $A_{S90}$ parameter
was fixed at the best-fitting value, as indicated (zero if only an
upper limit is shown). The $\Delta \chi^{2}$ criterion used
for these fixed parameters, except for $A_{S90}$, was
$2.706$, corresponding to 90\% confidence for one interesting parameter.
The $\Delta \chi^{2}$ criterion used for $A_{S90}$ was $12.60$ 
for the \suzaku fits (corresponding
to 68\% confidence for 11 interesting parameters), and
$9.27$ for the \bsax fits (corresponding to to 68\% confidence for 8 
interesting parameters).
The statistical errors for the remaining free parameters are shown in
the appropriate column in the table. No statistical errors
are given for the \fekalfa line flux and EW because these emission-line parameters
are determined completely by the other parameters of the \mytorus model
(but see \S\ref{coupledmytfits} for estimates). The 
continuum fluxes ($F_{\rm obs}$) and
luminosities ($L_{\rm obs}$) are all observed values and uncorrected for
absorption or other reprocessing effects.

\end{table}
\normalsize

\begin{figure}
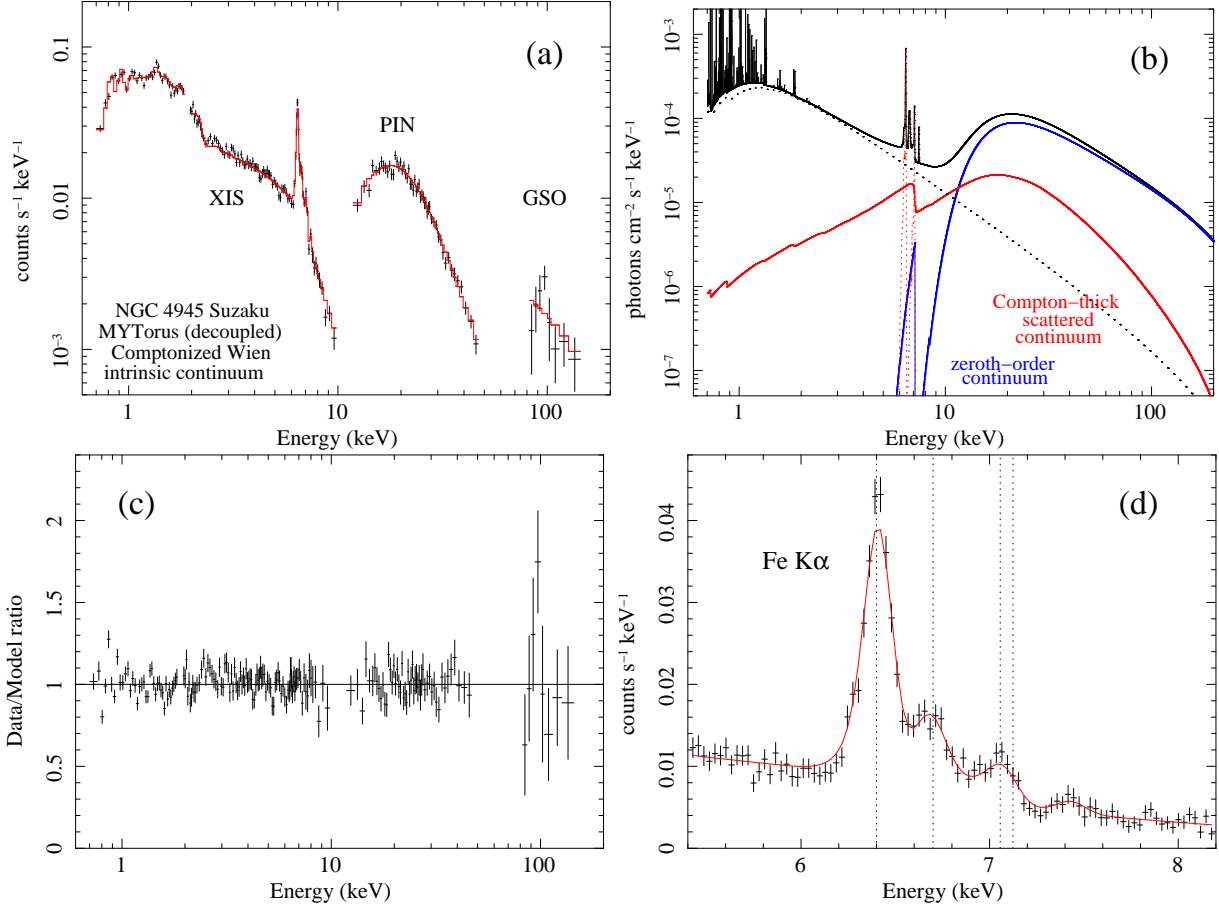

\centerline{
        \psfig{figure=f11a.ps,height=6cm,angle=270}
        \psfig{figure=f11b.ps,height=6cm,angle=270}
        }
\centerline{
        \psfig{figure=f11c.ps,height=6cm,angle=270}
        \psfig{figure=f11d.ps,height=6cm,angle=270}
        }
        \caption{\footnotesize Results of spectral fitting to the
NGC~4945 \suzaku data with the toroidal X-ray reprocessor model,
\mytorusp, with the Compton-scattered and zeroth-order continua decoupled 
(mimicking a patchy reprocessor). Unlike the coupled model fit shown in Fig.~\ref{fig:mytszcp},
the fit shown here has the zeroth-order continuum dominating over the scattered
continuum above $\sim 10$ keV.
The incident, intrinsic continuum for
the model fit shown here has a thermally Comptonized spectrum with $kT=50$~keV   
(see \tablemytdecoup and \S\ref{decoupledmytfits} for details).
(a) The counts spectra overlaid with
the folded best-fitting model.
The three data groups from left to right
correspond to data from the XIS, HXD/PIN, and the HXD/GSO.
The data from XIS0, 1, 2, and 3 are combined into a single spectrum.
For clarity, the data in some energy ranges is binned more coarsely than
the binning used for the actual spectral fitting.
(b) The best-fitting model photon spectrum. In addition to the
total spectrum (solid, black), some of the individual
model components are also shown. 
The dotted curve is the optically-thin
scattered component, the blue curve is the zeroth-order continuum,
and the red curve is the Compton-thick scattered component. 
(c) The ratios of the data to the best-fitting model.
(d) A close-up zoom of the Fe~K region in panel (a).
From left to right, the dotted lines correspond to the energies
of Fe~{\sc i}~K$\alpha$, \fexxv(r), Fe~{\sc i}~K$\beta$, and
the neutral Fe~K edge. The bump above the Fe~K edge is the \nika line. 
}
\label{fig:mytszzd}
\end{figure}

\subsection{Decoupled Model Setup}
\label{decoupledinterpretation}

In \S\ref{mytdecoupled} we described the motivation for applying
a Compton-thick reprocessor model in which the zeroth-order continuum
and the Compton-scattered continuum are decoupled.
In this section we apply such a ``decoupled'' \mytorus model, in which the
Compton-scattered continuum is composed of a face-on and an edge-on component, 
each of which can be varied independently of the the zeroth-order 
continuum. As explained in \S\ref{mytdecoupled} this setup can
mimic a clumpy, patchy
structure (not necessarily with a toroidal geometry),
in which a back-reflected, unobscured reflection spectrum
is able to dominate over the ``transmitted'' reflection spectrum
(see Fig.~\ref{fig:mytdecoupled}).
In our analysis of NGC~4945, except when otherwise explicitly stated,
coupling between the global and line-of-sight column density
was maintained (i.e., $N_{H,S} = N_{\rm H,Z} \equiv N_{\rm H}$).

As stated earlier,
using the \mytorus model in a decoupled mode most closely resembles
\adhoc models that use a disk-reflection component combined with line-of-sight
obscuration, so the decoupled fits will indicate how the disk-reflection models
relate to the broader set of spectral solutions that we have addressed in
the present paper (e.g., see discussion
in Yaqoob \& Murphy 2011b).

The model was set up as described in \S\ref{mytdecoupled},
with a Comptonized thermal spectrum for the intrinsic input continuum.
Preliminary fitting showed that $kT$ is poorly constrained,
so for each data set, fits with two representative values of $kT$ were performed
since tables with fixed values of $kT$ had to be used. 
Using the $N_{\rm H}$ versus $kT$ confidence contours as a guide,
we performed fits for $kT=22$~keV and $50$~keV for the \suzaku data,
and $kT=50$~keV and $100$~keV for the \bsax data. Some of the
parameters in the \bsax models were frozen as in the coupled fits 
(see \S\ref{coupledmytfits} for details). We did not apply the
decoupled \mytorus model to the \swift BAT spectrum because there were too
many free model parameters.

\subsection{Suzaku Results with the Decoupled Model}
\label{suzakudecoupledresults}

Since the present paper outlines the first detailed methodology for applying
the \mytorus model, we take this opportunity to assess the effect of
including or not including the GSO data in the spectral fitting.
We show these results for the decoupled models as opposed to the coupled
models because there are more free parameters in the former than the
latter.
NGC~4945 is amongst the brightest AGNs in the GSO bandpass, but for most
AGNs, the GSO data will be unusable due to excessive background-subtraction
systematics. We can use the NGC~4945 data to assess the impact on the derived
key physical parameters when the spectral fits are performed with XIS and PIN
data only, by comparing the results from the full-band, three-instrument fits.
Thus, a total of four fits were performed on the \suzaku data, and the results
are shown in \tablemytdcresultsp. For the three-instrument fit with
$kT=22$~keV, the folded model and counts spectra are shown
in Fig.~\ref{fig:mytszzd}(a).
The best-fitting photon spectrum is shown in Fig.~\ref{fig:mytszzd}(b),
and the data/model ratios are shown in Fig.~\ref{fig:mytszzd}(c).
It can be seen that an excellent fit is obtained, in the
1--50~keV range, with larger residuals apparent outside of this
energy range. Figure~\ref{fig:mytszzd}(d)
shows a zoomed view of the Fe~K region. As with the coupled \mytorus model, a good
fit to the \fekalfa and \fekbeta lines is obtained, and the Fe~K edge
region is also well-modeled. We see from Figure~\ref{fig:mytszzd}(b)
that in this model the zeroth-order continuum dominates over the Compton-scattered
continuum (above 10~keV), in contrast to the coupled model shown
in Figure~\ref{fig:mytszcp}(b), for which the converse is true.
The decoupled model with $kT=50$~keV is also dominated by the
zeroth-order continuum above 10~keV.

Formally, the best-fitting $\chi^{2}$ values and the associated null hypothesis
probabilities show that the fits should be statistically rejected at greater
than 98.8\% confidence, but we point out that the residuals are not
entirely statistical in origin. There are complex, energy-dependent residuals
due to the limitations of instrument calibration and despite that, the
overall residuals in the \suzaku 1--50~keV band are flat and less than $\sim 20\%$
for the \suzaku data.
Below $\sim 1$~keV, the optically-thin thermal line spectrum is likely
to be much more complex than the {\sc apec} model we have used, and we 
have also not accounted for contamination by numerous weak and unresolved X-ray
point sources (see Itoh \etal 2008, and references therein).

\subsubsection{Comparison of the Suzaku Results with and without the GSO Data}
\label{withwithoutgso}

If we compare the first pair of columns of results in \tablemytdcresults with each other,
and do the same for the second pair of columns, we see that omitting the
GSO data has a negligible effect on the best-fitting model parameters and their
statistical errors. In fact, for most parameters, the results are identical
for fits with and without the GSO, and this is true for both values of $kT$.
This outcome is encouraging because it means that a spectral-fitting
analysis of \suzaku data of 
AGN (the majority of which do not have useable GSO data) will not be impacted using
the XIS and PIN data only. Nevertheless, for NGC~4945 we refer
to the results obtained from the three-instrument fits in the remainder of the paper.
We see from \tablemytdcresults that the data cannot distinguish between the 
$kT=22$~keV and $kT=50$~keV models since the difference between
the two best-fitting $\chi^{2}$ values is only $3.0$, which, considering the
complexity of the data and the models, is negligible and insignificant.
However, we note that the
ratio of the GSO to XIS normalizations is very different for
the two solutions (see \tablemytdcresultsp). This ratio is $0.73^{+0.15}_{-0.14}$ and
$0.45^{+0.08}_{-0.09}$ for $kT=22$~keV and $kT=50$~keV respectively.
The higher value of the ratio is more in line with the corresponding 
ratio obtained for 3C~273 (see appendix), but the solution with the
lower value cannot be ruled out because of the background-subtraction systematics.
We also see that the differences between the best-fitting $\chi^{2}$ values
for the coupled \suzaku \mytorus fits in \tablemytcpplresults and the decoupled 
\suzaku \mytorus fits is not significant enough to determine whether
the coupled or decoupled \mytorus spectral fits can be ruled out (without
additional, independent information such as variability properties).

\subsection{BeppoSAX Results with the Decoupled Model}
\label{bepposaxdecoupled}

The \bsax counts spectra, best-fitting folded model, and the data/model 
residuals for the decoupled \mytorus spectral fit with $kT=50$~keV
are shown in Fig.~\ref{fig:bsaxmytsph}(b). The fit is excellent across the
entire bandpass, comparable to the fit with the BN11 spherical model
and not statistically better or worse. The decoupled \mytorus fit with
$kT=100$~keV has a lower best-fitting $\chi^{2}$ value, but again,
the difference is not statistically significant.

\subsection{Overall Assessment of the Parameters Derived with Decoupled Models}
\label{decoupledassessment}

\subsubsection{Effect of Comptonizing Plasma Temperature on the Plasma Optical Depth}
\label{effectofktontau}

We examined the effect of the Comptonizing plasma temperature, $kT$, on the
spectral-fitting results, since this temperature had to
be fixed for each fit. The largest effect of using different values of $kT$
is on the optical depth of the Comptonizing plasma, $\tau$, and this is to be 
expected, because the steepness of the intrinsic X-ray continuum (below the high-energy
roll-over) is determined by the Compton-$y$ parameter, which involves the product
$(kT)\tau$. The optical depth is as high as $\sim 2$ for the
\suzaku fit with
$kT=22$~keV, and as low as $0.15$ for the \bsax fit with
$kT=100$~keV.

\subsubsection{Column Density}
\label{decouplednhresults}

The effect of different values of $kT$
on the column density of the X-ray reprocessor, $N_{\rm H}$, is
small, only of the order of 5\%. For the \suzaku data,
the lower value of $kT=22$~keV gives
a lower value of $N_{\rm H}$ of $4.00^{+0.10}_{-0.07} \times \rm 10^{24} \ cm^{-2}$,
and the higher value of $kT=50$~keV gives a higher value of 
$N_{\rm H}$ of $4.21^{+0.07}_{-0.07} \times \rm 10^{24} \ cm^{-2}$.

The values of $N_{\rm H}$ obtained from the \bsax data are higher (nearly
$5 \times \rm 10^{24} \ cm^{-2}$), but the statistical errors are of
the order of 20\%, so the column densities are statistically consistent with those
obtained from the \suzaku data.

\subsubsection{Decoupled Column Densities}
\label{independentnhresults}

We checked whether the line-of-sight column density is required to
be significantly different to the global average column density.
This was done by decoupling the
column density associated with the zeroth-order continuum ($N_{\rm H,Z}$)
from that for the
Compton-scattered continuum ($N_{\rm H,S}$, associated with $A_{S00}$ and $A_{S90}$).
For both the \suzaku and \bsax data, the new line-of-sight column
densities ($N_{\rm H,Z}$) were statistically consistent with the original values
($N_{\rm H}$) before the column densities were decoupled.
Although the statistical errors are larger, the one-parameter, 90\% errors
are less than 10\%.
Moreover, for both the \suzaku and \bsax data, the
global column densities ($N_{\rm H,S}$) were statistically consistent with the
corresponding line-of-sight column densities ($N_{\rm H,Z}$).
In fact, the column
density of the material responsible for the Compton-scattered continuum
cannot be much less than $10^{24} \ \rm cm^{-2}$ because there would be
too much curvature in the resulting spectrum below 10~keV.
Indeed, we obtained one-parameter lower
limits at 90\% confidence on $N_{\rm H,S}$ of
$2.6 \times \rm 10^{24} \ cm^{-2}$ for the \suzaku data (for $kT=22$~keV), and
$1.06 \times 10^{24} \ \rm cm^{-2}$ for the \bsax data (for $kT=100$~keV).
On the other hand, an upper limit on $N_{\rm H,S}$
could not be obtained, with both the \suzaku and \bsax data being consistent with the highest value
tested ($10^{25} \rm \ cm^{-2}$).

\subsubsection{Compton-Scattered Continua}
\label{decoupledcsresults}

The parameter $A_{S90}$, which measures the relative magnitude of the
edge-on Compton-scattering component, was $5.2^{+3.8}_{-3.3} \times 10^{-2}$ for
the \suzaku fit with
$kT=22$~keV, and was consistent with 0.0 for $kT=50$~keV, with
an upper limit of $3.4 \times 10^{-2}$. 
The \bsax data could not constrain $A_{S90}$, which was
consistent with zero and we obtained an upper limit of $0.23$.
Note that for fits to both the \suzaku and \bsax data, $A_{S90}$
was fixed at its best-fitting value during the error analysis for the other 
parameters but was allowed to float to estimate its statistical errors.
For the relative normalizations
of the back-side Compton-scattered continuum component, we obtained
$A_{S00} = 1.67^{+0.19}_{-0.17} \times 10^{-2}$ and 
$A_{S00} = 1.39^{+0.13}_{-0.11} \times 10^{-2}$ for the \suzaku fits
with $kT=22$~keV and
$kT=50$~keV respectively. The difference between the two normalizations is
only $\sim 13\%$. For the \bsax fits, $A_{S00}$ is a factor of $\sim 7$ to $8$ lower,
at $2.1^{+2.8}_{-1.3} \times 10^{-3}$ and $1.8^{+2.4}_{-1.1} \times 10^{-3}$ for
$kT=50$~keV and $kT=100$~keV respectively.

Note that $A_{S90}$ and $A_{S00}$ cannot be directly
compared to each other because they are relative normalizations associated with
continua that have very different shapes and magnitudes. For similar
values of $A_{S90}$ and $A_{S00}$, the Compton-scattered continuum associated
with the former parameter
is much weaker than that associated with the latter parameter (e.g., see Fig.~\ref{fig:dfzdspectra}). 
We will see in \S\ref{lumratios} that in all of the decoupled \mytorus fits, the
zeroth-order continuum dominates over the Compton-scattered continuum for
both the \suzaku and \bsax data.

\subsubsection{Optically-thin Scattered Continuum}
\label{decoupledthincresults}

The parameter $f_{s}$, which corresponds to the fraction of the intrinsic continuum
scattered by an extended, optically-thin zone is $1.50^{+0.18}_{-0.10} \times 10^{-3}$ and
$0.94^{+0.05}_{-0.05} \times 10^{-3}$ for the \suzaku fits with
$kT=22$~keV and $kT=50$~keV
respectively. For the \bsax fits with $kT=50$~keV and $kT=100$~keV,
we obtained much lower values of $f_{s}$ of $2.9^{+4.8}_{-2.1} \times 10^{-4}$
and $2.0^{+3.4}_{-1.3} \times 10^{-4}$ respectively.
As explained in \S\ref{coupledmytfits}, the value of $f_{s}$ is highly model-dependent
because it depends on the intrinsic continuum luminosity, which can differ
by over an order of magnitude for different models.

\subsubsection{Soft X-ray Emission and Absorption Components}
\label{decoupledsoftxrayresults}

The remaining model parameters, namely the temperature and normalization
of the soft X-ray thermal component, $N_{\rm H1}$, $N_{\rm H2}$, and all
of the emission-line parameters (including the \fekalfa line
fluxes) do {\it not} show any discernible 
dependence on the value of $kT$ adopted. This is to be expected since
all of the parameters mentioned are associated with the spectrum below
$\sim 10$~keV. For the same reason, these parameter values
are similar to the corresponding
values obtained for the coupled \mytorus fit (see \S\ref{coupledmytfits} and \tablemytcpplresultsp) so
we do not discuss them again here.

\section{Variability, Intrinsic Luminosities, and Eddington Luminosity Ratios}
\label{lumratios}

\begin{table}
\caption[NGC~4945 Intrinsic to Observed Luminosity Ratios]
{NGC~4945 Intrinsic to Observed Luminosity Ratios}
\begin{center}
\begin{tabular}{llllccccc}
\hline
& & & & & & & & \\
Type & Data & Model & Continuum & $kT$ 
& $L_{\rm intr}/L_{\rm obs}$ & $L_{\rm intr}/L_{\rm obs}$
 & $L_{\rm intr}/L_{\rm obs}$ & $L_{\rm intr}/L_{\rm obs}$ \\
& & & & (keV) & (0.7--2 keV) & (2--10 keV) & (10--100 keV) & (14--195 keV) \\
& & & & & & & & \\
\hline
& & & & & & & & \\
Zeroth-order & \swift BAT & \mytorus & \comptt & $36^{+62}_{-10}$ & \ldots & \ldots & \ldots & $13.7$ \\
& & & & & & & & \\
Coupled & \suzaku & \mytorus & {\sc powerlaw} & \ldots & $186.2$ & $93.4$ & $4.0$ & $3.7$ \\
Coupled & \suzaku & \mytorus & \comptt & $50\pm4$~keV & $195.7$ & $102.1$ & $4.3$ & $3.7$ \\ 
Coupled & \suzaku & {\sc sphere} & {\sc powerlaw} & \ldots & $36.2$ & $21.2$ & $1.07$ & $1.01$ \\
Coupled & \bsax & \mytorus & \comptt & $50$~keV & \ldots & $1337.7$ & $24.7$ & $18.1$  \\
Coupled & \bsax & \mytorus & \comptt & $80$~keV & \ldots & $1588.2$ & $27.0$ & $19.4$ \\
Coupled & \bsax & \mytorus & \comptt & $100$~keV & \ldots & $1764.3$ & $28.0$ & $19.1$ \\
Coupled & \bsax & {\sc sphere} & {\sc powerlaw} & \ldots & $46.3$ & $1.43$ & $1.35$ \\
Coupled & \swift BAT & \mytorus & {\sc powerlaw} & \ldots & & & & $3.8$ \\
Coupled & \swift BAT & \mytorus & \comptt & $50^{+16}_{-17}$ & \ldots & \ldots & \ldots & $3.54$ \\
Coupled & \swift BAT &{\sc sphere} & {\sc powerlaw} & \ldots & \ldots & \ldots & \ldots & $1.22$ \\
& & & & & & & & \\
Decoupled & \suzaku & \mytorus & \comptt & $22$~keV & $1012.1$ & $469.4$ & $16.3$ & $13.5$ \\
Decoupled & \suzaku & \mytorus & \comptt & $50$~keV & $1650.0$ & $736.0$ & $22.0$ & $16.0$ \\
Decoupled & \bsax & \mytorus & \comptt & $50$~keV & \ldots & $3122.2$ & $44.1$ & $30.6$ \\
Decoupled & \bsax & \mytorus & \comptt & $100$~keV & \ldots & $4534.0$ & $51.9$ & $34.4$ \\
& & & & & & & & \\
\hline
\end{tabular}
\end{center}
The ratios of intrinsic ($L_{\rm intr}$) to observed 
($L_{\rm obs}$) model luminosities for the spectral fits presented
in \tablebatresultsp, \tablemytcpplresultsp, and \tablemytdcresultsp
(see \S\ref{lumratios} for details). For the coupled \mytorus fits to the \bsax
data, luminosity ratios are given for two values of the 
intrinsic continuum plasma temperature ($kT=50$ and $100$~keV) in 
addition to the fit with $kT=80$~keV shown in \tablemytcpplresultsp. 
\end{table}

\begin{table}
\caption[NGC~4945 Intrinsic Luminosities and Eddington Ratios]
{NGC~4945 Intrinsic Luminosities and Their Ratios to the Eddington Luminosity}
\begin{center}
\begin{tabular}{llllccccc}
\hline
& & & & & & & & \\
Type & Data & Model & Continuum & $kT$ 
& $L_{\rm intr}$ & $L_{\rm intr}$
 & $L_{\rm intr}/L_{\rm Edd}$ & $L_{\rm intr}/L_{\rm Edd}$ \\
& & & & (keV) & (2--10 keV) & (2--195 keV) & (2--195 keV) & (15--100 keV) \\
& & & &       & ($10^{42} \ \rm erg \ s^{-1}$) & ($10^{42} \ \rm erg \ s^{-1}$) \\
& & & & & & & & \\
\hline
& & & & & & & & \\
Zeroth-order & \swift BAT & \mytorus & \comptt & $36^{+62}_{-10}$ & $19.8$ & $50.3$ & $0.285$ & $0.123$ \\
& & & & & & & & \\
Coupled & \suzaku & \mytorus & {\sc powerlaw} & \ldots & $2.86$ & $16.7$ & $0.0946$ & $0.0459$ \\
Coupled & \suzaku & \mytorus & \comptt & $50\pm4$~keV & $3.10$ & $15.2$ & $0.0860$ & $0.0446$ \\ 
Coupled & \suzaku & {\sc sphere} & {\sc powerlaw} & \ldots & $0.64$ & $4.72$ & $0.0267$ & $0.0130$ \\
Coupled & \bsax & \mytorus & \comptt & $50$~keV & $50.0$ & $126.8$ & $0.717$ & $0.301$ \\
Coupled & \bsax & \mytorus & \comptt & $80$~keV & $59.4$ & $144.0$ & $0.814$ & $0.326$ \\
Coupled & \bsax & \mytorus & \comptt & $100$~keV & $65.6$ & $153.5$ & $0.868$ & $0.337$ \\
Coupled & \bsax & {\sc sphere} & {\sc powerlaw} & \ldots & $1.76$ & $7.1$ & $0.0401$ & $0.0185$  \\
Coupled & \swift BAT & \mytorus & {\sc powerlaw} & \ldots & $2.95$ & $11.1$ & $0.0628$ & $0.0286$ \\
Coupled & \swift BAT & \mytorus & \comptt & $50^{+16}_{-17}$ & $2.53$ & $10.1$ & $0.0572$ & $0.02884$ \\
Coupled & \swift BAT & {\sc sphere} & {\sc powerlaw} & $\ldots$ & $0.78$ & $3.41$ & $0.0193$ & $0.00904$ \\
& & & & & & & & \\
Decoupled & \suzaku & \mytorus & \comptt & $22$~keV & $14.3$ & $53.7$ & $0.304$ & $0.162$ \\
Decoupled & \suzaku & \mytorus & \comptt & $50$~keV & $22.3$ & $87.6$ & $0.495$ & $0.239$ \\
Decoupled & \bsax & \mytorus & \comptt & $50$~keV & $116.8$ & $250.4$ & $1.416$ & $0.520$ \\
Decoupled & \bsax & \mytorus & \comptt & $100$~keV & $169.1$ & $329.4$ & $1.863$ & $0.605$ \\
& & & & & & & & \\
\hline
\end{tabular}
\end{center}
Absolute luminosities $L_{\rm intr}$, in the 2--10~keV and 2--195~keV bands, corresponding to the
intrinsic to observed luminosity ratios shown in \tablelumratiosp.
For each spectral fit, the ratios of the 2--195~keV and 15--100~keV luminosities
to the Eddington luminosity, $L_{\rm Edd}$, are also shown.
\end{table}

\begin{figure}
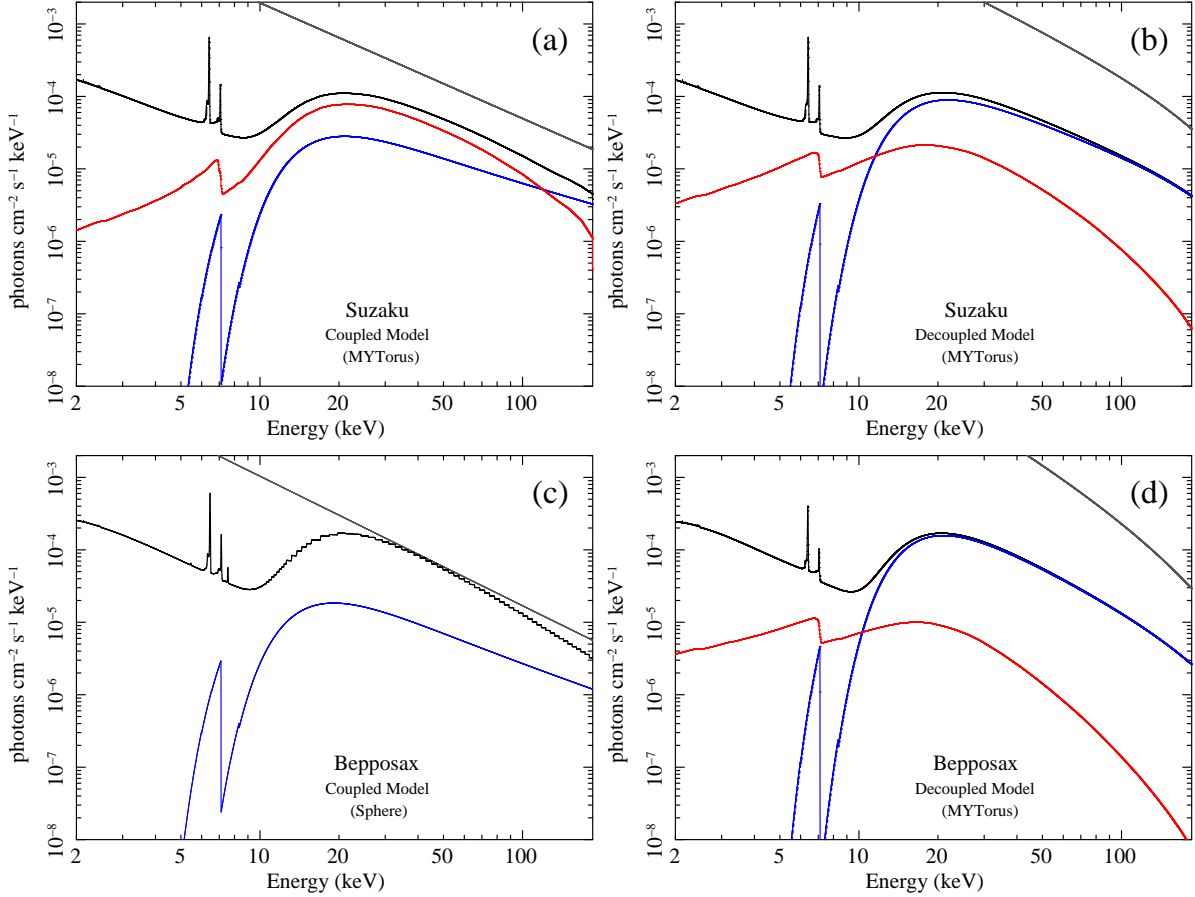

\centerline{
 	\psfig{figure=f12a.ps,height=6cm,angle=270}	
        \psfig{figure=f12b.ps,height=6cm,angle=270}
	}
\centerline{
        \psfig{figure=f12c.ps,height=6cm,angle=270}
        \psfig{figure=f12d.ps,height=6cm,angle=270}        
        }
        \caption{\footnotesize An illustration of how the intrinsic continuum 
derived for NGC~4945 (grey curves) compares to
the total observed spectrum (black curves), and the relative contributions of the zeroth-order
continuum (blue), and the Compton-scattered continuum (red). Shown are two degenerate solutions
(for coupled and decoupled models) to the \suzaku data in (a) and (b), and to the \bsax
data in (c) and (d). In these examples, the Compton-scattered continuum dominates over the
zeroth-order continuum in the region of the Compton hump for the coupled models, whilst the
converse is true for the decoupled models. It can be seen that, since the intrinsic continuum
must be a certain amount higher than the zeroth-order continuum for a given column density,
the intrinsic continuum is highest when the zeroth-order continuum dominates the observed
spectrum. Note that the spherical model for the coupled solutions to the \bsax data does not
allow separation of the Compton-scattered continuum, so only 
the zeroth-order continuum contribution is shown, which was calculated using the \mytorus
zeroth-order table for the required column density and intrinsic photon index. The parameters
of the coupled models are given in \tablemytcpplresults (see also \S\ref{coupledmytfits}),
and those for the decoupled models are given in \tablemytdcresults (see also \S\ref{decoupledmytfits}).
}
\label{fig:cmpcppdecoup}
\end{figure}

\begin{figure}
\centerline{
 \psfig{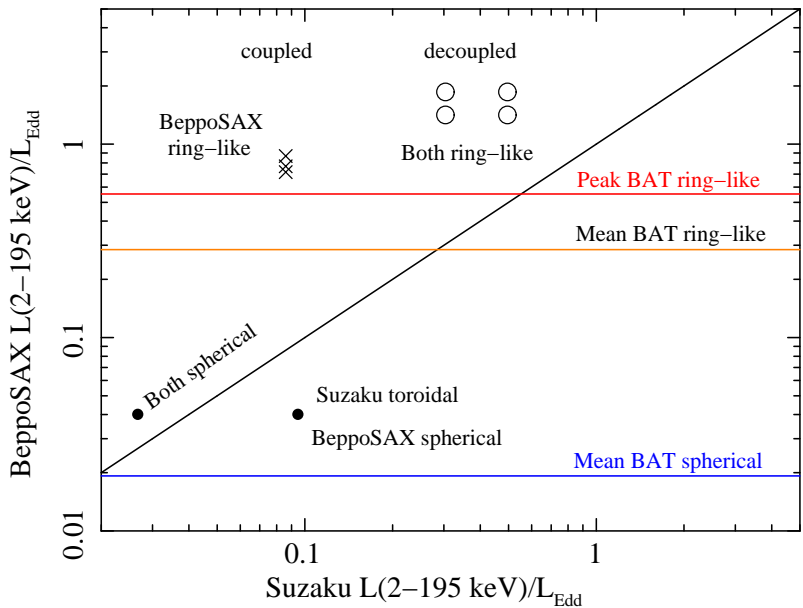}}
\caption{\footnotesize The ratios of the 2--195 keV intrinsic luminosities to the Eddington luminosity,
$L_{\rm Edd}$,
for the \bsax spectral fits versus the Eddington ratios for the \suzaku fits.
The solid line corresponds to equality of the \bsax and \suzaku ratios. Since $L_{\rm Edd}$
is constant, the diagram also shows whether various solutions require the intrinsic
luminosity during the \suzaku observation to be lower or higher than than that during the 
\bsax observation. Points below and above the line imply higher and lower luminosities
during the \suzaku observation
respectively. The filled circles correspond to the coupled models in which the
Compton-scattered continuum dominates over the zeroth-order continuum
(second and fourth columns of results in \tablemytcpplresultsp). The three
crosses correspond
to the coupled \bsax solutions in which the zeroth-order continuum dominates,
for three values of $kT$ (rows 3 to 5 in \tablemytdecoupp).
Open circles correspond to the decoupled model solutions: each solution
for \suzaku with $kT=22$ and $50$~keV is plotted against the
two solutions for \bsax ($kT=50$ and $100$~keV). These decoupled
models also have the zeroth-order continuum dominating at high energies,
and such solutions imply a ring-like, geometrically-thin geometry for the
X-ray reprocessor. The red horizontal line shows the maximum possible
value of $L_{\rm intr}/L_{\rm Edd}$ inferred from the peak value of the
66-month \swift BAT lightcurve in Fig.~\ref{fig:batlc}, obtained from
fitting the \swift BAT mean spectrum with a zeroth-order continuum model
and renormalizing the intrinsic luminosity to the lightcurve peak. The
mean \swift BAT Eddington ratio is shown for comparison (brown line),
as is the lowest Eddington ratio obtained from fitting the mean spectrum
with a spherical model. See \tableloverledd and \S\ref{lumratios}
for details. 
}
\label{fig:loverleddszvssax}
\end{figure}

\subsection{Intrinsic to Observed Luminosity Ratios}
\label{luminosityratios}

In \tablelumratios we have collected together all of the intrinsic to observed luminosity
ratios (in four energy bands) for the various spectral fits described in previous sections.
As would be expected, for a given data set,
the luminosity ratios for models in which the zeroth-order continuum dominates
over the Compton-scattered continuum (above $\sim 10$~keV) are larger than those 
for models in which the 
Compton-scattered continuum dominates (see discussion in \S\ref{coupledmytfits}).
The differences in intrinsic luminosity between the two types of model
can be substantial. In the 2--10~keV band the luminosity ratios span
more than three orders of magnitude for the different models. In the
higher energy bands the luminosity ratios are smaller but still can
be as great as a factor of 25. Comparing the luminosity ratios
for \suzaku and \bsaxp, we see that the ratios are larger
for the \bsax data (the difference between
the \suzaku and \bsax luminosity ratios depends on the energy band and the model,
and can be over an order of magnitude).

\subsection{Absolute Luminosities}
\label{absoluteluminosities}

\tableloverledd shows absolute intrinsic luminosities for the various spectral fits
in two energy bands, 2--10~keV and 2--195~keV. For the latter, extrapolation
of the model over energy gaps in the data was necessary. 
Fig.~\ref{fig:cmpcppdecoup} helps to visualize how the intrinsic continuum relates
to other components of the model for the 
different classes of solutions for both the \suzaku and \bsax data.
In each of the panels in Fig.~\ref{fig:cmpcppdecoup}, the grey curves correspond to the
intrinsic continua, the red curves to the Compton-scattered continua
and the blue curves to the zeroth-order continua. The intrinsic spectrum for the coupled \mytorus model for the
\suzaku data in Fig.~\ref{fig:cmpcppdecoup}(a) can be directly compared with the intrinsic spectrum for the
decoupled \mytorus model in Fig.~\ref{fig:cmpcppdecoup}(b) for the same data.
A similar comparison between coupled and decoupled models that were
found to be degenerate solutions for the \bsax data can be seen in 
in Fig.~\ref{fig:cmpcppdecoup}(c) and Fig.~\ref{fig:cmpcppdecoup}(d) respectively.
For the sake of clarity, we do not show all the possible solutions in Fig.~\ref{fig:cmpcppdecoup},
but instead illustrate with examples covering the important different scenarios. 

\subsection{Variability of the Intrinsic Luminosity}
\label{lumvariability}
 
For the decoupled models, which are the ones relevant for NGC~4945,
the intrinsic continuum luminosity is required to decrease
in going from the \bsax observation to the \suzaku observation.
The amount of this decrease depends on the model parameters and 
the energy band, but it can more than a factor of 2, and even an
order of magnitude in the 2--10~keV band. Yet the {\it observed} luminosities
for the \bsax and \suzaku data are within a factor of 2 of each other
(see \tablemytcpplresultsp, \tablemytdcresultsp, and Fig.~\ref{fig:saxbatoverlay}), with the 
observed luminosity during the \suzaku observation being less than,
or similar to, the corresponding observed luminosity during the \bsax observation.

\subsection{Ratios of the Intrinsic Luminosity to the Eddington Luminosity}
\label{eddingtonratios}
 
\tableloverledd also gives intrinsic continuum luminosities ($L_{\rm intr}$) as a ratio of
the Eddington luminosity, $L_{\rm Edd}$, which for the black-hole mass adopted here
(see \S\ref{strategy}), is $1.77 \times 10^{44} \rm \ erg \ s^{-1}$.
The ratios $L_{\rm intr}/L_{\rm Edd}$ (hereafter, equivalently, $L/L_{\rm Edd}$),
are shown in \tableloverledd for intrinsic luminosities
in the 2--195~keV band, which is likely to be within a factor of $\sim 2$--4
of the bolometric luminosity (e.g., see Vasudevan \etal 2010). However, the 2--195~keV 
luminosities involve
some extrapolation of the models for different data sets, so we have
also computed $L/L_{\rm Edd}$ in the 15--100~keV band, since this is
covered by the \suzakup, \bsaxp, and \swift BAT data. The consistency
of the behavior of $L/L_{\rm Edd}$ in the two energy bands is evident from
\tableloverleddp. As we might expect, the models in which the
zeroth-order continuum dominates have the largest value of $L/L_{\rm Edd}$,
and all of these models have a 2--195~keV $L/L_{\rm Edd}$ ratio greater than $0.28$.
For the case of the decoupled \bsax models, $L/L_{\rm Edd}$ exceeds unity.
Since the bolometric value of $L/L_{\rm Edd}$ is higher than the
corresponding ratio for
the 2--195~keV band, all of the decoupled models present a difficulty.

Previous X-ray studies of NGC~4945 
also deduced a high value for $L/L_{\rm Edd}$ (e.g., Itoh \etal 2008;
Fukazawa \etal 2011; Marinucci \etal 2012).
This is because the decoupled models we have applied have
much in common with disk-reflection models.
A value of $R \sim 3 \times 10^{-3}$,
(the relative ``reflection fraction'' in the {\sc pexrav} model), was
obtained by Itoh \etal (2008), and this is consistent with the high-energy
spectrum being dominated by the zeroth-order continuum. 
From their analysis of the \bsax data, Guainazzi \etal (2000) 
came to similar conclusions from applying the same kind of model
({\sc pexrav} and {\sc cabs}), namely that there is no evidence
for Compton reflection ($R \ll 1$), and that the intrinsic luminosity is
a significant fraction of the Eddington luminosity. As expected,
our Compton-thick column densities and inferred intrinsic luminosities
from the decoupled \mytorus fits are similar to the corresponding
values obtained by Guainazzi \etal (2000). However, as
we have pointed out, the method of using {\sc pexrav}
and {\sc cabs} may be biased towards finding the solutions dominated
by the zeroth-order continuum and these may not be appropriate for
all Compton-thick AGNs.
At the other extreme, \tableloverledd shows that coupled models in
which the Compton-scattered continuum dominates give
much lower values of $L/L_{\rm Edd}$. Specifically, the spherical model
fitted to the \suzakup, \bsaxp, and \swift BAT data gave values of
$\sim 0.027$, $\sim 0.040$, and $\sim 0.019$ respectively for the 2--195~keV $L/L_{\rm Edd}$ ratio.
However, these solutions are not relevant for NGC~4945 because
of the independent variability of the high-energy continuum
with respect to the \fekalfa line (the spectrum above 10~keV
cannot be dominated by the Compton-scattered continuum).
Nevertheless, we discuss the solutions for possible relevance to
other AGNs.
These Compton-thick  
solutions with a spherical geometry are missed by the standard scheme of modeling
Compton-thick AGNs because the Compton-scattered continuum from
a fully-covering reprocessor can have a shape that is more like the
zeroth-order continuum (but with a different column density) than
a disk-reflection continuum. Fitting the X-ray spectrum from a
fully-covered Compton-thick source with disk-reflection can lead
to the erroneous conclusion that the ``reflection fraction'' ($R$)
is negligible, when in fact the spectrum is {\it dominated} by the
Compton-scattered continuum. However, every AGN should be assessed
on a case-by-case basis because the parameter $R$ in disk-reflection models
is in general allowed to float with no constraint. Indeed, BN11
concluded that in their sample of obscured AGNs, the 
disk-reflection models gave lower intrinsic luminosities than toroidal
models. 
 
\subsubsection{A Comparison of the Eddington Ratio and Column Density for Different Models}
\label{cmpeddrationh}

In the case of NGC~4945, there is a further problem with the high values of $L/L_{\rm Edd}$
obtained for some of the models. We illustrate this in Fig.~\ref{fig:loverleddszvssax},
which shows the 2--195~keV $L/L_{\rm Edd}$ ratios obtained from
the \bsax data versus the  2--195~keV $L/L_{\rm Edd}$ ratios obtained from the \suzaku data, for
comparable models. The diagonal solid line corresponds to an equal
ratio for the two data sets. Points that lie below this line require
that the intrinsic luminosity of NGC~4945 during the \bsax observation
was less than that during the \suzaku observation, whereas points
that lie above the line indicate a change in luminosity in the
opposite sense (i.e., larger luminosities during the \bsax
observation). 
The problem arises when we consider the fact that if we fit the
time-averaged 58-month \swift BAT spectrum with a model that
consists of {\it only} the zeroth-order continuum, then we
obtain an absolute upper limit on the intrinsic continuum
luminosity (averaged over 58-months). We can then obtain
a normalization factor from the \swift BAT light curve 
(Fig.~\ref{fig:batlc}) and estimate the maximum intrinsic
luminosity that was attained during the 66-month period covered
by the light curve (by dividing the highest peak flux by the mean flux).
Thus, we can estimate the highest value of the 2--195~keV 
$L/L_{\rm Edd}$ ratio that NGC~4945 attained during the period
covered by the \swift BAT lightcurve, and this is shown as a red
line in Fig.~\ref{fig:loverleddszvssax}. For reference, the value of $L/L_{\rm Edd}$
is also shown for the mean level of the \swift BAT light curve
(brown line), and also for the spherical model fit to the \swift
BAT spectrum (blue line; see also \tablebatresultsp). The problem is that {\it all} of
the models in which the zeroth-order continuum dominates
require that the intrinsic luminosity during the
\bsax observation was higher than it ever was during the
\swift BAT 66-month monitoring. While this is not impossible, it
should be remembered that these solutions for the \bsax data
require super-Eddington luminosities.

\begin{figure}
\centerline{
 \psfig{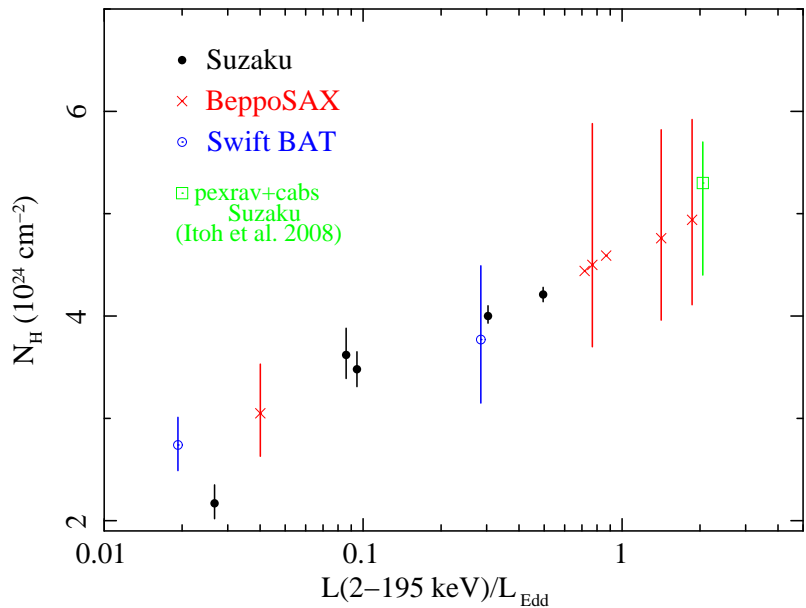}}
\caption{\footnotesize The column density of the Compton-thick X-ray reprocessor
obtained from various spectral fits reported in
\tablebatresultsp, \tablemytcpplresultsp, and \tablemytdcresultsp, versus
the corresponding ratio of the 2--195~keV intrinsic continuum luminosity to the Eddington luminosity
(see \tableloverleddp). The diagram shows at glance the range of $N_{\rm H}$ 
(a factor greater than 2) spanned
by the spectral fits over the different models that were applied to
three data sets (from \suzakup, \bsaxp, and \swift BAT). The diagram also
shows at a glance the corresponding range in the intrinsic Eddington luminosity
ratio, which is very large, about 2 orders of magnitude.
As might be expected, spectral solutions with higher values of $N_{\rm H}$
generally have higher intrinsic broadband luminosities.
Note that statistical errors are not shown for
two of the \bsax points (red crosses) because only best-fitting parameters were obtained
for these coupled model fits, which have $kT=50$ and $100$~keV. The statistical errors
are similar to those for the corresponding coupled model fit with $kT=80$~keV (see \tablemytcpplresultsp).
}
\label{fig:nhvsloverledd}
\end{figure}

We also see from \tablelumratios and \tableloverledd
that whereas key model parameters such as $N_{\rm H}$ were
not sensitive to the assumed plasma temperature ($kT$) in the decoupled
fits, the inferred intrinsic to observed luminosity ratio in the 2--10~keV band {\it does}
depend on the plasma temperature. The difference is greatest between the $kT=22$~keV
and $kT=50$~keV \suzaku fits, the latter yielding a ratio $\sim 60\%$ larger
than the former. For the \bsax data, the difference is only $\sim 20\%$ between
the fits with $kT=50$~keV and $kT=100$~keV. The dependence
of the intrinsic to observed luminosity ratios on $kT$ is less for
the higher energy bands, for all fits and data sets. This is because changes in the
shape of the spectrum produce larger flux amplification at low energies if the
spectrum pivots at high energies. Considering all of the models and all of the data sets,
the possible range in intrinsic luminosity from
the NGC~4945 spectral fits alone spans over two orders of magnitude
in the 2--10~keV band, and just under two orders of magnitude in the 14--195~keV band.
The derived values of $N_{\rm H}$ from the various fits span the range
$\sim 2$--$6 \times 10^{24} \ \rm cm^{-2}$ (including statistical errors),
and are plotted against the 
corresponding values of the 2--195~keV Eddington ratio in Fig.~\ref{fig:nhvsloverledd}.
As might be expected, $N_{\rm H}$ is greater for greater values of $L/L_{\rm Edd}$.
For comparison, we used the parameters of the {\sc pexrav}/{\sc cabs} model applied
by Itoh \etal (2008) to estimate the 2--195~keV $L/L_{\rm Edd}$ ratio, and show this
in Fig.~\ref{fig:nhvsloverledd} (green data point). It can be seen that this value
of $L/L_{\rm Edd}$ is greater than 2 and is higher than any of the values we
obtained from any of our models applied to any of the three data sets. 

\subsection{Interpretation of the Allowed Solutions for NGC~4945}
\label{solutioninterpretation}

In summary, since the fact that the spectrum above 10~keV in NGC~4945
must be dominated by the zeroth-continuum is robust, values
of $L/L_{\rm Edd}$ close to or greater than unity are unavoidable.
The most straightforward geometrical interpretation of our results is that
the central X-ray source is obscured by a single clumpy
distribution, with the line of sight obscured by material in the
same distribution (its components must be larger than the X-ray source). 
The size of the distribution would have to be on a scale of
tens of parsecs if it is to be identified with the spatially-resolved
\fekalfa line-emitting region reported in Marinucci \etal (2012).
The covering factor must be small enough that the Compton-scattered
continuum does not dominate over the zeroth-order continuum above 10~keV.
Our spectral fitting results indicate that the covering factor is
of the order of $\sim 10\%$ or less.
A separate, more compact, ring-like structure in addition to the
extended distribution is not ruled out (its covering factor would have
to be small enough to make a negligible contribution to the \fekalfa line).

So far we have only considered an isotropic X-ray continuum
source. However, there is a viable and intriguing alternative possibility that
alleviates the need for a central source with such an enormous
luminosity radiating near or beyond the Eddington limit. That is,
NGC 4945 could harbor a beamed AGN, embedded in a patchy or clumpy Compton-thick
shell that could have a large covering factor. The X-ray emission in
such a scenario would need to be collimated along the line of sight so that the
true ratio of $L/L_{\rm Edd}$ could be an order of magnitude less than
unity. The Compton-scattered continuum would be swamped by the direct
zeroth-order continuum, regardless of the geometry and covering factor
of the Compton-thick circumnuclear matter. Beaming is not uncommon in type~1
AGNs and surrounding such a source with a Compton-thick shell would
produce an object like NGC~4945. If the true intrinsic luminosity of NGC~4945 is
closer to $10^{43} \ \rm erg \ s^{-1}$ than $10^{44} \ \rm erg \ s^{-1}$,
and if the X-ray emission were {\it not} beamed along our line of sight,
the Compton-thick obscuration would make it difficult to detect the source in
hard X-ray surveys. In other words, there could be a large number of
beamed Compton-thick AGNs with intrinsic luminosities of the order of $10^{43} \ \rm erg \ s^{-1}$
or less
that have not been detected because of an unfavorable beaming direction.

The beaming scenario is consistent with all of the X-ray observational
constraints in NGC~4945 and actually provides a more natural interpretation
of the data than an isotropic X-ray source surrounded by a Compton-thick
structure that cannot have a covering factor that is too large or too small.
If NGC~4945 is a beamed Compton-thick AGN, no fine-tuning of the
covering factor is required. Independent evidence of beamed emission has also
been found in the radio band by Lenc \& Tingay (2009), who reported the 
discovery of a jet-like structure at 2.3 GHz, with apparent dimensions of
$\sim 5$~pc in length by $\sim 1.5$~pc in width. The alignment of the
putative jet could not be constrained uniquely, but Lenc \& Tingay (2009)
derived $\beta \cos{\theta}>0.52$, where $\beta \equiv (v/c)$ is the
speed of the jet material, and $\theta$ is the angle between
the direction of motion and the line of sight. Therefore, alignment of
the putative jet emission along the line of sight is not ruled out if $\beta >0.52$
($\beta$ is unknown).
Moreover, the detection of NGC~4945 in the GeV band by Fermi (Lenain \etal 2011)
lends further support to the beaming scenario for the intrinsic X-ray continuum.

\section{Summary}
\label{summary}

\begin{figure}
\centerline{
        \psfig{figure=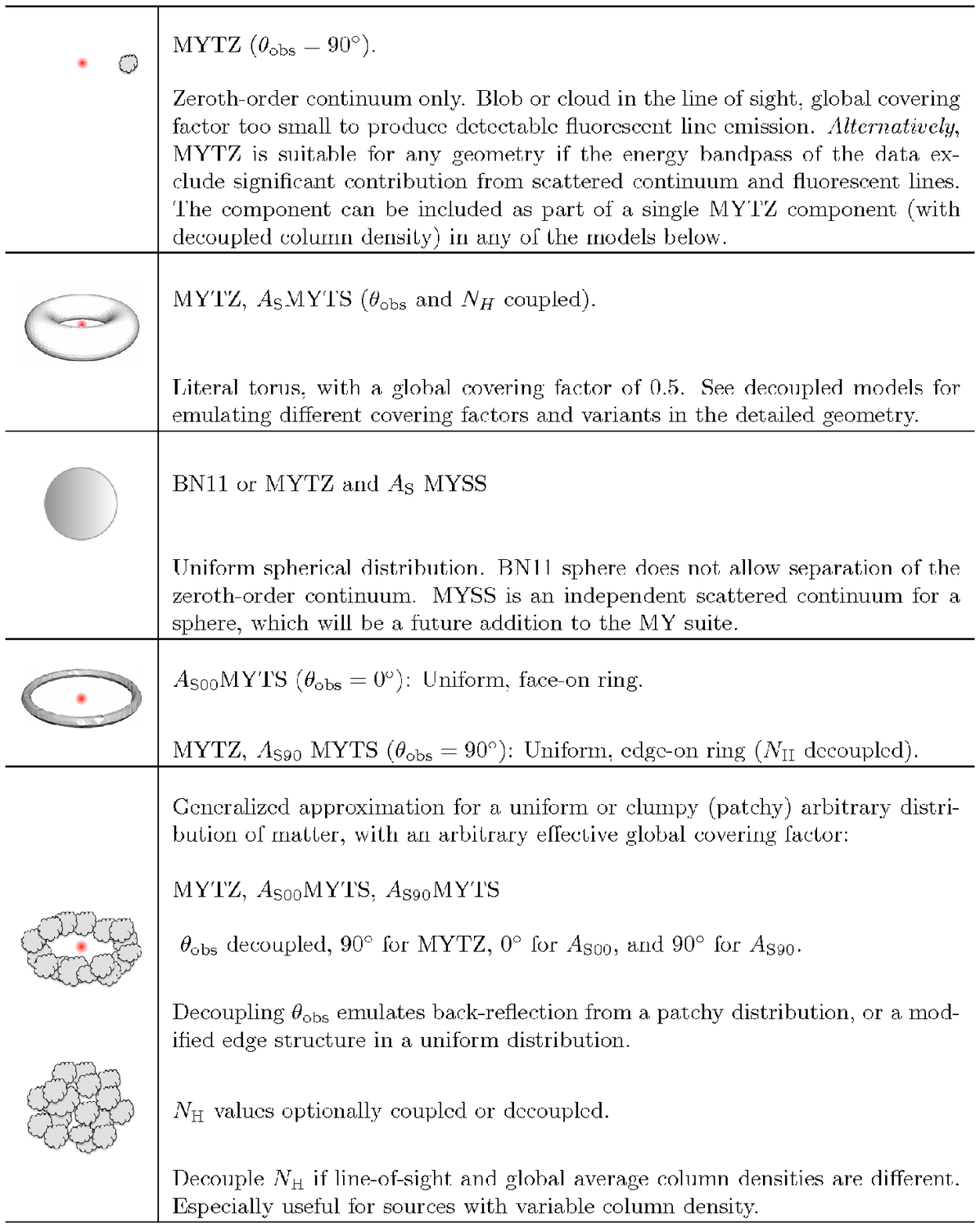,width=15cm,angle=0}
        }
        \caption{\footnotesize A schematic summary for modeling the
X-ray absorption and reprocessing in AGNs for different physical
configurations of the obscuring matter. The X-ray continuum source is shown as
a red, filled circle.}
\label{fig:summaryscheme}
\end{figure}

We have presented a detailed methodology for applying the \mytorus X-ray
reprocessor model to X-ray spectra of AGN, and we have presented the results of
applying the model to noncontemporaneous \suzakup, \bsaxp, and \swift BAT data
for the Seyfert 2 galaxy NGC~4945. We have also applied the toroidal and
spherical models of Brightman and Nandra (2011) to these data.
We have given detailed procedures to implement several different geometrical
configurations for the X-ray reprocessor, some of which are exact, and some
are approximations. Although the different modes of applying the \mytorus 
suite of models to emulate structures that are different to the
standard, uniform torus are approximations in some cases, the models overcome
some important restrictions imposed by disk-reflection models.
The different schemes, which are not just restricted
to Compton-thick AGNs, are summarized 
and illustrated in Fig.~\ref{fig:summaryscheme}. 
Below we summarize our general conclusions and then summarize the
specific conclusions for NGC~4945.
\\

\par\noindent
(i) The current scheme of using a disk-reflection spectrum to model
the Compton-scattered continuum assumes an infinite column density so
such models cannot access the rich variety of
spectral shapes that result from finite column densities. Disk-reflection
models may also not be able to access spectral solutions with
the full range of intrinsic continuum luminosities that are allowed
by the data. \\
(ii) In sources that are Compton-thick in the line of sight, 
that component of the continuum
diminished by matter in the line of sight (the zeroth-order continuum)
will not leave a prominent signature below $\sim 10$~keV (in the
form of the structure in the Fe~K band, and continuum curvature below it).
This may give rise to ambiguity in the possible model solutions, even
for high signal-to-noise data. Conversely, for Compton-thin sources
with column densities in the range $\sim  10^{23}$ to $\sim 10^{24} \rm \ cm^{-2}$,
the spectrum below 10~keV will have specific contributions from
the Compton-scattered continuum, zeroth-order continuum, and
fluorescent line emission, leading to less ambiguity in allowed
spectral solutions. (Sources with column densities less than
$10^{23}$ are of course much simpler to model.) \\
(iii) In Compton-thick sources it is possible that the dominant
spectral components in the $\sim 5-10$~keV band are due to X-ray reflection
and fluorescence from the back side of obscuring matter located
on the ``far-side'' of a clumpy or patchy structure. In other words,
spectral components from the near-side of the structure could be
so suppressed that even if only a few per cent or less of the
far-side reflection is unobscured, it could still dominate the
spectrum in the Fe~K band. \\
(iv) In cases in which significant variability above $\sim 10$~keV has
been established, the spectral solution, or solutions, adopted must be consistent
with the variability data. Specifically, either the zeroth-order continuum
must dominate over the Compton-scattered continuum in the pertinent energy band,
or else the entire reprocessing structure must have a light-crossing
time that is sufficiently small to be consistent with the variability.
Which continuum component dominates in a given energy band is a complex function
of the geometry, global covering factor, and column density (both in and out
of the line of sight). Each data set must be assessed on a case-by-case basis. \\
(v) If there is no variability information, one must explore all permissible
spectral solutions, which might be associated with very different intrinsic
continuum luminosities. We have emphasized the need for models to allow the zeroth-order
continuum, Compton-scattered continuum, and the \fekalfa line flux to be independently varied,
in order to accommodate spectral variability and/or light-crossing time
delays, which should be considered even if only a single observation
is available. Currently, only the \mytorus model allows separability
of the different model components. \\
(vi) In general, an X-ray reprocessor model that fits a spectrum
and is dominated by the
zeroth-order continuum in a given energy band
has the largest possible intrinsic luminosity in that energy band
because any Compton-scattered continuum from material out of the line of sight
can only reduce the required intrinsic luminosity to produce the same flux.
On the other hand a fully-covering distribution of matter gives
a spectrum above $\sim 10$~keV that is dominated by the Compton-scattered
continuum if the material is Compton-thick. Therefore, spectral fits
with only a zeroth-order continuum model and with a spherical model can
yield upper and lower limits on the intrinsic continuum luminosity respectively
(they will be equal in the Compton-thin limit). \\
(vii) We have described an important extension to the \mytorus model that
allows the use of a thermally Comptonized intrinsic continuum. These spectra
have a natural high-energy rollover, characterized by the plasma temperature, $kT$.
Currently, the \mytorus suite of models is the only one publicly available that allows
an intrinsic continuum that is not a simple power law. We found that despite
the poor spectral resolution of the \swift BAT spectra, it is possible in some
situations for even \swift BAT data to be able distinguish between the
power-law and thermally Comptonized intrinsic continua. 

We now summarize below the specific results and conclusions from applying
our methods to \suzakup, \bsaxp, and \swift BAT data for NGC~4945.
\\

\par\noindent
(i) Whereas the noncontemporaneous \suzaku and \bsax NGC 4945 data cover
a period of time of the order of a day, the \swift BAT spectrum 
from the ongoing all-sky survey was time-averaged
over 58 months. The data sets were therefore fitted independently. \\
(ii) For Comptonized thermal plasma models of the intrinsic continuum, the temperature
of the Comptonizing plasma, $kT$, is difficult to constrain precisely because it
is model-dependent. In general, the \suzaku data are consistent with $kT$ in
the range 20--50~keV, whilst the \bsax data have a preference for $kT$ in the
range 50-100~keV. The \swift BAT constraints on $kT$ are looser and consistent
with those from both \suzaku and \bsaxp. \\
(iii) We found that even for the \suzaku data (which cover the most spectral
features and have the best spectral resolution in the Fe~K band),
degenerate spectral solutions exist.
Even applied to a single spectrum, these degenerate solutions
can yield intrinsic luminosities that differ by an order of magnitude or more.
The different models applied to the different data sets for NGC~4945
cover a range in intrinsic luminosity that spans two
orders of magnitude and nearly a factor of 3 in column density
(from $\sim 2 \times 10^{24}$ to $6 \times 10^{24} \ \rm cm^{-2}$).  Yet the
{\it observed} luminosities of the \bsax and \suzaku spectra do
not differ by more than a factor of 2. \\
(iv) The 66-month \swift BAT light curve shows
large-amplitude variability on a timescale of months, covering a dynamic
range of a factor of $\sim 8$ from minimum to maximum flux.
Marinucci \etal (2012) presented independent evidence that the
\fekalfa line flux is spatially extended over a region of $\sim 30$~pc,
and does not respond to continuum variability. This means that only
those spectral solutions that we obtained in which the zeroth-order
continuum dominates above 10~keV are permissible for NGC~4945. \\
(v) The last inference above also means that the Compton-scattered
continuum must be sufficiently suppressed so that it never dominates
above $\sim 10$~keV, and our modeling quantifies this constraint. Our
results also show that the \fekalfa line and the Compton-scattered
continuum must be dominated by emission from matter illuminated on
the ``back side.'' Thus, one possibility for the structure of the X-ray 
reprocessor is a clumpy medium with a global covering factor
that is small enough to prevent the Compton-scattered continuum
from dominating the high-energy spectrum (the exact value is model-dependent). 
Two separate structures cannot be ruled out, such as
a ring-like (edge-on) compact structure responsible for the line-of-sight
extinction, along with a larger, extended clumpy medium. However,
our modeling shows that the line-of-sight column density
is similar to that out of the line of sight. \\
(vi) The requirement that the spectrum above 10~keV has to be
dominated by the zeroth-order continuum also requires a bolometric
luminosity that is of the order of or greater than the Eddington luminosity.
However, this is based on isotropic intrinsic X-ray emission.
Previous studies have also noted the problematic Eddington ratio, and
indeed, NGC 4945
was excluded from the sample in a study of bolometric corrections
from X-ray and infrared data by Vasudevan \etal (2010). \\
(vii) Another possibility arises if the intrinsic X-ray continuum is beamed
or collimated along our line of sight. This would alleviate the need for
fine-tuning the covering factor of the extended, clumpy medium because 
the Compton-scattered continuum and fluorescent line emission from matter
that is not close to the beaming direction would be relatively weak.
If the individual clumps are much larger than the X-ray source, the
same distribution could provide the line-of-sight extinction. Beaming
also relieves the need for an intrinsic X-ray luminosity that is close
to, or exceeds the Eddington luminosity. The actual intrinsic 
luminosity could easily be an order of magnitude less if the
beaming scenario is correct. The beaming scenario appears to be supported
by the detection of a jet-like structure in the radio band (Lenc \& Tingay 2009),
and the detection of NGC~4945 in GeV band by Fermi (Lenain \etal 2011).

Our results also serve to demonstrate that weak, heavily obscured
AGNs found in deep X-ray surveys may have column densities and
intrinsic luminosities that are different to those inferred from
modeling the spectra with disk-reflection spectra.
In particular, the
X-ray spectrum of a fully-covered
Compton-thick source resembles the spectrum from a Compton-thick
source with a negligible covering factor more than it resembles
a disk-reflection spectrum. Thus, sources that are dominated
by a Compton-scattered (reflection) spectrum might be
mistakenly classified as showing no evidence for reflection,
as testified by a negligible value of $R$.
Given the expected spectral degeneracies, it may not be possible
even in principle to infer the source composition of the cosmic X-ray background
without directly resolving it at high energies. 

Finally, it should be clear from our results that, aside from the complex
dependence of the \fekalfa line flux on $N_{\rm H}$ and geometry
(e.g., Yaqoob \etal 2010), the possible large range in the intrinsic
continuum luminosity for those sources that admit
degenerate models means that the \fekalfa line flux
cannot be used as a simple proxy for the intrinsic continuum luminosity.
Nor can the Fe abundance be trivially deduced from the depth of the Fe~K
edge because the edge depth has a complex dependence on the column density
and geometry of the X-ray reprocessor.

\section{Appendix: Analysis of a Suzaku Observation of 3C~273}
\label{appendix}

In this section we utilize data from a \suzaku observation of 3C 273 in order
to derive cross-normalization factors for the XIS, HXD/PIN, and HXD/GSO
instruments. Although the \suzaku Guest Observer Facility (GOF) has published
XIS:PIN cross-normalization 
ratios\footnote{ftp://legacy.gsfc.nasa.gov/suzaku/doc/xrt/suzakumemo-2008-06.pdf},
these ratios depend on the energy bands of the XIS and HXD data that are used for
fitting, and the actual
spectrum of the source. The given cross-normalizations were derived from
\suzaku spectra of the Crab nebula. The Crab X-ray spectrum is very steep,
with a photon index greater than 2, and the high count rate can induce
pile-up problems in the CCDs. 
In the HXD/PIN and HXD/GSO bands, the detailed 
Crab spectrum is actually unknown and must
be {\it assumed} (all previous X-ray instruments have been calibrated
with the Crab spectrum, based on assumptions about its spectrum).
For \suzakup, the calibration {\it assumes} a broken power-law 
continuum, and {\it assumes} values for the photon indices of the
two continuum components, and the break energy.
Therefore, using instrument cross-normalizations
derived from a ``simple'' point source such as 3C~273 is more appropriate for use
with the analysis of AGN \suzaku spectra, making use of matching energy
bands for fitting the 3C~273 data and the target source.
For the HXD/GSO, the cross-normalization with the XIS  is even more uncertain
than it is for HXD/PIN:XIS. Again, the cross-normalization depends on
the exact energy bands used for spectral fitting, and the spectrum of the
source. The statement by the \suzaku GOF does not provide sufficient
quantitative information, saying only that, 
``there is a cross-normalization problem at the 20\% level (1.0:0.80, PIN:GSO).''
\footnote{http://heasarc.gsfc.nasa.gov/docs/suzaku/analysis/watchout.html}
Although energy-dependent ``ancillary response files'' (ARFs) 
have been released for use with the GSO to empirically compensate for
remaining calibration residuals, these files were derived from fitting
Crab data, and
from fitting over a much broader energy band than would be typical for AGN \suzaku data.
Again, the parameters of the Crab spectrum were {\it assumed}, and the
official \suzaku GOF description of the GSO ARF files states that
the uncertainty in the Crab parameters ``is still being studied.''
\footnote{http://heasarc.gsfc.nasa.gov/docs/suzaku/analysis/gso\_newarf.html}
Therefore, we do not employ the empirical GSO ARF files for either
3C 273 or NGC 4945. Instead, we employ only the default response matrices for the
HXD/PIN and HXD/GSO, and derive the HXD/PIN:XIS and
HXD/GSO:XIS cross-normalization ratios directly
from the 3C 273 data, and then adopt those ratios for fitting the \suzaku NGC 4945 data.

\suzaku observed the bright quasar 3C~273 ($z=0.158$) in the period 2007, June 30
to 2007, July 1 for a duration of $\sim 85$~ks (from which net exposure times
after data screening and cleaning were $\sim 54.9$~ks and $\sim 48.0$~ks for
the XIS and HXD respectively). 
The observation was made in ``HXD nominal'' mode, which refers to one of the
default aimpoints of the telescope and detector system that optimizes
the count rate in the HXD, as opposed to the XIS.
The data were first reprocessed with version
1.01 of the pipeline reprocessing software, {\sc aepipeline},
used in conjunction with the \suzaku {\sc ftools} version 16, which were part
of the general HEAsoft (version 6.9) release
\footnote{http://heasarc.gsfc.nasa.gov/docs/suzaku/analysis/suzaku\_ftools.html}.
The latest calibration data base (CALDB) versions in effect at the time
of data processing and reduction were 
``XIS CALDB 20100709'', ``HXD CALDB 20100812'' and ``XRT CALDB 20080709.''
For the HXD/GSO, the gain calculation and gain history file changed
significantly after 2010, August 20
\footnote{http://heasarc.gsfc.nasa.gov/docs/suzaku/analysis/gso\_newgain.html}, 
and the new implementation was
employed in the current data reprocessing.

Default data screening and cleaning criteria were employed in the pipeline
reprocessing.\footnote{These can be found at
http://heasarc.gsfc.nasa.gov/docs/suzaku/processing/v231225.html.}
The HXD
background files released for the 3C~273 observation
were used in conjunction with the \suzaku {\sc ftools} to generate
background spectra.\footnote{
For the HXD/PIN: {\tt ae702070010\_hxd\_pinbgd.evt}; for the HXD/GSO:
{\tt ae702070010\_hxd\_gsobgd.evt}.} For the HXD/PIN, 
files corresponding to the ``tuned'' version
of the background were used\footnote{http://heasarc.gsfc.nasa.gov/docs/suzaku/analysis/pinbgd.html},
and for the HXD/GSO, version 2.0 of the
background files were employed
\footnote{http://heasarc.gsfc.nasa.gov/docs/suzaku/analysis/gsobgd.html}.

For the XIS, data were available only from XIS0, XIS1, and XIS3 since
XIS2 was already no longer available for the 3C 273 observation.
On-source data were extracted from the screened and cleaned event fits from
a circular region with a radius of 2.60', from each detector.
Background spectra were extracted from several rectangular regions 
that were sufficiently distant from 3C 273 and the calibration
sources. The on-source spectra from the three XIS detectors were
combined into a single spectrum, and the combined background
spectrum from the three XIS detectors was subtracted in order
to obtain the final XIS spectrum of 3C 273. 
The background-subtracted XIS count rate averaged over the
observation and the three detectors is $4.582\pm0.005$ ct/s
in the 0.5--10~keV band.
The XIS background constitutes only 0.9\% and 4.7\% of the total on-source
count rate in the 0.5--10~keV and 8--10~keV bands respectively
(the relative contribution of the
background is highest at the high-energy end of the bandpass).
Spectral fitting was performed over identical energy ranges to those
used for the NGC~4945 analysis in the present paper. These
ranges were 0.7--9.82~keV (XIS, excluding the 
poorly calibrated 1.83--1.93~keV range), 
11.6--47.7~keV (HXD/PIN), and
80--165~keV (HXD/GSO). The background-subtracted count rates in these bands are:
$4.260 \pm 0.005$, $0.436\pm0.004$, and $0.148\pm0.018$ for
the XIS, HXD/PIN, and HXD/GSO respectively. In these energy bands,
3C 273 contributes $99.4\%$, $48.7\%$, and $1.8\%$ of the total
on-source count rate. The source contribution in the GSO is
very small compared to the background, so systematic errors in the
GSO background model (up to 2\% 
\footnote{http://heasarc.gsfc.nasa.gov/docs/suzaku/analysis/gsobgd.html}
of the nominal predicted background count rate) could have a significant
effect on the background-subtracted spectrum. However, we did not
apply any systematic adjustments or systematic errors for the baseline
spectral fitting so that we could examine the true residuals
to the default background model.

For the XIS, one of the default bin sizes, $\sim 30$~eV (512-channel spectrum),
was used, and the default binning for the HXD/PIN (375~eV per bin, 256 channels)
was used. For the GSO, the spectrum must be made with energy bin widths
that match the predicted model background spectrum, and these have a variable,
predetermined width.\footnote{http://heasarc.gsfc.nasa.gov/docs/suzaku/analysis/gso\_newgain.html} 
The GSO spectrum was then further binned by a factor of 2 to
improve the signal-to-noise ratio.

Response matrices
({\sc ``rmf''} files), and telescope effective area files 
(``ARFs'') were made using the \suzaku {\sc ftools} {\sc xisrmfgen} and
{\sc xissimarfgen} respectively. The three ``rmf'' and three ``ARF'' files
were all combined, using the appropriate weighting (according to
the count rates for each XIS), into a single response file for
the combined XIS background-subtracted spectrum of 3C~273.
The HXD response files used were {\tt ae\_hxd\_pinhxnome1\_20080129.rsp}
and {\tt ae\_hxd\_gsohxnom\_20100524.rsp} for the PIN and GSO respectively.

\begin{figure}
\centerline{
        \psfig{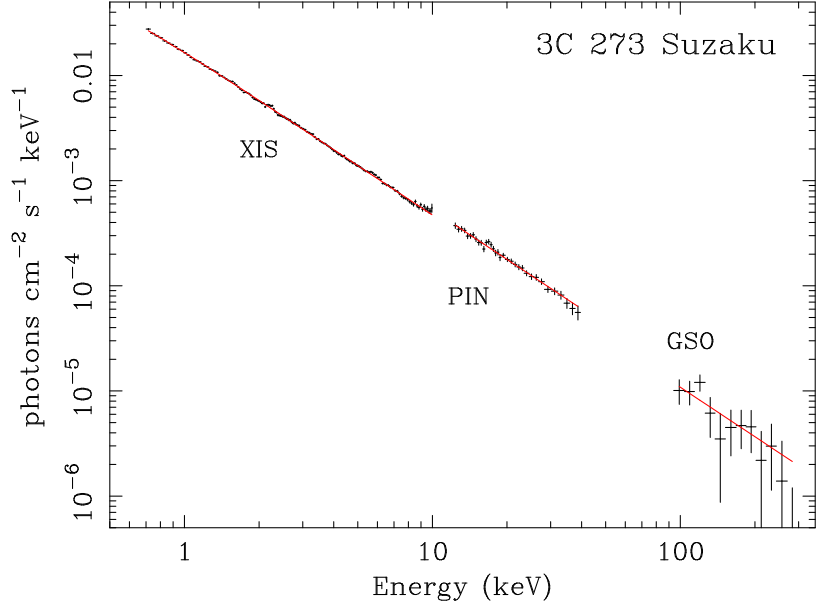}}
\caption{\footnotesize 3C~273 \suzaku unfolded photon spectrum, 
obtained from fitting the data with a
simple power-law continuum model and Galactic absorption, 
with free HXD/PIN:XIS and HXD/GSO:XIS relative normalization ratios
(see text for details). The data from XIS0, 1, and 3 are combined into a single spectrum.
Note that the data were fitted over narrower energy ranges
than shown, in order to match the NGC~4945 energy ranges that were fitted
(see text for details). These energy ranges were 0.7--9.82~keV (XIS, excluding the  
poorly calibrated 1.83--1.93~keV range),  11.6--47.7~keV (HXD/PIN), and
80--165~keV (HXD/GSO). }
\label{fig:qsoufspec}
\end{figure}

The 3C~273 XIS, HXD/PIN and HXD/GSO data were fitted in the energy ranges
mentioned above, with a simple power-law continuum with Galactic absorption.
A column density of $1.79 \times 10^{20} \ \rm cm^{-2}$ was used in
the Galactic absorption model (Dickey \& Lockman 1990). There were a
total of four free parameters: the photon index, $\Gamma$, the overall
normalization, the normalization of the HXD/PIN model relative 
to the XIS model ($C_{\rm PIN:XIS}$), and the normalization of the
HXD/GSO model relative to the XIS ($C_{\rm GSO:XIS}$).
The best-fitting model and unfolded photon spectrum are shown in
Fig.~\ref{fig:qsoufspec}. The data are shown over wider energy bands
for each instrument than were actually fitted in order to show as much
of the highest quality data as possible.

\begin{figure}
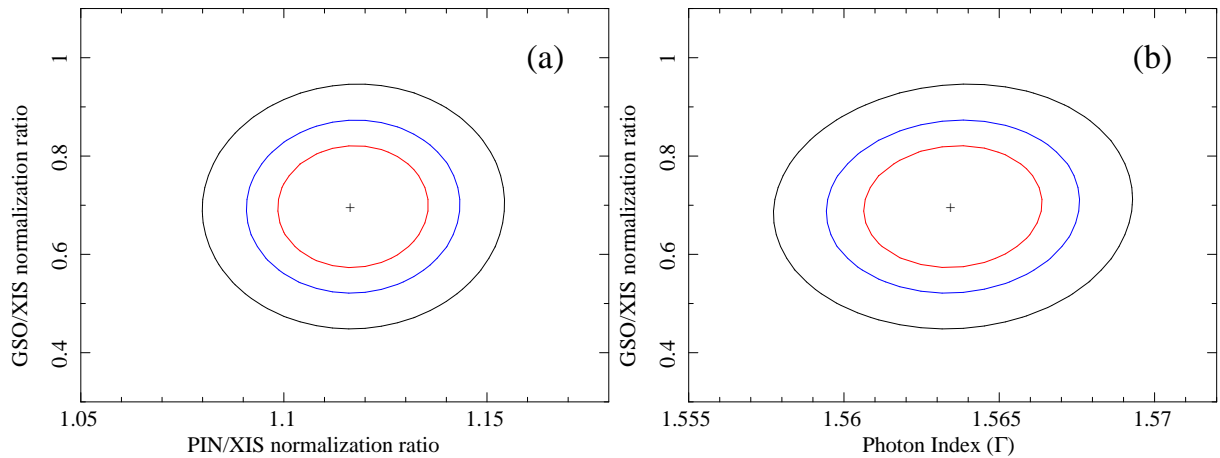

\centerline{
        \psfig{figure=f17a.ps,height=6cm,angle=270}
	\psfig{figure=f17b.ps,height=6cm,angle=270}
	}
        \caption{\footnotesize Contours corresponding to 68\% (red lines), 90\% (blue lines),
and 99\% (black lines) confidence for two interesting parameters, obtained from a simple
power-law, three-instrument fit to 3C~273 
\suzaku data, with Galactic absorption. (a) HXD/GSO:XIS 
versus HXD/PIN:XIS relative 
normalizations ($C_{\rm GSO:XIS}$ and $C_{\rm PIN:XIS}$ respectively). 
(b) HXD/GSO:XIS normalization ($C_{\rm GSO:XIS}$)
versus the power-law photon index, $\Gamma$.}
\label{fig:qsocontours}
\end{figure}

We obtained $\chi^{2} = 508.8$ for 416 degrees of freedom, and
best-fitting parameter values of $C_{\rm PIN:XIS} = 1.12^{+0.02}_{-0.03}$,
$C_{\rm GSO:XIS}=0.70^{+0.14}_{-0.14}$, and $\Gamma=1.564^{+0.003}_{-0.004}$
(errors are 90\% confidence for one interesting parameter).
In Fig.~\ref{fig:qsocontours}(a) we show contours corresponding to 68\%, 90\%,
and 99\% confidence for $C_{\rm GSO:XIS}$ versus $C_{\rm PIN:XIS}$. 
In Fig.~\ref{fig:qsocontours}(b) we show contours for the same confidence levels,
for $C_{\rm GSO:XIS}$ versus $\Gamma$.

\vspace{5mm}
Acknowledgments \\
The author thanks an anonymous referee for diligent reading of the
manuscript and helping to improve the paper.
The author also thanks A. Marinucci and G. Risaliti for very helpful discussions.
The author is indebted to NASA's research programs that have made
this work possible and is grateful for partial support from NASA grants NNG04GB78A and
NNX09AD01G. This research has made use of data and software provided by 
the High Energy Astrophysics Science Archive Research Center (HEASARC), 
which is a service of the Astrophysics Science Division at NASA/GSFC and the 
High Energy Astrophysics Division of the Smithsonian Astrophysical Observatory.

\bsp
\label{lastpage}

\end{document}